\documentclass[preprint,12pt]{elsarticle}

\usepackage{graphicx}
\usepackage{amssymb}
\biboptions{compress}

\journal{Phot.~Nano.~Fund.~Appl.}

\usepackage{amsmath}
\usepackage{url}
\usepackage{color}
\usepackage{tensor}
\usepackage{mathrsfs}
\usepackage{nicefrac}
\usepackage{tikz-cd}
\usepackage{tikz}
\usetikzlibrary{decorations.markings}
\usetikzlibrary{matrix,arrows}
\usepackage{array}
\usepackage[normalem]{ulem}
\usepackage{upgreek}

\renewcommand{\vec}[1]{\boldsymbol{#1}}
\newcommand{\tsr}[1]{\overset\leftrightarrow{#1}}
\newcommand{\ext}{_{\rm ext}}
\newcommand{\tot}{_{\rm tot}}
\newcommand{\ind}{_{\rm ind}}
\newcommand{\itl}{_{\rm int}}
\newcommand{\h}{\hspace{1pt}}
\newcommand{\hh}{\hspace{0.5pt}}
\newcommand{\mh}{\hspace{-1pt}}
\renewcommand{\j}{\mathrm i}
\newcommand{\de}{\mathrm d}
\newcommand{\e}{\mathrm e}

\newcommand{\raisemath}[1]{\mathpalette{\raisemith{#1}}}
\newcommand{\raisemith}[3]{\raisebox{#1}{$#2#3$}}

\newcommand{\EE}{_{\raisemath{-2pt}{EE}}}
\newcommand{\EB}{_{\raisemath{-2pt}{EB}}}
\newcommand{\BE}{_{\raisemath{-2pt}{BE}}}
\newcommand{\BB}{_{\raisemath{-2pt}{BB}}}

\newcommand{\p}{^{\,\mathrm p}}

\newcommand{\lar}[1]{\textnormal{\mbox{\large $#1$}}}

\DeclareMathAlphabet{\mathbbmsl}{U}{bbm}{m}{sl}

\numberwithin{equation}{section}

\begin{document}

\begin{frontmatter}

%% Title, authors and addresses

%% use the tnoteref command within \title for footnotes;
%% use the tnotetext command for the associated footnote;
%% use the fnref command within \author or \address for footnotes;
%% use the fntext command for the associated footnote;
%% use the corref command within \author for corresponding author footnotes;
%% use the cortext command for the associated footnote;
%% use the ead command for the email address,
%% and the form \ead[url] for the home page:
%%
%% \title{Title\tnoteref{label1}}
%% \tnotetext[label1]{}
%% \author{Name\corref{cor1}\fnref{label2}}
%% \ead{email address}
%% \ead[url]{home page}
%% \fntext[label2]{}
%% \cortext[cor1]{}
%% \address{Address\fnref{label3}}
%% \fntext[label3]{}

\title{Functional Approach to Electrodynamics of Media}

%% use optional labels to link authors explicitly to addresses:
%% \author[label1,label2]{<author name>}
%% \address[label1]{<address>}
%% \address[label2]{<address>}

\author[vienna,freiberg]{R.~Starke}
\ead{Ronald.Starke@physik.tu-freiberg.de}

\author[heidelberg]{G.A.H.~Schober\corref{cor1}}
\ead{G.Schober@thphys.uni-heidelberg.de}

\cortext[cor1]{Corresponding author.}

\address[vienna]{Department of Computational Materials Physics, University of Vienna, \\ Sensengasse 8/12, 1090 Vienna, Austria}
\address[freiberg]{Institute for Theoretical Physics, TU Bergakademie Freiberg, Leipziger Stra\ss e 23, \\ 09599 Freiberg, Germany}
\address[heidelberg]{Institute for Theoretical Physics, University of Heidelberg, Philosophenweg 19, \\ 69120 Heidelberg, Germany  \vspace{-10pt}}

\begin{abstract}
In this article, we put forward a new approach to electrodynamics of materials. Based on the identification of induced electromagnetic fields as the microscopic counterparts of polarization and magnetization, we systematically employ the mutual functional dependencies of induced, external and total field quantities. This allows for a unified, relativistic description of the electromagnetic response without assuming the material to be composed of electric or magnetic dipoles. Using this approach, we derive universal (material-independent) relations between electromagnetic response functions such as the dielectric tensor, the magnetic susceptibility and the microscopic conductivity tensor. Our formulae can be reduced to well-known identities in special cases, but more generally include the effects of inhomogeneity, anisotropy, magnetoelectric coupling and relativistic retardation. If combined with the Kubo formalism, they would also lend themselves to the ab initio calculation of all linear electromagnetic 
response functions. \bigskip
\end{abstract}

\begin{keyword}
%% keywords here, in the form: keyword \sep keyword
Polarization \sep Magnetization \sep First-principles calculation \sep Bi-anisotropy \sep Multiferroics \sep Relativity

%% MSC codes here, in the form: \MSC code \sep code
%% or \MSC[2008] code \sep code (2000 is the default)

\end{keyword}

\end{frontmatter}

%%
%% Start line numbering here if you want
%%
% \linenumbers

%% main text

\newpage
\ \vspace{-2.4cm}
\tableofcontents

\section{Introduction} \label{sec_intro}

{\itshape Standard approach and its limitations.}---Electrodynamics of media aims at describing the response of a material probe to external electromagnetic perturbations \cite{Fliessbach, Landau, Jackson, Griffiths}. For this purpose, Maxwell's equations alone are not enough, because these relate only the electromagnetic fields to their sources, but do not describe the reaction of currents and charges in the material to the electromagnetic fields. Therefore, the usual approach to electrodynamics of media is to introduce additional field quantities besides the electric field~$\vec E$ and the magnetic induction~$\vec B$. These additional fields are the polarization~$\vec P$ and the magnetization~$\vec M$, as well as the displacement field~$\vec D$ and the magnetic field~$\vec H$. They are interrelated by
\begin{align}
 \vec D & = \varepsilon_0 \vec E + \vec P , \label{eq_fields_1} \\[3pt]
 \vec H & = \frac{1}{\mu_0} \vec B - \vec M , \label{eq_fields_2}
\end{align}
where $\varepsilon_0$ and $\mu_0$ denote the vacuum permittivity and permeability respectively. The fields $\vec P$ and $\vec M$ are interpreted as electric and magnetic dipole densities inside the material. The so-called {\itshape macroscopic Maxwell equations} are then written in the following form:
\begin{align}
 \nabla \cdot \vec D & = \rho\ext \,, \label{eq_maxwell_1} \\[5pt]
 \nabla \times \vec H - \frac{\partial \vec D}{\partial t} & = \vec j\ext \,, \label{eq_maxwell_2} \\[5pt]
 \nabla \cdot \vec B & = 0 \,, \label{eq_maxwell_3} \\[5pt]
 \nabla \times \vec E + \frac{\partial \vec B}{\partial t} & = 0 \,, \label{eq_maxwell_4}
\end{align}

\vspace{2pt} \noindent
where $\rho\ext$ and $\vec j\ext$ are the external charge and current densities. Traditionally, these were identified with the free sources as opposed to those bound in the medium \cite{Jackson, Griffiths}. In modern treatments, however, the sources are split into external and induced contributions~\cite{Melrose, Martin, Ashcroft, Kittel, Bruus, Giuliani, SchafWegener}; see in particular the remarks in Ref.~\cite[pp.~261~f.]{Fliessbach}. The general equations \eqref{eq_maxwell_1}--\eqref{eq_maxwell_4} are complemented by material-specific {\itshape constitutive relations}, which in the simplest case read
\begin{align}
 \vec P & = \varepsilon_0 \h \chi_{\rm e} \h \vec E , \label{eq_con_1} \\[3pt]
 \vec M & = \chi_{\rm m} \h \vec H , \label{eq_con_2}
\end{align}
with electric and magnetic susceptibilities $\chi_{\rm e}$ and $\chi_{\rm m}$. Taken together, the equations \eqref{eq_fields_1}--\eqref{eq_con_2} can be used to determine the electric and magnetic fields inside the material medium in terms of the external sources.  They are usually supposed to hold on a macroscopic scale, i.e., $\vec E$ and $\vec B$ represent suitable averages of the microscopic fields, while $\vec P$ and $\vec M$ are regarded as average dipole moments of macroscopic volume elements \cite{Fliessbach, Landau, Jackson, Griffiths, Wachter}. We will henceforth refer to this approach as the {\itshape standard approach} to electrodynamics of media. Traditional textbooks (such as \cite{Jackson, Griffiths}) set up this approach and derive
Eqs.~\eqref{eq_maxwell_1}--\eqref{eq_maxwell_4} based on the 
assumption of microscopic dipoles induced by the electromagnetic fields, as for example in the Clausius--Mossotti model \cite{Mossotti}. Such simplified models have been ubiquitous in the description of the electromagnetic response of solids for more than a century since Maxwell's original work in the eighteen sixties~\cite{Maxwell65, Maxwell73}.

However, two modern developments in solid-state physics have put severe limitations on the standard approach: the quest for an  ab initio description of material properties \cite{Martin, Hafner08, Hafner10, Bluegel} and the discovery of new materials with exotic electromagnetic responses. As for the first point, the Modern Theory of Polarization \cite{Resta10, Resta07, Vanderbilt} has raised conceptual questions about the interpretation of $\vec P$ and $\vec M$ as electric and magnetic dipole densities. It has been shown that the localized polarizable units of the Clausius-Mossotti model are in stark contrast to the actual delocalized electronic charge distributions of real materials, and hence this model fails in most cases to describe polarization effects adequately \cite{Resta07}. More fundamentally, it was argued that the polarization of a crystalline solid cannot, even in principle, be defined as a bulk quantity in terms of periodic electronic 
charge 
distributions, and instead the polarization {\itshape change} in a typical measurement setup was defined in terms of a macroscopic charge flow in the interior of the sample \cite{Resta10, Resta07}. Regarding the second point, with the advent of metamaterials \cite{Sihvola, Zheludev, Capolino} and in particular bi-anisotropic media with magnetoelectric coupling \cite{Marques, Li, Wegener}, it became increasingly important to generalize the constitutive relations \eqref{eq_con_1}--\eqref{eq_con_2} and to reformulate the theory of electromagnetic responses in a more systematic way \cite{Cho08, Cho10}. Since often the properties of these materials are determined by nanoscale resonant inclusions (``meta-atoms'') \cite{Cube}, macroscopic averaging procedures have also come under intense investigation \cite{Alu, Chipouline, Vinogradov}. Further systems with magnetoelectric coupling include the prototype material Cr$_2$O$_3$ \cite{Dzyaloshinskii, Fechner, Fechner14}, single-phase multiferroics \cite{Spaldin, Picozzi,
 Tokura, Mochizuki}, 
composite 
multiferroics and multiferroic interfaces \cite{Wang, Rondinelli, Fechner10, FechnerPRL} as well as systems with Rashba spin-orbit coupling \cite{Barnes, Ojanen, Duckheim}. Topological insulators show the topological magnetoelectric effect~\cite{Zhang, Essin, Essin2}, which becomes observable if the surface response is separated from the bulk contribution by breaking the time-reversal symmetry \cite{Gregor, Mulligan}, for example, through the introduction of ferromagnetism \cite{Liu, Abanin, Nomura}.

\bigskip \noindent
{\itshape Functional Approach.}---To overcome the limitations of the standard approach, we will develop in this paper the Functional Approach to electrodynamics of media. This approach identifies induced electromagnetic fields as the microscopic counterparts of macroscopic polarizations and interprets them as functionals of the external perturbations. 
Thus, we start from the following identifications  {\itshape on a microscopic scale}:\,\footnote{The analogy between $\vec P(\vec x)$, $\vec M(\vec x)$ 
and electromagnetic fields generated by internal sources has been called ``a deceptive parallel'' in Ref.~\cite[Sec.~4.3.2, 6.3.2]{Griffiths}, because
it implies in particular that $\nabla \cdot \vec M(\vec x)=0$. The bar magnet with 
uniform magnetization parallel to its axis is considered an example where this condition does not hold. There, the magnetization is assumed to be constant inside a finite cylinder and zero outside, and hence a nonvanishing 
divergence appears at the top and bottom surfaces. However, the so defined field~$\vec M(\vec x)$ is not obser\-{}vable itself and indeed only serves to determine by $\nabla \times \vec M(\vec x) = \vec j(\vec x)$ the macroscopic current, which is localized at the lateral surface of the cylinder. Any curl-free vector field can be added to $\vec M(\vec x)$ without changing this information, fact which has been referred to as ``gauge freedom'' of the magnetization in Refs.~\cite{Smith, Hirst, Chatel}. A careful examination of the procedure described in Ref.~\cite{Griffiths} shows that the real, observable magnetic field is given in terms of the current by $\nabla \times \vec B(\vec x) = \mu_0 \h \vec j(\vec x)$ and $\nabla \cdot \vec B(\vec x) = 0$, or in terms of the uniform magnetization by the transverse part $\vec B(\vec x) = \mu_0 \vec M_{\mathrm T} (\vec x)$. Similarly, in electrostatics the observable electric field is given in terms of the uniform polarization of a macroscopic body by the longitudinal 
part
$\vec E(\vec x) = -\vec P_{\mathrm L} (\vec x)/\varepsilon_0$. Such problems actually do not concern electrodynamics of media but the determination of electromagnetic fields as generated by macroscopic surface currents or charges. Electrodynamics of media concerns the question of how these surface charges or currents can be induced by external fields.}
\begin{align}
 \vec P(\vec x, t) & = -\varepsilon_0\vec E\ind(\vec x, t) \,, \label{eq_PE} \\[2pt]
 \vec D(\vec x, t) & = \varepsilon_0\vec E\ext(\vec x, t) \,, \label{eq_DE} \\[2pt]
 \vec E(\vec x, t) & = \vec E\tot(\vec x, t) \,, \label{eq_EE}
\end{align}

\noindent
and \vspace{-2pt}
\begin{align}
 \vec M(\vec x, t) & = \vec B\ind(\vec x, t)/\mu_0 \,, \label{eq_MB} \\[2pt]
 \vec H(\vec x, t) & = \vec B\ext(\vec x, t)/\mu_0 \,, \label{eq_HB} \\[2pt]
 \vec B(\vec x, t) & = \vec B\tot(\vec x, t) \,. \label{eq_BB}
\end{align}

\noindent
These identifications are common practice in semiconductor physics (cf.~\cite[p.~33, footnote 14]{SchafWegener}) and in electronic structure physics (cf.~\cite[Appendix A.2]{Kaxiras}). The quantities on the right hand side of Eqs.~\eqref{eq_PE}--\eqref{eq_BB} refer to the induced, the external and the total microscopic fields, 
respectively (see Sec.~\ref{sec_func} and cf.~\cite{Fliessbach, Smith}). By {\itshape microscopic} fields we mean that these are 
derived from microscopic charge and current
distributions, which in turn are derived from continuous quantum
(many-body) wave functions. Explicitly, the induced fields are related to the induced charges 
and currents by the microscopic Maxwell equations
\begin{align}
 -\nabla \cdot \vec P(\vec x, t) & = \rho_{\rm ind}(\vec x, t) \,, \label{eq_maxwell_induced_1} \\[5pt]
 \nabla \times \vec M(\vec x, t) + \frac{\partial}{\partial t}\vec P(\vec x, t) & = \vec j_{\rm ind}(\vec x, t) \,, \label{eq_maxwell_induced_2} \\[5pt]
 \nabla \cdot \vec M(\vec x, t) & = 0 \,, \label{eq_maxwell_induced_3} \\[5pt]
 -\nabla \times \vec P(\vec x, t) + \frac{1}{c^2} \frac{\partial}{\partial t} \vec M(\vec x, t) & = 0 \,. \label{eq_maxwell_induced_4}
\end{align}
Analogously, the external fields are related to the external sources by the following equations (cf. \cite[p.~4]{Melrose}):
\begin{align}
 \nabla \cdot \vec D(\vec x, t) & = \rho_{\rm ext}(\vec x, t) \,, \\[5pt]
 \nabla \times \vec H(\vec x, t) - \frac{\partial}{\partial t}\vec D(\vec x, t) & = \vec j_{\rm ext}(\vec x, t) \,, \\[5pt]
 \nabla \cdot \vec H(\vec x, t) & = 0 \,, \\[5pt]
 \nabla \times \vec D(\vec x, t) + \frac{1}{c^2} \frac{\partial}{\partial t} \vec H(\vec x, t) & = 0 \,.
\end{align}

\vspace{2pt} \noindent
External and induced sources are not necessarily
located ``outside'' or ``inside'' the material probe respectively. 
Instead, this distinction implies that the induced fields are generated by the degrees of freedom of the medium, whereas the external fields are associated with degrees of freedom which do not belong to the medium. In particular, this applies to external sources which can be \linebreak controlled experimentally \cite{Smith}. 
Thus, we use the term {\itshape external fields} in precisely the same sense as in classical or quantum mechanics, 
where one considers particles moving in external fields as opposed to particles moving in the fields generated by themselves.

In view of the above identifications, Eqs.~\eqref{eq_fields_1}--\eqref{eq_fields_2} hold as exact identities in our approach simply by the definition of the total fields as the sum of the external and the induced fields. (This is in contrast to other approaches such as the multipole theory \cite{Raab}, where Eqs.~\eqref{eq_fields_1}--\eqref{eq_fields_2} are regarded as first order approximations.) Moreover, Eqs.~\eqref{eq_maxwell_1}--\eqref{eq_maxwell_2} are none other than the inhomogeneous Maxwell equations relating the external fields to the external sources, while Eqs.~\eqref{eq_maxwell_3}--\eqref{eq_maxwell_4} are the homogeneous Maxwell equations for the total fields (cf.~\cite[Eqs.~(2.124)]{SchafWegener}). 
In particular, these equations hold in general on a microscopic scale, and they do not need to be derived as in the standard approach (based on induced electric and magnetic dipoles, cf.~\cite{Jackson, Griffiths}). Macroscopically averaged fields defined by
\begin{equation} \label{eq_average}
 \langle \vec E \rangle(\vec x, t) = \int \! \mathrm d^3 \vec x' \, f(\vec x - \vec x') \vec E(\vec x', t)
\end{equation}
(with $f(\vec x)$ being smooth, localized at $\vec x = 0$ and ranging over distances large compared to atomic dimensions in the material) satisfy again Maxwell equations with the averaged sources, 
because the averaging procedure commutes with the partial derivatives \cite{Wachter}. In particular, this implies that any fundamental relation between the microscopic
electromagnetic fields which is derived from the Maxwell equations will hold for the macroscopically averaged fields as well, a conclusion which applies especially to the universal response relations derived in Sec.~\ref{sec_univ}.

Our main focus in this article is on the microscopic electromagnetic response functions, which are represented by functional derivatives
of induced
electric and magnetic fields with respect to external perturbations. It is well-known that 
these are not independent of each other, and hence it is possible to formulate electrodynamics of media as a {\itshape single susceptibility theory} \cite{Cho10}, where all linear electromagnetic response functions are derived from a single response tensor. For example, in magneto-optics it has been suggested to introduce an effective permittivity tensor relating the $\vec D$ and $\vec E$ fields while setting $\vec H = \vec B/\mu_0$ identically \cite{Prokhorov, Agranovich04, Agranovich09}. 
By contrast, here we stick to the usual definition of electromagnetic response functions and relate each of them to the microscopic conductivity tensor, a possibility which has already been mentioned in Ref.~\cite{Melrose} (see also \cite{Smith, Buhmann12}). Starting from the model-independent definitions \eqref{eq_PE}--\eqref{eq_BB}, we will thereby obtain {\itshape universal response relations}, which can be used to study the electromagnetic response not only of solids, but also liquids, single atoms or 
molecules, or even the vacuum of quantum electrodynamics (see Sec.~\ref{subsec_hedin}).

\bigskip \noindent
{\itshape Comparison to the literature.}---Our main motivation for the identifications \eqref{eq_PE}--\eqref{eq_BB} comes from microscopic condensed matter physics in general
and ab initio electronic structure theory in particular \cite{Martin, Ashcroft, Kittel, Bruus, Giuliani, Botti}. There, it is common practice to introduce a microscopic dielectric function (or permittivity) $\varepsilon_{\mathrm r}=\varepsilon_{\mathrm r}(\vec x,t;\vec x',t')$ by means of the defining equation
\begin{equation} \label{eq_defeps}
\varphi\tot=\varepsilon_{\mathrm r}^{-1} \h \varphi\ext \,.
\end{equation}
Here, $\varphi\tot$ and $\varphi\ext$ denote the total and external scalar potentials respectively, and the products
refer to non-local integrations in space and time. The total potential is in turn defined as the sum
\begin{equation}
\varphi\tot=\varphi\ext+\varphi\ind \,. 
\end{equation}
It has been noted already (see, e.g., \cite[Sec.~6.4]{Bruus} and \cite[Sec.~4.3.2]{MartinRothen}) that these definitions are {\it analogous} to
electrodynamics of media provided that one uses the identifications \eqref{eq_DE} and \eqref{eq_EE}. The Functional Approach to electrodynamics
of media promotes these analogies to {\it definitions} of the microscopic fields $\vec D(\vec x, t)$ and $\vec E(\vec x, t)$, and generalizes them via the further identifications \eqref{eq_PE} and \eqref{eq_MB}--\eqref{eq_BB}. Consequently, the microscopic dielectric tensor will be defined by (see~Sec.~\ref{sec_univ})
\begin{equation}
 \vec E\tot = (\tsr{\varepsilon_{\mathrm r}}\h)^{-1} \vec E\ext \,,
\end{equation}
which obviously generalizes \eqref{eq_defeps}. We further note that in 
the textbook of Flie\ss bach \cite{Fliessbach}, Eqs.~\eqref{eq_PE}--\eqref{eq_BB} are deduced for the macroscopically averaged fields in the special case of homogeneous, isotropic media,  while in our approach these equations represent the most general definition of microscopic (electric and magnetic) polarizations.

Microscopic approaches to electrodynamics of media have already been described by Hirst \cite{Hirst} and in the so-called premetric approach by Truesdell and Toupin \cite{Truesdell} as well as Hehl and Obukhov \cite{Hehl1, Hehl2} (see~also the textbook by Kovetz \cite{Kovetz}). While we have in common the microscopic interpretation of Maxwell's equations in media, the Functional Approach differs from the above approaches in other respects: (i) By distinguishing between induced and external (instead of bound and free) charges and currents, it is independent of any assumption about the medium. (ii) While the premetric approach distinguishes between ``excitations'' ($\vec D$ and $\vec H$) and ``field strengths'' ($\vec E$ and $\vec B$) \cite{Hehl_depl}, we only consider electromagnetic fields produced by different (induced, external or total) sources. Besides the electromagnetic fields, there are no further fields in our approach, which would be of a physically or mathematically different nature, or which 
would transform differently under coordinate changes. (iii) We provide by Eqs.~\eqref{eq_PE} and \eqref{eq_MB} a {\itshape unique} definition of the microscopic polarization and magnetization without referring to the constitutive relations of the material. Therefore, the Functional Approach is also suitable for deriving universal response relations, as we will show in Sec.~\ref{sec_univ}. (iv) While in the premetric approach the linear electromagnetic response is described by a
fourth-rank constitutive tensor with 36 components \cite{Hehl1, Hehltensor}, we rely on the functional dependence of the induced current on the external vector potential (see Sec.~\ref{subsec_fund}). In fact, we will show that the 9 spatial components of the current response tensor are sufficient to describe the linear response of any material. Moreover, we will derive in Sec.~\ref{subsec_fsrt} \h a closed expression for the constitutive tensor in terms of the current response tensor.

Finally, we note that our approach is consistent with the Modern Theory of Polarization \cite{Resta10, Resta07, Vanderbilt}. This approach relies on a fundamental equation for the {\itshape change} in macroscopic polarization in terms of a transient current through the sample given by (cf.~also \cite[Eq.~(22.4)]{Martin} and \cite{Keldysh, Kootstra, Berger})
\begin{equation} \label{eq_MTP}
 \vec P(\Delta t)-\vec P(0) = \int_0^{\Delta t} \mh \mathrm d t \,\h \vec j\ind(t) \,,
\end{equation}
where
\begin{equation}
\vec j\ind(t) = \frac 1 V \int \! \mathrm d^3 \vec x \,\h \vec j\ind(\vec x, t) \,, \smallskip
\end{equation}
and $V$ denotes the volume of the sample. Our identification of induced electromagnetic fields with electric and magnetic polarizations is indeed consistent with this approach, 
because Eq.~\eqref{eq_MTP} can be derived classically from the Maxwell equations \eqref{eq_maxwell_induced_1}--\eqref{eq_maxwell_induced_4} for the microscopic induced 
fields: First, as the Modern Theory of Polarization focuses on the change of polarization due to {\itshape charge transport,} we may restrict attention to the longitudinal part 
of the current. Differentiating Gauss' law for the induced electric field with respect to time and using the continuity equation yields
\begin{equation}
 -\nabla \cdot \frac{\partial}{\partial t} \h \vec P(\vec x, t) = \frac{\partial}{\partial t} \, \rho\ind(\vec x, t) = -\nabla \cdot (\vec j\ind)_{\mathrm L}(\vec x, t) \,.
\end{equation}
This shows directly the identity of the longitudinal parts
\begin{equation}
\frac{\partial}{\partial t} \h \vec P_{\mathrm L}(\vec x,t) = (\vec j\ind)_{\mathrm L}(\vec x,t) \,,
\end{equation}
from which one concludes Eq.~\eqref{eq_MTP} by spatial integration over the sample volume.
The importance of the formula \eqref{eq_MTP} lies in the fact that it can be reexpressed---assuming a periodic current distribution in a crystalline solid---through 
the current of a single unit cell~\cite{Resta07},
\begin{equation}
\vec j\ind(t) = \frac{1}{V_{\rm cell}} \int_{\rm cell} \! \mathrm d^3 \vec x \,\h \vec j\ind(\vec x, t) \,.
\end{equation}
Hence, the polarization change can be defined as a bulk quantity, which in turn can be inferred from the knowledge of lattice-periodic Bloch wave functions. Therefore, this quantity is also accessible from the results of modern ab initio computer simulations \cite{VASP, Wien2k, QE-2009}.

\bigskip \noindent
{\itshape Organization of the paper.}---We start in Sec.~\ref{sec_pre} by reviewing some aspects of classical electrodynamics which are necessary for this paper.

In Sec.~\ref{sec_green}, we analyze the initial value problem for both the scalar wave equation and the Maxwell equations. Subsequently, we construct the most general form of the tensorial electromagnetic Green function and discuss its special forms which correspond to special gauge conditions.

Using the tensorial Green function, we express in Sec.~\ref{sec_CanFun} the electromagnetic vector potential as a functional of the electric and magnetic fields. By the help of this {\it canonical functional}, 
we then define the notion of {\it total functional derivatives} with respect to electric or magnetic fields. Sections \ref{sec_green}--\ref{sec_CanFun} are technical in nature, but the resulting formulae for the vector potential are of central importance for our formulation of electrodynamics of media.

In Sec.~\ref{sec_func}, we introduce the general electromagnetic response formalism based on the functional dependence 
of the induced four-current on the external four-potential. We propose alternative response functionals and thereby establish the connection to the Schwinger--Dyson equations in quantum electrodynamics and the Hedin equations in electronic structure theory. Furthermore, we derive a closed expression for the fourth rank constitutive tensor used in the premetric approach by Hehl and Obhukov \cite{Hehl1, Hehltensor} in terms of the fundamental response functions.

In Sec.~\ref{sec_univ}, we first argue that physical response functions have to be identified with total functional derivatives.
Based on this, we derive universal (i.e.~model- and material-independent) analytical expressions for the microscopic dielectric tensor, the magnetic susceptibility and the magnetoelectric coupling coefficients in terms of the optical conductivity. We further discuss the intricate problem of expanding the induced electric and magnetic fields in terms of the external fields.

In the closing Sec.~\ref{sec_emp}, we investigate the various empirical limiting cases of our universal response relations. In particular, we show that these reduce to well-known identities in a non-relativistic limit and in the special case of homogeneous, isotropic materials.

\section{Classical electrodynamics} \label{sec_pre}

\subsection{Notations and conventions} \label{sec_pre_notations}

{\itshape Metric tensor.}---For the Minkowski metric we choose 
 \begin{equation} \label{eq_metric}
 \eta_{\mu\nu} = \eta^{\mu\nu} = \mathrm{diag}(-1,\,1,\,1,\,1) \,. 
 \end{equation}
 This convention is particularly useful in condensed matter physics, as one can directly read out a spatial three-tensor $\chi_{ij}$ from a Minkowski four-tensor~$\chi\indices{^\mu_\nu}$ 
without distinguishing between covariant and contravariant indices (i.e. $\chi_{ij} = \chi\indices{^i_j}$\h). Consequently, all spatial tensors will be written
with covariant (lower) indices. Furthermore, we adopt the convention of summing over all doubly appearing indices (even if both are lower case).

\bigskip \noindent
{\itshape Fourier transformation.}---With $x^\mu = (ct, \vec x)^{\mathrm T}$ and $k_\mu =\eta_{\mu\nu}k^\nu= (-\omega/c, \, \vec k)$, we have
\begin{equation}
k x \, \equiv \, k_\mu x^\mu \, = \, -\omega t + \vec k \! \cdot \! \vec x \,. \medskip
\end{equation}
We define the Fourier transform of a field quantity $\rho(x) = \rho(\vec x, t)$ and its inverse as
\begin{align}
\rho(\vec k, \omega) & = \int\! \frac{\mathrm d^3 \vec x}{(2\pi)^{\nicefrac 3 2}} \int \! \frac{c \, \de t}{(2\pi)^{\nicefrac 1 2}} \, \h \rho(\vec x, t) \, \e^{\j \omega t - \j \vec k \cdot \vec x} \,, \label{eq_ftrho} \\[5pt]
\rho(\vec x, t) & = \int \! \frac{\mathrm d^3 \vec k}{(2\pi)^{\nicefrac 3 2}} \int \! \frac{\de \omega}{c \, (2\pi)^{\nicefrac 1 2}} \, \h \rho(\vec k, \omega) \, \e^{-\j \omega t + \j \vec k \cdot \vec x} \,.
\end{align}
With the relativistic volume elements \,$\de^4 x = \de^3 \vec x \, \de x^0$ \h and \h$\de^4 k = \de^3 \vec k \, \de k^0$, we can write these equations equivalently as
\begin{align}
\rho(k) & = \int \! \frac{\de^4 x}{(2\pi)^2} \, \h \rho(x) \, \e^{-\j k x} \,, \\[5pt]
\rho(x) & = \int \! \frac{\de^4 k}{(2\pi)^2} \, \h \rho(k) \, \e^{\j k x} \,.
\end{align}
In particular, under Fourier transformation $\partial_\mu$ is mapped to $\j k_\mu$\hh. 
For a response relation as 
\begin{equation}
 \rho(x) = \int \! \de^4 x' \, \chi(x, x') \h \varphi(x') \,, \smallskip
 \end{equation}
we require ``covariance under Fourier transformation'', i.e., the Fourier transformed quantities should obey the analogous relation
\begin{equation} \label{eq_ft}
 \rho(k) = \int \! \de^4 k' \, \chi(k, k') \h \varphi(k') \,.
\end{equation}
This implies that the response function transforms as
\begin{align}
 \chi(k, k') & = \int \! \frac{\de^4 x}{(2\pi)^2} \int \! \frac{\de^4 x'}{(2\pi)^2} \, \h \e^{-\j k x} \, \chi(x, x') \, \e^{\j k' x'} \,, \\[5pt]
 \chi(x, x') & = \int \! \frac{\de^4 k}{(2\pi)^2} \int \! \frac{\de^4 k'}{(2\pi)^2} \, \h \e^{\j k x} \, \chi(k, k') \, \e^{-\j k' x'} \,.
\end{align}
Our conventions are in accord with a functional chain rule,
\begin{align}
 \chi(k, k') = \frac{\delta\rho(k)}{\delta\varphi(k')}
 & = \int \! \de^4 x \int \! \de^4 x' \ \frac{\delta \rho(k)}{\delta \rho(x)} \, \frac{\delta \rho(x)}{\delta \varphi(x')} \, \frac{\delta \varphi(x')}{\delta \varphi(k')} \\[5pt]
 & = \int \! \de^4 x \int \! \de^4 x' \ \frac{\e^{-\j k x}}{(2\pi)^2} \, \chi(x, x') \, \frac{\e^{\j k' x'}}{(2\pi)^2} \,.
\end{align}
The above general formulae may simplify on physical grounds: we always assume {\itshape homogeneity in time} 
and sometimes also {\itshape homogeneity in space}. The response function can then be written as
\begin{equation}
\chi(x, x') = \chi(\vec x-\vec x', t-t') \,,
\end{equation}
or in the Fourier domain
\begin{equation}
\chi(k, k') = \chi(\vec k,\omega) \, \delta^3(\vec k-\vec k') \, \delta(\omega/c - \omega'/c) \,. \smallskip
\end{equation}
With $\vec r = \vec x - \vec x'$ and $\tau = t - t'$, the Fourier transformation then reads
\begin{align}
\chi(\vec k, \omega) & = \int \! \de^3 \vec r \int \! c \, \de \tau \,\h \chi(\vec r,\tau) \, \e^{{\rm i}\omega\tau-{\rm i}\vec k\cdot\vec r} \,, \label{eq_ftchi} \\[5pt]
\chi(\vec r, \tau) & = \int \! \frac{\de^3 \vec k}{(2\pi)^3} \int \frac{\de\omega}{2\pi c} \,\h  \chi(\vec k, \omega) \, \e^{-{\rm i}\omega\tau+{\rm i}\vec k\cdot\vec r} \,. \label{eq_ftchiinv}
\end{align}
In particular, we note that the response function $\chi(\vec k,\omega)$ does not Fourier transform in the same way as the field quantity $\rho(\vec k,\omega)$ (compare Eqs.~\eqref{eq_ftrho} and \eqref{eq_ftchi}). 
Furthermore, the relation
\begin{equation}
\rho(\vec x, t) = \int \! \de^3 \vec x' \! \int \! c \, \de t' \, \chi(\vec x - \vec x', t - t') \h \varphi(\vec x',t')
\end{equation}
is equivalent to
\begin{equation} \label{eq_homresp}
\rho(\vec k,\omega) = \chi(\vec k,\omega) \h \varphi(\vec k,\omega) \,, \smallskip
\end{equation}
which shows that the Fourier covariance is lost if one works with the reduced transform $\chi(\vec k, \omega)$.
Finally, we note that in the limit $\omega \to 0$, \,$\chi(\vec k, 0)$ relates the {\itshape static} parts $\rho(\vec k, 0)$ and $\varphi(\vec k, 0)$ of the respective field quantities, whereas the response function itself
\begin{equation}
\chi(\vec x - \vec x', t - t') = \delta(c \h t - c \h t')\int \! \frac{\de^3\vec k}{(2\pi)^3} \, \chi(\vec k,0) \, \e^{{\rm i}\vec k\cdot(\vec x-\vec x')}
\end{equation}
is not static (time-independent) but {\itshape instantaneous}.

\bigskip \noindent
{\itshape Projections and rotations.}---Longitudinal and transverse projection operators are defined in Fourier space by
\begin{align}
 (P_{\mathrm L} )_{ij}(\vec k) & = \frac{k_i \hh k_j}{|\vec k|^2} \,, \label{eq_defPL} \\[3pt]
 (P_{\mathrm T} )_{ij}(\vec k) & = \delta_{ij} - \frac{k_i \hh k_j}{|\vec k|^2} \,. \label{eq_defPT}
\end{align}
Any vector field $\vec E(\vec k)$ can be decomposed into its longitudinal and transverse parts \vspace{-5pt}
\begin{equation}\label{decompFS}
 \vec E(\vec k) = \tsr P_{\mathrm L} (\vec k) \h \vec E(\vec k) + \tsr P_{\mathrm T}(\vec k) \h \vec E(\vec k) \equiv \vec E_{\mathrm L} (\vec k) + \vec E_{\mathrm T}(\vec k) \,,
\end{equation}
where
\begin{equation}
\vec E_{\mathrm L} (\vec k) = \frac{\vec k \h (\vec k \cdot \vec E(\vec k))}{|\vec k|^2} \,,
\end{equation}
and
\begin{equation}
\vec E_{\mathrm T}(\vec k) = \frac{|\vec k|^2 \vec E(\vec k) - \vec k (\vec k \cdot \vec E(\vec k))}{|\vec k|^2} = -\frac{\vec k \times (\vec k \times \vec E(\vec k))}{|\vec k|^2} \,.
\end{equation}
These two parts are orthogonal with respect to the euclidean inner product,
\begin{equation}
\vec E^*_{\mathrm L} (\vec k) \cdot \vec E_{\mathrm T}(\vec k) = 0 \,.
\end{equation}
In real space, transverse vector fields are divergence free, $\nabla \cdot \vec E_{\mathrm T}(\vec x) = 0$, 
whereas longitudinal vector fields can be represented as the gradient of a scalar function, $\vec E_{\mathrm L} (\vec x) = -\nabla \varphi(\vec x)$.
They are orthogonal with respect to the inner product
\begin{equation}
\int \! \de^3\vec x \, \vec E_{\mathrm L}(\vec x)\cdot\vec E_{\mathrm T}(\vec x)=0 \smallskip
\end{equation}
(where we assume that $\vec E(\vec x)$ is real), and the decomposition (\ref{decompFS}) translates into the Helmholtz vector theorem
\begin{equation}
\vec E(\vec x)=-\frac{1}{4\pi}\nabla\int\!\de^3\vec x'\,\frac{(\nabla'\cdot\vec E)(\vec x')}{|\vec x-\vec x'|}
+\frac{1}{4\pi}\nabla\times\int\!\de^3\vec x'\,\frac{(\nabla'\times\vec E)(\vec x')}{|\vec x-\vec x'|} \,.
\end{equation}
In addition to the longitudinal and transverse projectors, 
we define the {\itshape transverse rotation operator} (cf.~\cite[Eq.~(A.3)]{Gregor})
\begin{equation} \label{eq_defRT}
 (R_{\mathrm T})_{ij}(\vec k) = \epsilon_{i\ell j} \, \frac{k_\ell}{|\vec k|} \,,
\end{equation}
which acts on a vector field $\vec E(\vec k)$ by
\begin{equation}
 \tsr R_{\mathrm T}(\vec k) \h \vec E(\vec k) = \frac{\vec k \times \vec E(\vec k)}{|\vec k|} \,.
\end{equation}
The three vectors $P_{\mathrm L} \h \vec E$, $P_{\mathrm T} \h \vec E$ and $R_{\mathrm T} \h \vec E$ are orthogonal, and the operators $P_{\mathrm L}$, $P_{\mathrm T}$ and $R_{\mathrm T}$ generate an algebra with the multiplication table shown in Table~\ref{table}.

\renewcommand\arraystretch{1.9}
\begin{table}[t]
\begin{center}
\begin{tabular}{|>{\centering\arraybackslash}m{0.9cm}|>{\centering\arraybackslash}m{0.9cm}>{\centering\arraybackslash}m{0.9cm}>{\centering\arraybackslash}m{0.9cm}|}
\hline
$\cdot$ & $\tsr P_{\mathrm L}$ & $\tsr P_{\mathrm T}$ & $\tsr R_{\mathrm T}$ \\
\hline
$\tsr P_{\mathrm L}$ & $\tsr P_{\mathrm L}$ & 0 & 0 \\
$\tsr P_{\mathrm T}$ & 0 & $\tsr P_{\mathrm T}$ & $\tsr R_{\mathrm T}$ \\
$\tsr R_{\mathrm T}$ & 0 & $\tsr R_{\mathrm T}$ & $-\tsr P_{\mathrm T}$ \\
\hline
\end{tabular}
\end{center}
\caption{Multiplication table for the longitudinal projection operator $P_{\mathrm L} (\vec k)$, the transverse projection operator $P_{\mathrm T}(\vec k)$ and the transverse rotation operator $R_{\mathrm T}(\vec k)$. These operators acting in the three-dimensional euclidean space are defined by Eqs.~\eqref{eq_defPL}, \eqref{eq_defPT} and \eqref{eq_defRT}. \label{table}}
\end{table}
\renewcommand\arraystretch{1.0}

\subsection{Fields, potentials and sources} \label{sec_pre_fields}

The Maxwell equations relate the electric and magnetic fields $\{\vec E, \vec B\}$ to the charge and current densities $\{\rho, \vec j \}$ as
\begin{align}
 \nabla \cdot \vec E & = \rho / \varepsilon_0 \,, \label{eq_max_1} \\[5pt]
 \nabla \times \vec B - \frac{1}{c^2} \frac{\partial \vec E}{\partial t} & = \mu_0 \h \vec j \,, \label{eq_max_2} \\[5pt]
 \nabla \cdot \vec B & = 0 \,, \label{eq_max_3} \\[5pt]
 \nabla \times \vec E + \frac{\partial \vec B}{\partial t} & = 0 \,, \label{eq_max_4}
\end{align}
where $c^2=1/(\varepsilon_0\mu_0$). These equations are also referred to as Gauss's law, Amp\`{e}re's law (with Maxwell's correction), Gauss's law for magnetism and Faraday's law, respectively. There is a general consensus that these equations are universally valid on a microscopic scale~\cite{Fliessbach, Martin, SchafWegener, Bertlmann, Sexl, Misner, Itzykson}. In the context of electrodynamics of media, see in particular the remarks in Ref.~\cite[pp.~3~f.]{Melrose}. The sources on the right hand side of Eqs.~\eqref{eq_max_1} and \eqref{eq_max_2} have to satisfy the continuity equation on grounds of consistency,
\begin{equation} \label{eq_contintuity}
 \frac{\partial \rho}{\partial t} + \nabla \cdot \vec j = 0 \,.
\end{equation}
Electromagnetic potentials $\{\varphi, \vec A\}$ are introduced by
\begin{align}
\vec E & = -\nabla \varphi - \frac{\partial \vec A}{\partial t} \,, \label{eq_gaugepot_1} \\[3pt]
\vec B & = \nabla \times \vec A \,. \label{eq_gaugepot_2}
\end{align}
These are only determined up to gauge transformations:
two potentials $\{\varphi, \vec A\}$ and $\{\varphi', \vec A'\}$ (defined on the whole three-dimensional space)
lead to the same electric and magnetic fields $\{\vec E, \vec B\}$ 
if and only if they differ by a pure gauge, i.e., if there is a real function $f$ such that
\begin{align}
\varphi' & = \varphi - \frac{\partial f}{\partial t} \,, \label{eq_gaugetrafo_1} \\[5pt]
\vec A' & = \vec A + \nabla f \,. \label{eq_gaugetrafo_2}
\end{align}
A relativistic formulation of electrodynamics is obtained by introducing the four-current density \smallskip
\begin{equation}
j^\mu = (c \rho, \h j_1, \h j_2, \h j_3)^{\mathrm T} \,,
\end{equation}
the four-potential \smallskip
\begin{equation}
A^\mu = (\varphi / c, \h A_1, \h A_2, \h A_3)^{\mathrm T} \,, \smallskip
\end{equation}
and the field strength tensor (cf.~\cite{Misner}, which also uses the metric convention \eqref{eq_metric}),
\begin{equation} \label{eq_FEB}
 F^{\mu\nu} = \left( \begin{array}{cccc} 0 & E_1 / c & E_2 / c & E_3 / c \\[2pt]
 -E_1 / c & 0 & B_3 & -B_2 \\[2pt]
 -E_2 / c & -B_3 & 0 & B_1 \\[2pt]
 -E_3 / c & B_2 & -B_1 & 0
 \end{array} \right). \smallskip
\end{equation}
The electric and magnetic fields can be gained back by from the field strength tensor by \vspace{-3pt}
\begin{align}
E_i & = c \h F^{0i} \,, \label{eq_EF} \\[3pt]
B_i & = \frac{1}{2} \h \epsilon_{ik\ell} F^{k\ell} \,. \label{eq_BF}
\end{align}
In a relativistic notation, the continuity equation \eqref{eq_contintuity} 
can be written as
\begin{equation}
\partial_\mu \h j^\mu = 0 \,,
\end{equation}
while the gauge tranformation \eqref{eq_gaugetrafo_1}--\eqref{eq_gaugetrafo_2} reads
\begin{equation}
A'^\mu = A^\mu + \partial^\mu \mh f \,.
\end{equation}
The Maxwell equations \eqref{eq_max_1}--\eqref{eq_max_4} can also be written in a manifestly covariant form, see \cite[Sec.~2.3]{Wachter}.

\subsection{Functional interdependencies} \label{sec_interdep}

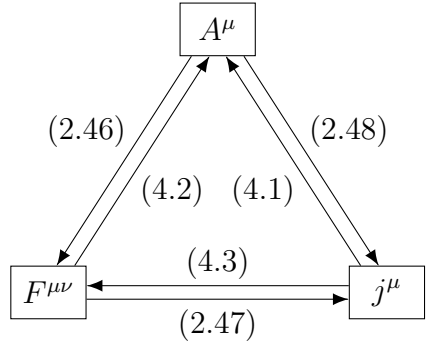
\begin{figure}
\begin{center}
\begin{tikzpicture}
 \tikzstyle{myrect} = [rectangle, draw, minimum width = 1cm, minimum height = 0.7cm];
 \node [myrect] (A) at (2.25cm, 3.5cm) {$A^\mu$};
 \node [myrect] (F) at (0cm, 0cm) {$F^{\mu\nu}$};
 \node [myrect] (j) at (4.5cm, 0cm) {$j^\mu$};
 \tikzstyle{myline} = [decoration = {markings, mark = at position 1 with {\arrow[scale = 1.5]{latex}}},
 postaction = {decorate},
 shorten >= 0.4pt, -];
 \draw[myline, transform canvas = {xshift = -0.12cm}] (A) to node[xshift = -0.5cm, yshift = 0.4cm] {\eqref{eq_AF}} (F);
 \draw[myline, transform canvas = {xshift = 0.12cm}] (F) to node[xshift = 0.4cm, yshift = -0.4cm] {\eqref{eq_FA}} (A);
 \draw[myline, transform canvas = {yshift = -0.09cm}] (F) to node[yshift = -0.35cm] {\eqref{eq_Fj}} (j);
 \draw[myline, transform canvas = {yshift = 0.09cm}] (j) to node[yshift = 0.3cm] {\eqref{eq_jF}} (F);
 \draw[myline, transform canvas = {xshift = 0.12cm}] (A) to node[xshift = 0.5cm, yshift = 0.4cm] {\eqref{eq_Aj}} (j);
 \draw[myline, transform canvas = {xshift = -0.12cm}] (j) to node[xshift = -0.4cm, yshift = -0.4cm] {\eqref{eq_jA}} (A);
\end{tikzpicture}
\end{center} \vspace{-0.2cm}
\caption{Mutual relationships between the electromagnetic four-potential $A^\mu$, the field strength tensor $F^{\mu\nu}$ and the four-current density $j^\mu$. The arrow labels refer to the equation numbers in the text. \label{fig_triangle}}
\end{figure}

We now discuss in how far any electromagnetic system can be 
described equivalently by its four-current $j^\mu$, by its four-potential $A^\mu$, or by
its field strength tensor $F^{\mu\nu}$. For this purpose,
we relate any of these quantities to the remaining two (see Fig.~\ref{fig_triangle}):
Given the four-potential, the field strength tensor is determined by
\begin{equation} \label{eq_AF}
 F^{\mu\nu}[A^\lambda] = \partial^\mu \mh A^\nu - \partial^\nu \mh A^\mu = \left( \eta\indices{^\nu_\lambda} \partial^\mu - \eta\indices{^\mu_\lambda} \partial^\nu \right) A^\lambda \,,
\end{equation}
with $\eta\indices{^\mu_\nu} = \eta^{\mu\alpha}\eta_{\alpha\nu} = \mathrm{diag}(1,\h1,\h1,\h1)$. This formula is equivalent to the defining equations \eqref{eq_gaugepot_1}--\eqref{eq_gaugepot_2} for the potentials.

Given the field strength tensor, the four-current is determined by the inhomogeneous Maxwell equations, which read in a manifestly covariant form
\begin{equation} \label{eq_Fj}
 j^\mu[F^{\nu\lambda}] = \frac{1}{\mu_0} \h \partial_\lambda F^{\mu\lambda} = \frac{1}{2 \mu_0} \left( \eta\indices{^\mu_\nu} \partial_\lambda - \eta\indices{^\mu_\lambda} \partial_\nu \right) F^{\nu\lambda} \,. \smallskip
\end{equation}
Combining Eqs.~\eqref{eq_AF} and \eqref{eq_Fj}, we further obtain the four-current in terms of the four-potential,
\begin{equation} \label{eq_Aj}
 j^\mu[A^\nu] = \frac{1}{\mu_0} \left( \eta\indices{^\mu_\nu} \Box + \partial^\mu \partial_\nu \right) A^\nu \,,
\end{equation}
where
\begin{equation}
\Box = -\partial_\lambda \partial^\lambda = \frac{1}{c^2} \frac{\partial^2}{\partial t^2} - \Delta \medskip
\end{equation}
is called the d'Alembert operator. Eq.~\eqref{eq_Aj} can also be regarded as the equation of motion for the four-potential in terms of the four-current. The solution to this differential equation yields the four-potential $A^\mu[j^\nu]$ in terms of the sources. Expressing the sources in terms of the field strength tensor
then leads to the four-potential $A^\mu[F^{\nu\lambda}]$ as a functional of the field strength tensor. We will construct the most general solution to the differential equation \eqref{eq_Aj} and derive explicit expressions for the functionals $A^\mu[j^\nu]$ and $A^\mu[F^{\nu\lambda}]$ in the following sections \ref{sec_green} and \ref{sec_CanFun}.

It remains to express the electric and magnetic fields as functionals of the sources, which is equivalent to finding the general solution of the Maxwell equations. By applying the d'Alembert operator on both sides of Eq.~\eqref{eq_AF} and using Eq.~\eqref{eq_Aj}, we obtain the wave equation for the components of the field strength tensor,
\begin{equation} \label{eq_jFalt}
\Box F^{\mu\nu} = \mu_0 \h (\partial^\mu j^\nu-\partial^\nu j^\mu) \,.
\end{equation}
This is equivalent to the equations of motion for the decoupled electric and 
magnetic fields in terms of the charge and current densities,
\begin{align}
\Box \vec E & = -\frac{1}{\varepsilon_0} \nabla\rho- \mu_0 \frac{\partial \vec j}{\partial t} \,, \label{eom_E} \\[5pt]
\Box \vec B & = \mu_0 \nabla \times \vec j \,. \label{eom_B}
\end{align}
The solutions to these equations formally yield the functional $F^{\mu\nu}[j^\lambda]$. This, however, requires a more precise
discussion for the following reasons: (i) The wave equations require initial conditions to be fixed, and (ii) they follow from the Maxwell equations, but are not equivalent to them. This means, there are solutions to the wave equations which do not obey the Maxwell equations. The initial value problem for both the wave equation and the Maxwell equations will therefore be discussed in the next section.

\section{Electromagnetic Green functions} \label{sec_green}

\subsection{Initial value problem for scalar wave equation} \label{subsec_scalar}

In order to determine the functional $F^{\mu\nu}[j^\lambda]$ explicitly from Eq.~\eqref{eq_jFalt},
we first study the initial value problem of the inhomogeneous scalar wave equation 
\begin{equation} \label{eq_wave_equation}
\Box\phi(x) = \mu_0 \h g(x). \smallskip
\end{equation}
Thus, we seek the uniquely determined solution to this equation with given
initial conditions
\begin{align}
 \phi(t_0, \vec x) & = \phi_0(\vec x) \,, \label{eq_initial_one} \\[5pt]
 (\partial_t \phi)(t_0, \vec x ) & = \phi_1(\vec x ) \,. \label{eq_initial_two}
\end{align}

\pagebreak \noindent
For this purpose, we introduce the {\it scalar Green function} $\mathbbmsl D_0$ and the {\it propagator} $\mathbbmsl U_0$.
The former is a distribution obeying
\begin{equation} \label{eq_scalargreen}
\Box\mathbbmsl D_0(x - x') = \mu_0 \h \delta^4(x - x') \equiv \frac{\mu_0}{c} \, \delta^3(\vec x - \vec x') \h \delta(t - t') \,,
\end{equation}
whereas the latter is subject to
\begin{equation}
\Box\mathbbmsl U_0(x - x') = 0 \label{eq_prop1}
\end{equation}
and the initial conditions
\begin{align}
\mathbbmsl U_0(\vec x - \vec x', \tau = 0) & \, = \, 0 \,, \label{eq_prop2} \\
\frac{1}{c} \h (\partial_t\mathbbmsl U_0)(\vec x-\vec x', \tau=0) & \, = \, \delta^3(\vec x-\vec x') \,, \label{eq_prop3}
\end{align}
where $\tau = t - t'$. The equation \eqref{eq_scalargreen} for the Green function has infinitely many solutions,
the most important of which are the {\it retarded} Green function $\mathbbmsl D^+$ and the {\it advanced} Green function $\mathbbmsl D^-$. These are given by
\begin{equation}
 \mathbbmsl D^\pm_0(\vec x - \vec x', t- t') = \frac{\mu_0}{4\pi c} \, \frac{\delta(t- t' \mp |\vec x - \vec x'| / c)}{|\vec x - \vec x'|} \,,
\end{equation}
or in the Fourier domain
\begin{equation}\label{eq_scGF_FT}
 \mathbbmsl D^\pm_0(\vec k, \omega) = \frac{\mu_0}{-(\omega/c\pm\j\eta)^2 + |\vec k|^2} \, \equiv \frac{\mu_0}{k^2} \,,
\end{equation}
with $\eta \to 0^+$. (The infinitesimal $\eta$ will be suppressed in the sequel.) 
With these definitions, the unique solution to the equations \eqref{eq_prop1}--\eqref{eq_prop3} for the propagator can be written as (cf.~\cite[Eq.~(8.34)]{MartinRothen})
\begin{equation}
\mathbbmsl U_0 = \frac{1}{\mu_0} \h ( \mathbbmsl D^+_0 - \mathbbmsl D^-_0 ) \,.
\end{equation}
The general solution of the inhomogeneous wave equation \eqref{eq_wave_equation} with initial conditions \eqref{eq_initial_one}--\eqref{eq_initial_two} is now given by the Duhamel formula (cf.~\cite[Sec.~1.3]{Ikawa} and \cite[Sec.~3.3]{Scharf})
\begin{equation}
\begin{aligned} \label{eq_solCauchy}
\phi(\vec x,t) & = \int_{t_0}^\infty \! c \, \de t'\int \! \de^3 \vec x' \, \mathbbmsl D^+_0(\vec x - \vec x', t - t') \, g(\vec x',t') \\[5pt]
& \quad + \frac{1}{c} \h \int \! \de^3 \vec x' \, (\partial_t \mathbbmsl U_0)(\vec x - \vec x', t - t_0) \, \phi_0(\vec x', t_0) \\[5pt]
& \quad + \frac{1}{c} \h \int \! \de^3 \vec x' \, \mathbbmsl U_0(\vec x - \vec x', t - t_0) \, \phi_1(\vec x', t_0) \,.
\end{aligned}
\end{equation}
Note that the field $\phi$ naturally decomposes into a retarded part and a vacuum solution, $\phi = \phi_{\rm ret} + \phi_{\rm vac}$, 
which satisfy $\Box\phi_{\rm ret}=\mu_0 g$ and $\Box\phi_{\rm vac}=0$, respectively. The retarded part is 
given in terms of the sources by the retarded Green function, while the vacuum part is given in terms of the initial conditions by the propagator.

\subsection{Initial value problem for Maxwell equations} \label{subsec_Maxwell_initial}

Formally, the decoupled equations \eqref{eom_E}--\eqref{eom_B} for the electric and magnetic fields $\{\vec E,\vec B\}$ are inhomogeneous wave equations.
In principle, the corresponding initial value problem can therefore be solved by the method of the previous subsection.
Concretely, this means that the electric and magnetic fields are uniquely determined by the formula \eqref{eq_solCauchy}
once we are given appropriate initial conditions for these fields and their first time-derivatives.
However, in the case of the Maxwell equations, the initial value problem is more complicated due to the fact that these initial conditions cannot be prescribed arbitrarily: generally, the initial conditions at time $t = t_0$ have to be given such that they fulfill the equations (cf.~\cite[p.~29]{Ikawa})
\begin{align}
\nabla\cdot\vec E_0 & = \rho_0/\varepsilon_0 \,, \label{eq_constraint_1} \\[4pt]
\nabla\times\vec E_0 & = -(\partial_t\vec B)_0 \,, \\[4pt]
\nabla\cdot\vec B_0 & = 0 \,, \\[4pt]
\nabla\times\vec B_0 & = \mu_0 \h \vec j_0+ (\partial_t\vec E)_0 / c^2 \,, \label{eq_constraint_4}
\end{align}
where $\vec E_0(\vec x) = \vec E(\vec x, t_0)$, \,$(\partial_t\vec E)_0(\vec x) = (\partial_t \vec E)(\vec x, t_0)$, etc.
These equations are formally identical to the Maxwell equations evaluated at time $t=t_0$,
but they have the meaning of {\it constraints}, namely, they constrain the freedom
of choosing initial conditions for the Maxwell equations. Any solution $\{\vec E(\vec x, t), \vec B(\vec x, t)\}$ of the inhomogeneous wave equations \eqref{eom_E}--\eqref{eom_B} with initial conditions at time $t = t_0$ fulfilling Eqs.~\eqref{eq_constraint_1}--\eqref{eq_constraint_4} is also a solution of the Maxwell equations for all later times $t > t_0$\h.
In particular, the above constraints imply that only the transverse fields $\vec E_{\rm T}$
and $\vec B$ can be prescribed at the initial time. The longitudinal part of the magnetic field has to vanish, while the longitudinal part of the electric field
is completely determined in terms of the charges and therefore does not carry {\itshape independent degrees of freedom.}
Furthermore, the initial condition for $\partial_t\vec B$ is determined by the initial condition for $\vec E_{\rm T}$, and the initial condition for $\partial_t \vec E$ is determined by the initial condition for $\vec B$ and the spatial current.

These consideration show that the electromagnetic fields are in general not uniquely determined by the sources and therefore do not constitute functionals of the sources alone. Instead, they also depend on the initial conditions.
Fortunately, when we deal with electromagnetic perturbations of a sample which are produced in the laboratory, 
we can assume that all observable (external and induced) quantities, i.e., the charge and current densities as well as the corresponding fields, vanish before some initial switching-on. Thus, if we choose $t_0$ sufficiently early (before the external perturbation is switched on) we are entitled to choose identically vanishing initial conditions $\vec E_0 \equiv 0$, $(\partial_t\vec E)_0 \equiv 0$, etc. 
The solutions of the wave equations \eqref{eq_jFalt} or \eqref{eom_E}--\eqref{eom_B} can then be deduced componentwise in terms of the respective sources. 
In particular, we obtain the functional
\begin{align}
 F^{\mu\nu}[j^\lambda] & = \mathbbmsl D_0 \, ( \partial^\mu j^\nu - \partial^\nu j^\mu ) \\[5pt]
 & = (\partial^\mu \mathbbmsl D_0 \, \eta\indices{^\nu_\lambda} - \partial^\nu \mathbbmsl D_0 \, \eta\indices{^\mu_\lambda} ) \, j^\lambda \,, \label{eq_FscDj}
\end{align}
where we have used partial integration, and where
\begin{equation} \label{eq_ret_gf}
 \mu_0 \h \Box^{-1} = \mathbbmsl D_0 \h \equiv \h \mathbbmsl D_0^+
\end{equation}
denotes the retarded Green function of the scalar wave equation. We stress that this choice of the Green function is dictated by the underlying initial-value problem. The electromagnetic fields defined in this way are uniquely determined by the sources and hence constitute functionals of them. 

\subsection{Tensorial Green function}\label{sec_pre_propagators}

We now turn to the explicit functional $A^\mu[j^\nu]$ for the potentials in terms of the sources.
Consider the equation of motion \eqref{eq_Aj} of the four-potential in terms of the four-current,
\begin{equation} \label{eq_eom}
 \left( \eta\indices{^\mu_\nu} \Box + \partial^\mu \partial_\nu \right) A^\nu = \mu_0 \h j^\mu \,.
\end{equation} 
Again we assume that the initial conditions for the electric and magnetic fields vanish, and hence we seek a {\it tensorial Green function} $(D_0)\indices{^\mu_\nu}$ for the four-potential that is {\itshape retarded}, i.e.,
\begin{equation}
 (D_0)\indices{^\mu_\nu}(\vec x - \vec x', t - t') = 0 \quad \textnormal{if} \ \, t < t' \,.
\end{equation}
Given the current $j^\mu$, the Green function should yield by
\begin{equation} \label{eq_ansatz}
 A^\nu = (D_0)\indices{^\nu_\mu} \, j^\mu
\end{equation}
a solution $A^\nu$ to the equation of motion \eqref{eq_eom}. The general solution to this inhomogeneous differential equation  can then be obtained by adding to the particular solution \eqref{eq_ansatz} the general solution of the corresponding homogeneous differential equation.
With vanishing initial fields, the solutions of the homogeneous equation are precisely given by the {\itshape pure gauges,}
\begin{equation}
 A^\nu = \partial^\nu \hspace{-1pt} f \,,
\end{equation}
as
\begin{equation}
 \left( \eta\indices{^\mu_\nu} \Box + \partial^\mu \partial_\nu \right) \partial^\nu \hspace{-1pt} f = 0 \,. \medskip
\end{equation}
We will now first construct the most general form of a tensorial Green function, and in the following subsection discuss special expressions which are obtained by gauge fixing conditions.

In analogy to the scalar case, one could expect that the tensorial Green function satisfies the equation
\begin{equation}
 \left( \eta\indices{^\mu_\lambda} \Box + \partial^\mu \partial_\lambda \right) (D_0)\indices{^\lambda_\nu}(x - x') \stackrel{?}{=} \eta\indices{^\mu_\nu} \h \delta^4(x - x') \,.
\end{equation}
However, it turns out that this equation has no solutions (see \cite[Sec.~3.4]{Bertlmann} and \cite[Sec.~4.3]{Bogoliubov}). The key to overcome this problem is to restrict oneself to {\itshape physical} four-currents which satisfy the continuity equation $\partial_\mu j^\mu = 0$. Hence, only for such currents the Green function should yield by Eq.~\eqref{eq_ansatz} a four-potential which solves the equation of motion \eqref{eq_eom}. To put this program into practice, we define projection operators on the Minkowski longitudinal and transverse parts in Fourier space by (cf.~\cite[p.~36]{Bogoliubov})
\begin{align}
 (P_{\mathrm L} )\indices{^\mu_\nu}(k) & = \frac{k^\mu k_\nu}{k^2} \,, \\[3pt]
 (P_{\mathrm T} )\indices{^\mu_\nu}(k) & = \eta\indices{^\mu_\nu} - \frac{k^\mu k_\nu}{k^2} \,,
\end{align}
with $k^2 = k_\mu k^\mu$. (These projection operators must not be confused with their euclidean counterparts introduced in Sec.~\ref{sec_pre_notations}.) 
The Minkowski projection operators are orthogonal with respect to the Minkowski inner product, and hence any vector field $C^\mu(k)$ can be decomposed into its longitudinal and transverse parts
$(C_{\mathrm L} )^\mu = (P_{\mathrm L} )\indices{^\mu_\nu} \, C^\nu$ and $(C_{\mathrm T} )^\mu = (P_{\mathrm T} )\indices{^\mu_\nu} \, C^\nu$, such that
\begin{align}
 (C_{\mathrm L})^\mu + (C_{\mathrm T})^\mu & = C^\mu \,, \\[5pt]
 (C_{\mathrm L})_\mu \, (C_{\mathrm T})^\mu & = 0 \,. \medskip
\end{align}
In real space, transverse vector fields have vanishing four-divergence, \linebreak $\partial_\mu (C_{\mathrm T} )^\mu(x) = 0$, whereas longitudinal
vector fields can be represented as a four-gradient $(C_{\mathrm L} )^\mu(x) = (\partial^\mu \hspace{-1pt} f)(x)$. With these definitions, the continuity equation is equi\-{}valent to the statement that $j^\mu$ is a Minkowski-transverse four-vector, i.e., $P_{\rm L} \h j = 0$ and $P_{\rm T} \h j = j$. Moreover, the equation of motion \eqref{eq_eom} can be written in Fourier space as
\begin{equation}
k^2 \, (P_{\mathrm T} )\indices{^\mu_\nu}(k) \, A^\nu(k) = \mu_0 \h j^\mu(k) \,,
\end{equation}
or in an abridged notation
\begin{equation}
 k^2 \h P_{\mathrm T} \h A = \mu_0 \h j \,.
\end{equation}
Putting the ansatz \eqref{eq_ansatz} into this equation leads to the condition
\begin{equation} \label{eq_require}
k^2 \h P_{\mathrm T} \h D_0 \h j = \mu_0 \h j \,,
\end{equation}
which should hold for any physical (Minkowski-transverse) four-current $j$. This implies the {\itshape defining equation} for the free tensorial Green function:
\begin{equation} \label{eq_condition}
k^2 \h P_{\mathrm T} \h D_0 \h  P_{\mathrm T} = \mu_0 \h P_{\mathrm T} \,.
\end{equation}
In order to construct the most general solution to this equation, we note that any tensorial operator $D_0$ can be decomposed into four independent parts as
\begin{equation} \label{eq_decompose}
 D_0 = P_{\mathrm T} \h D_0 \h P_{\mathrm T} + P_{\mathrm L} \h D_0 \h P_{\mathrm T} + P_{\mathrm T} \h D_0 \h P_{\mathrm L} + P_{\mathrm L} \h D_0 \h P_{\mathrm L} \,,
\end{equation}
and these four parts contain $3 \times 3 = 9$, \,$1 \times 3 = 3$, \,$3 \times 1 = 3$ and $1 \times 1 = 1$ independent component functions, respectively. Evidently, Eq.~\eqref{eq_condition} fixes only the first operator as
\begin{equation}
 P_{\mathrm T} \h D_0 \h P_{\mathrm T} = \frac{\mu_0}{k^2} \h P_{\mathrm T} \,,
\end{equation}
whereas the remaining operators can be chosen arbitrarily. We note that this solution is only formal in that it involves the singular expression $1/k^2$, the inverse Fourier transform of which is not well-defined. (The same problem arises if one tries to define the Minkowski projection operators in real space.) In accordance with our previous considerations, we interpret any contribution of the form $1/k^2$ as a retarded Green function in the sense of Eq.~\eqref{eq_scGF_FT}, hence
\begin{equation} \label{eq_this_fixes}
  P_{\mathrm T} \h D_0 \h P_{\mathrm T} = \mathbbmsl D_0 \h P_{\mathrm T} \,.
\end{equation}
We conclude that in constructing the most general tensorial Green function $(D_0)\indices{^\mu_\nu}$, 
one may choose three arbitrary operators on the right hand side of Eq.~\eqref{eq_decompose}, 
whereas the first contribution is fixed by Eq.~\eqref{eq_this_fixes}, with $\mathbbmsl D_0 = \mu_0 \h \Box^{-1}$ being the retarded Green function of the scalar wave equation.

An explicit expression for the tensorial Green function is now given by
\begin{equation}\label{eq_gen_GF}
 (D_0)\indices{^\mu_\nu}(k) = \mathbbmsl D_0(k) \left( \eta\indices{^\mu_\nu} - \frac{k^\mu k_\nu}{k^2} \right) + k^\mu \tilde f_\nu(k) + \tilde g^\mu(k) \h k_\nu + k^\mu \h \tilde h(k) \h k_\nu \,,
\end{equation}
where the functions $\tilde f_{\nu}$, $\tilde g^\mu$ and $\tilde h$ can be chosen arbitrarily up to the constraints of Minkowski-transversality
\begin{equation}
 \tilde f_\nu(k) \, k^\nu = k_\mu \, \tilde g^\mu(k) = 0 \,.
\end{equation}
Conversely, these functions can be reconstructed from a given tensorial Green function by (suppressing momentum dependencies in the notation)
\begin{align}
 \tilde f_\nu(k) & = \frac{k_\lambda}{k^2} \, (D_0)\indices{^\lambda_\rho} \, (P_{\mathrm T})\indices{^\rho_{\nu}} \,, \\[5pt]
 \tilde g^\mu(k) & = (P_{\mathrm T})\indices{^\mu_\lambda} \, (D_0)\indices{^\lambda_\rho} \, \frac{k^\rho}{k^2} \,, \\[5pt]
 \tilde h(k) & = \frac{k_\lambda}{k^2} \, (D_0)\indices{^\lambda_\rho} \, \frac{k^\rho}{k^2} \,.
\end{align}
In order to obtain a more symmetric form of Eq.~\eqref{eq_gen_GF}, we rewrite it in terms of dimensionless parameter functions $f_\nu$, $g^\mu$ and $h$ as
\begin{equation} \label{eq_gen_GF_dimless}
 (D_0)\indices{^\mu_\nu}(k) = \mathbbmsl D_0(k) \left( \eta\indices{^\mu_\nu} + \frac{c k^\mu}{\omega} \, f_\nu(k) + g^\mu(k) \, \frac{c k_\nu}{\omega} + \frac{c k^\mu}{\omega} \, h(k) \, \frac{c k_\nu}{\omega} \right).
\end{equation}
These new functions are related to the previously defined functions by
\begin{align}
 f_\nu(k) & = \frac{\omega}{c} \, \frac{k^2}{\mu_0} \, \tilde f_\nu(k) \,, \\[5pt]
 g^\mu(k) & = \frac{\omega}{c} \, \frac{k^2}{\mu_0} \, \tilde g^\mu(k) \,, \\[5pt]
 h(k) & = \frac{\omega^2}{c^2} \left( \h \frac{k^2}{\mu_0} \, \tilde h(k) - \frac{1}{k^2} \right).
\end{align}
Again, $f_\nu$ and $g^\mu$ are Minkowski-transverse in the sense that
\begin{equation}
 f_\nu(k) \, k^\nu = k_\mu \, g^\mu(k) = 0 \,,
\end{equation}
and consequently they can be written as
\begin{align}
 f_\nu(\vec k, \omega) & = \left( -\frac{c \h \vec k \cdot \vec f(\vec k, \omega)}{\omega}\,, \ \vec f^{\rm T}(\vec k, \omega) \right), \\[3pt]
 g^\mu(\vec k, \omega) & = \left( \frac{c \h \vec k \cdot \vec g(\vec k, \omega)}{\omega}\,, \ \vec g^{\rm T}(\vec k, \omega) \right)^{\!\!\rm T}.
\end{align}
Our formula \eqref{eq_gen_GF_dimless} for the tensorial Green function generalizes the formulae of Ref.~\cite[Sec.~VIII. \S 77]{Berestetskii}. 
The seven dimensionless functions $\vec f$, $\vec g$ and $h$ can be chosen arbitrarily, and for each choice of them we obtain by Eq.~\eqref{eq_gen_GF_dimless} a tensorial electromagnetic Green function which satisfies Eq.~\eqref{eq_condition}.

\subsection{Gauge fixing} \label{sec_gaugefixing}

Special forms of the tensorial Green function correspond to special choices of the free parameter functions in Eq.~\eqref{eq_gen_GF_dimless}. For example, the {\it Lorenz Green function} (cf.~\cite[Eq.~(41) with $\alpha=0$]{JacksonRMP}) reads
\begin{equation} \label{eq_LorGreen}
 (D_0)\indices{^\mu_\nu}(k) = \mathbbmsl D_0(k) \, (P_{\mathrm T} )\indices{^\mu_\nu}(k) \,,
\end{equation}
and is obtained from
\begin{equation}
 \vec f(k) = \vec g(k) = 0 \,, \qquad h(k) = -\frac{\omega^2}{c^2 k^2} \,.
\end{equation}
With this definition, $A^\mu = (D_0)\indices{^\mu_\nu} \h j^\nu$ implies the Lorenz gauge condition~\cite{JacksonRMP}
\begin{equation}
\partial_\mu A^\mu=0 \,.
\end{equation}
In relativistic quantum field theory, one mainly uses the {\it Feynman Green function} (cf.~\cite[Eq.~(41) with $\alpha = 1$]{JacksonRMP} and \cite[Eq.~(77.8)]{Berestetskii}),
\begin{equation} \label{eq_Feynman}
 (D_0)\indices{^\mu_\nu}(k) = \mathbbmsl D_0(k) \, \eta\indices{^\mu_\nu} \,,
\end{equation}
which corresponds to $\vec f(k) = \vec g(k) = h(k) = 0$. By contrast, in non-relativistic quantum field theory the {\it Coulomb Green function} \cite[Eqs.~(77.12)--(77.13)]{Berestetskii}
\begin{equation} \label{eq_CoulGF}
 (D_0)\indices{^\mu_\nu}(\vec k, \omega) = \left( \begin{array}{cc} \mu_0 / |\vec k|^2 & 0 \\[5pt]
 0 & \mathbbmsl D_0(\vec k, \omega) \, \tsr P_{\mathrm T}(\vec k)
 \end{array} \right) \smallskip
\end{equation}
is frequently used. This requires the more involved choice
\begin{align}
 & \vec f(\vec k, \omega) = \vec g(\vec k, \omega) = \frac{\omega \, c\vec k}{c^2 |\vec k|^2 - \omega^2} \, \frac{\omega^2}{c^2|\vec k|^2} \,, \\[7pt]
 & h(\vec k, \omega) = - \frac{\omega^2}{c^2 |\vec k|^2 - \omega^2} \left( 1 + \frac{\omega^2}{c^2 |\vec k|^2} \right) .
\end{align}
It implies the Coulomb gauge of the electromagnetic potential $A = D_0 \h j$, which reads
\begin{equation}
\nabla\cdot\vec A = 0 \,. \medskip
\end{equation}
Finally, for electrodynamics of materials the {\it temporal Green function}
\begin{equation} \label{eq_weylgreen}
 (D_0)\indices{^\mu_\nu}(\vec k, \omega) = \mathbbmsl D_0(\vec k, \omega) \left( \! \begin{array}{cc} 0 & 0 \\[5pt]
 -c\vec k / \omega & \tsr 1
 \end{array} \right)
\end{equation}
turns out to be particularly useful. This is obtained from
\begin{align}
 \vec f(\vec k, \omega) & = \frac{\omega \, c \vec k}{c^2 |\vec k|^2 - \omega^2} \,, \\[8pt]
 \vec g(\vec k, \omega) & = 0 \,, \\[5pt]
 h(\vec k, \omega) & = -\frac{\omega^2}{c^2 |\vec k|^2 - \omega^2} \,.
\end{align}
In this case the electromagnetic potential satisfies
\begin{equation} \label{eq_temporal}
 A^0 \equiv \varphi/c = 0 \,,
\end{equation}
which is the temporal gauge condition. In the following, we will employ this last gauge condition, because it facilitates the expression of the vector potential $\vec A$ in terms of the electric and magnetic fields $\{\vec E, \vec B\}$ (see Sections~\ref{subsec_Hamgauge} and \ref{sec_pre_total}).

\section{Canonical functional} \label{sec_CanFun}

\subsection{Temporal gauge} \label{subsec_Hamgauge}

We now come back to the problem of finding explicit expressions for 
the four-potential as a functional of the electric and magnetic fields, $A^\mu=A^\mu[F^{\nu\lambda}]$. The tensorial Green function being fixed, 
the general solution to the equation of motion for the four-potential can be written as
\begin{equation} \label{eq_jA}
 A^\mu[j^\nu] = (D_0)\indices{^\mu_\nu} \, j^\nu + \partial^\mu \mh f \,,
\end{equation}
with an arbitrary pure gauge $\partial^\mu \mh f$. 
In particular, this means that the four-potential is determined in terms of the four-current only up to gauge transformations.
Combining Eq.~\eqref{eq_jA} with Eq.~\eqref{eq_Fj} and by partial integration, the four-potential can be expressed in terms of the field strength tensor as
\begin{equation}
 A^\mu[F^{\nu\lambda}] = \frac{1}{2\mu_0} \h \left( \partial_\lambda (D_0)\indices{^\mu_\nu} - \partial_\nu (D_0)\indices{^\mu_\lambda} \h \right) F^{\nu\lambda} + \partial^\mu \mh f \,. \label{eq_FA}
\end{equation}
On the other hand, combining Eq.~\eqref{eq_jA} with Eq.~\eqref{eq_AF} yields the fields in terms of the sources,
\begin{equation} \label{eq_jF}
F^{\mu\nu}[j^\lambda] = \big( \partial^\mu (D_0)\indices{^\nu_\lambda} - \partial^\nu (D_0)\indices{^\mu_\lambda} \h \big) \h j^\lambda.
\end{equation}
This last functional does not depend on the choice of the tensorial Green function, and by taking the Feynman Green function \eqref{eq_Feynman} one sees that it is equivalent to Eq.~\eqref{eq_FscDj}. Hence it corresponds to the retarded solution of Maxwell's equations with vanishing initial conditions.

We now rewrite the functional \eqref{eq_FA} explicitly in terms of the electric and magnetic fields using the temporal gauge condition \eqref{eq_temporal}. First note that $\varphi = 0$ implies
\begin{align}
\vec E & = -\partial_t \vec A \,, \label{eq_weyl_E} \\[5pt]
\vec B & = \nabla \times \vec A \,. \label{eq_weyl_B}
\end{align}
Plugging these equations into the Maxwell equations, we find the equation of motion for the vector potential in the temporal gauge,
\begin{equation} \label{eq_tosolve}
\Box \vec A(\vec x, t) = \mu_0 \h \vec j(\vec x, t) + \frac 1 {\varepsilon_0} \h \int_{t_0}^{t} \! \de t' \, \nabla \rho(\vec x, t') \,.
\end{equation}
Here, the lower bound $t_0$ of the time integral is arbitrary: alternating the lower bound modifies the vector potential by a time-independent
gradient, which has no bearing on the electromagnetic fields.
Solving Eq.~\eqref{eq_tosolve} for the vector potential by means of the scalar Green function $\mathbbmsl D_0 = \mu_0 \h \Box^{-1}$ yields in Fourier space
\begin{equation} \label{eq_zw}
 \vec A(\vec k, \omega) = \mathbbmsl D_0(\vec k, \omega) \left( \vec j(\vec k, \omega) - \frac{c^2 \vec k}{\omega} \, \rho(\vec k, \omega) \right).
\end{equation}
This is equivalent to $A=D_0 \h j$ with the temporal Green function given by Eq.~\eqref{eq_weylgreen}. Further using the inhomogeneous Maxwell equations to express the charge and 
current densities in terms of the electric and magnetic fields, we obtain an expression $\vec A[\vec E, \vec B]$, which we call the {\itshape canonical functional in the temporal gauge}:
\begin{equation}
\begin{aligned}
 & \vec A(\vec k, \omega) = -\varepsilon_0 \h \mathbbmsl D_0(\vec k, \omega) \\[5pt]
 & \quad \, \times \frac{1}{\j\omega} \, \Big( \omega^2 \vec E(\vec k, \omega) - c^2 \vec k \h (\vec k \cdot \vec E(\vec k, \omega)) + \omega \, c^2 \vec k \times \vec B(\vec k, \omega) \Big) \, . \label{eq_can}
\end{aligned}
\end{equation}
To simplify this expression, we define the {\itshape electric solution generator}
\begin{align}
 \tsr {\mathbbmsl E}(\vec k, \omega) & = -\varepsilon_0 \h \omega^2 \h \mathbbmsl D_0(\vec k, \omega) \left( \left(1 - \frac{c^2 |\vec k|^2}{\omega^2} \right) \tsr P_{\mathrm L} (\vec k) + \tsr P_{\mathrm T}(\vec k) \right) \label{eq_OpE} \\[5pt]
 & = \tsr P_{\mathrm L} (\vec k) + \frac{\omega^2}{\omega^2 - c^2 |\vec k|^2} \, \tsr P_{\mathrm T}(\vec k) \,,
\end{align}
and the {\itshape magnetic solution generator}
\begin{align}
 \tsr {\mathbbmsl B}(\vec k, \omega) & = -\varepsilon_0 \h \omega^2 \h \mathbbmsl D_0(\vec k, \omega) \left( \frac{c|\vec k|}{\omega} \, \tsr R_{\mathrm T}(\vec k) \right) \label{eq_OpB} \\[5pt]
 & = \frac{\omega \, c|\vec k|}{\omega^2 - c^2 |\vec k|^2} \, \tsr R_{\mathrm T}(\vec k) \,.
\end{align}
(The reason for this labeling will become clear at the end of this subsection.) These are dimensionless operators, which are given in components by
\begin{align}
 \mathbbmsl E_{ij}(\vec k, \omega) & = \frac{\omega^2 \delta_{ij} - c^2 k_i k_j}{\omega^2 - c^2 |\vec k|^2} \,, \\[5pt]
 \mathbbmsl B_{ij}(\vec k, \omega) & = \frac{\omega \, \epsilon_{i\ell j} \h c \h k_\ell}{\omega^2 - c^2 |\vec k|^2} \,.
\end{align}
In terms of the electric and magnetic solution generators, the canonical functional \eqref{eq_can} can be written as
\begin{align}
 \vec A(\vec k, \omega) & = \frac{1}{\j\omega} \left( \tsr{\mathbbmsl E}(\vec k, \omega) \, \vec E(\vec k, \omega) + \tsr{\mathbbmsl B}(\vec k, \omega) \, c \h \vec B(\vec k, \omega) \right). \label{eq_can_tsr}
\end{align}

\pagebreak \noindent
Equivalently, it is given in component form by
\begin{equation} \label{eq_can_comp}
 A_i(\vec k, \omega) = \frac{1}{\j\omega} \, \frac{\omega^2 \delta_{ij} - c^2 k_i k_j}{\omega^2 - c^2 |\vec k|^2} \, E_j(\vec k, \omega) + \frac{1}{\j\omega} \, \frac{ \omega \, \epsilon_{i\ell j} \h c \h k_\ell}{\omega^2 - c^2 |\vec k|^2} \, c B_j(\vec k, \omega) \,.
\end{equation}
In particular, we read off the partial functional derivatives of the vector
potential with respect to the electric and magnetic fields,
\begin{align}
 \frac{\delta A_i(\vec k, \omega)}{\delta E_j(\vec k', \omega')} & = \frac{c}{\j\omega} \, \frac{\omega^2 \delta_{ij} - c^2 k_i k_j}{\omega^2 - c^2 |\vec k|^2}\,\delta^3(\vec k-\vec k') \h \delta(\omega-\omega') \,, \label{eq_partial_1} \\[5pt]
 \frac{1}{c} \h \frac{\delta A_i(\vec k, \omega)}{\delta B_j(\vec k', \omega')} & = \frac{c}{\j\omega} \, \frac{\omega \, \epsilon_{i\ell j} \h c \h k_\ell}{\omega^2 - c^2 |\vec k|^2}\,\delta^3(\vec k-\vec k') \h \delta(\omega-\omega') \,. \label{eq_partial_2}
\end{align}
Our formalism is consistent in that the same functional $\vec A[\vec E, \vec B]$ is obtained if we first use the continuity equation to eliminate $\rho$ in terms of $\vec j$ in Eq. \eqref{eq_zw},
\begin{equation}
 \vec A(\vec k, \omega) = \mathbbmsl D_0(\vec k, \omega) \left( \vec j(\vec k, \omega) - \frac{c^2 \vec k \h (\vec k \cdot \vec j(\vec k, \omega))}{\omega^2} \right), \smallskip
\end{equation}
and then express the current in terms of the electric and magnetic fields by Amp\`{e}re's law. The reason for this is that the continuity equation and Amp\`{e}re's law imply
\begin{equation}
 \rho(\vec k, \omega) = \frac{\vec k \cdot \vec j(\vec k, \omega)}{\omega} = \varepsilon_0 \h \j \vec k \cdot \vec E(\vec k, \omega) \,,
\end{equation}
which is the same expression for $\rho$ in terms of $\vec E$ as given by Gauss's law.

We conclude this subsection with a remark about the electric and magnetic solution generators defined in Eqs.~\eqref{eq_OpE} and \eqref{eq_OpB}. Apart from the canonical functional, these operators also appear in the equations of motion for the electric and magnetic fields in terms of the sources: Inverting Eqs.~\eqref{eom_E}--\eqref{eom_B} yields the
expressions in Fourier space
\begin{align}
 \vec E(\vec k, \omega) & = \mathbbmsl D_0(\vec k, \omega) \, \big( {-c^2} \h \j\vec k \h \rho(\vec k, \omega) + \j \omega \h \vec j(\vec k, \omega) \h \big) \,, \\[5pt]
 \vec B(\vec k, \omega) & = \mathbbmsl D_0(\vec k, \omega) \, \big( \h \j \vec k \times \vec j(\vec k, \omega) \h \big) \,.
\end{align}

\pagebreak \noindent
Using the continuity equation to eliminate $\rho$ in terms of $\vec j$ in the first equation, we find that these equations are equivalent to
\begin{align}
 \vec E(\vec k, \omega) & = \frac{1}{\varepsilon_0} \, \frac{1}{\j \omega} \, \tsr{\mathbbmsl E}(\vec k, \omega) \, \vec j(\vec k, \omega) \,, \label{eq_tsrE} \\[5pt]
 c \h \vec B(\vec k, \omega) & = \frac{1}{\varepsilon_0} \, \frac{1}{\j \omega} \, \tsr{\mathbbmsl B}(\vec k, \omega) \, \vec j(\vec k, \omega) \,. \label{eq_tsrB}
\end{align}
Thus, with the electric and magnetic solution generators one can formally solve the Maxwell equations in terms of the spatial current. The relevance of the above expressions will become clear in Sections~\ref{subsec_fsrt} and \ref{sec_comp}.

\subsection{Total functional derivatives} \label{sec_pre_total}

The expansion coefficients of $\vec E$ and $\vec B$ in the canonical functional, Eqs. \eqref{eq_partial_1}--\eqref{eq_partial_2}, correspond to
partial derivatives of the vector potential, because these are obtained by varying $\vec E$ and $\vec B$ independently.
Physically, however, $\vec E$ and $\vec B$ are not independent of each other: by Faraday's law, the magnetic field is related to the transverse part of the electric field as
\begin{equation}
\vec B = \frac{\vec k \times \vec E}{\omega} \,, \label{eq_BE}
\end{equation}
and conversely, the electric field can be decomposed as
\begin{equation}
\vec E = \vec E_{\mathrm L} + \vec E_{\mathrm T} = \vec E_{\mathrm L} -\omega \h \frac{\vec k\times\vec B}{|\vec k|^2} \, \label{eq_EB}.
\end{equation}
Therefore, it is also possible to express the vector potential $\vec A$ exclusively
in terms of $\vec E$, or in terms of $\vec E_{\mathrm L}$ and $\vec B$. To show this explicitly, 
we rewrite the canonical functional $\vec A[\vec E, \vec B]$ from Eq.~\eqref{eq_can} as
\begin{equation}
\begin{aligned}
 \vec A(\vec k, \omega) & = \frac{1}{\j \omega} \h \vec E_{\mathrm L} (\vec k, \omega) \\[5pt]
 & \quad + \frac{1}{\j\omega} \, \frac{1}{\omega^2 - c^2 |\vec k|^2} \, \Big( \omega^2 \vec E_{\mathrm T}(\vec k, \omega) + \omega \, c^2 \vec k \times \vec B(\vec k, \omega) \Big) \,. \label{eq_can2}
\end{aligned}
\end{equation}
Eliminating $\vec B$ in terms of $\vec E_{\mathrm T}$ or vice versa, we find 
\begin{align}
 \vec A[\vec E] & = \frac{1}{\j \omega} \, \vec E_{\mathrm L}+ \frac{1}{\j \omega} \, \vec E_{\mathrm T} = \frac{1}{\j \omega} \, \vec E \,, \label{eq_AE} \\[10pt]
 \vec A[\vec E_{\mathrm L},\vec B] & = \frac{1}{\j \omega} \, \vec E_{\mathrm L} + \frac{\j \vec k \times \vec B}{|\vec k|^2} \,. \label{eq_AELB}
\end{align}
In matrix notation, using the transverse rotation operator defined in Sec.~\ref{sec_pre_notations}, we can write these equations equivalently as
\begin{align}
 \vec A(\vec k, \omega) & = \frac{1}{\j\omega} \, \vec E(\vec k, \omega) \,, \label{eq_AEtot} \\[5pt]
 \vec A(\vec k, \omega) & = \frac{1}{\j\omega} \, \vec E_{\mathrm L} (\vec k, \omega) + \frac{1}{\j\omega} \left( -\frac{\omega}{c|\vec k|} \right) \tsr R_{\mathrm T}(\vec k) \, c \h \vec B(\vec k, \omega) \,. \label{eq_ABtot}
\end{align}
Finally, we can write them in component form as
\begin{align}
 A_i(\vec k, \omega) & = \frac{1}{\j\omega} \, E_i(\vec k, \omega) \,, \label{eq_AiEi} \\[5pt]
 A_i(\vec k, \omega) & = \frac{1}{\j\omega} \, (E_{\mathrm L} )_i(\vec k, \omega) + \frac{1}{\j\omega} \left( \frac{ \omega \, \epsilon_{i\ell j} \h c \h k_\ell}{- c^2 |\vec k|^2} \right) c B_j(\vec k, \omega) \,. \label{eq_AEL}
\end{align}
These relations can also be obtained directly from Eqs.~\eqref{eq_weyl_E}--\eqref{eq_weyl_B}, 
which read in Fourier space $\vec E = \j\omega \vec A$ and $\vec B = \j \vec k \times \vec A$\h: 
Inverting the first equation yields immediately \eqref{eq_AE}. Moreover, using the decomposition of $\vec A$ into longitudinal and transverse parts,
\begin{equation}
 \vec A = \frac{\vec k (\vec k \cdot \vec A)}{|\vec k|^2} - \frac{\vec k \times (\vec k \times \vec A)}{|\vec k|^2} = \frac{1}{\j\omega} \frac{\vec k (\vec k \cdot \vec E)}{|\vec k|^2} + \frac{\j \vec k \times \vec B}{|\vec k|^2} \,,
\end{equation}
and identifying the first term on the right hand side with $\vec E_{\mathrm L} /\j\omega$, we recover again~\eqref{eq_AELB}. From this alternative derivation we conclude
that Eqs.~\eqref{eq_AE} and \eqref{eq_AELB} are in fact generally valid and apply in particular to vacuum fields,
although the canonical functional \eqref{eq_can} had originally been constructed for retarded fields generated by sources. Vacuum fields have a conceptual relevance,
because in some situ\-{}ations it is an appropriate idealization to consider the external perturbations as free electromagnetic waves (although strictly speaking these do not have sources and hence cannot be produced in the laboratory).

We now define the {\it total functional derivatives} of the vector potential
with respect to the electric and magnetic fields as
\begin{align}
\frac{\mathrm d A_i}{\mathrm d E_j} & = \frac{\delta A_i}{\delta E_j} + \frac{\delta A_i}{\delta B_\ell} \, \frac{\delta B_\ell}{\delta E_j} \,, \label{eq_def_totE} \\[5pt]
\frac{\mathrm d A_i}{\mathrm d B_j} & = \frac{\delta A_i}{\delta B_j} + \frac{\delta A_i}{\delta E_\ell} \, \frac{\delta E_\ell}{\delta B_j} \,. \label{eq_def_totB}
\end{align}
Here, the functional derivatives of the magnetic field with respect to the electric field and vice versa are given by Eqs.~\eqref{eq_BE}--\eqref{eq_EB} as
\begin{equation}
 c \, \frac{\delta B_\ell(\vec k, \omega)}{\delta E_j(\vec k, \omega)} = \frac{c|\vec k|}{\omega} \, \epsilon_{\ell n j} \h \frac{k_n}{|\vec k|} \label{eq_BE_comp}
\end{equation}
and
\begin{equation}
 \frac{1}{c} \h \frac{\delta E_\ell(\vec k, \omega)}{\delta B_j(\vec k, \omega)} = 
 \frac{1}{c} \h \frac{\delta (E_{\mathrm T} )_\ell(\vec k, \omega)}{\delta B_j(\vec k, \omega)} =
 -\frac{\omega}{c|\vec k|} \, \epsilon_{\ell n j} \h \frac{k_n}{|\vec k|} \,. \medskip \label{eq_EB_comp}
\end{equation}
The importance of the total functional derivatives lies in the fact that they lead to physical response functions, as we will argue in Sec.~\ref{subsec_physical}. They can be calculated from Eqs.~\eqref{eq_def_totE}--\eqref{eq_def_totB} by using the formulae for the partial derivatives \eqref{eq_partial_1}--\eqref{eq_partial_2} derived in the previous subsection together with Eqs.~\eqref{eq_BE_comp}--\eqref{eq_EB_comp}. More easily, we can read them off directly from the representations \eqref{eq_AE}--\eqref{eq_AELB} of the vector potential. We thus obtain the following formulae:
\begin{align}
 \frac{\mathrm d A_i(\vec k, \omega)}{\mathrm d E_j(\vec k', \omega')} & = \frac{c}{\j\omega} \, \delta_{ij} \, \delta^3(\vec k - \vec k') \h \delta(\omega - \omega') \,, \label{eq_total_1} \\[5pt]
 \frac{1}{c} \h \frac{\mathrm d A_i(\vec k, \omega)}{\mathrm d B_j(\vec k', \omega')} & = \frac{c}{\j\omega} \left( \frac{ \omega \, \epsilon_{i\ell j} \h c \h k_\ell}{- c^2 |\vec k|^2} \right) \delta^3(\vec k - \vec k') \h \delta(\omega - \omega') \,. \label{eq_total_2}
\end{align}
In real space, they can be written as
\begin{align}
\frac{\mathrm d A_i(\vec x, t)}{\mathrm d E_j(\vec x', t')}&= -\frac{1}{c} \, \delta_{ij} \, \delta^3(\vec x - \vec x') \h \varTheta(t - t') \,,\label{eq_put_3}\\[5pt]
\frac{1}{c} \h \frac{\mathrm d A_i(\vec x,t)}{\mathrm d B_j(\vec x',t')} & = \frac{1}{4\pi c^2} \, \epsilon_{i \ell j} \, \frac{\partial}{\partial x^\ell} \h \frac{\delta(t - t')}{|\vec x - \vec x'|} \,, \label{eq_third}
\end{align}
where $\varTheta(t-t')$ denotes the Heaviside step function.

\subsection{Zero frequency limit} \label{sec_zero_freq}

We now come to a problem which we have ignored so far, namely the divergence of many formal expressions
in the limit $\omega\rightarrow 0$. While the zero frequency limit can be performed without any problems for quantities like
the scalar Green function $\mathbbmsl D_0(\vec k,\omega)$ (as long as $\vec k\neq 0$), the same
does not apply to the operator $1/\j\omega$ which connects, for example, the vector potential $\vec A(\vec k, \omega)$ to the electric field $\vec E(\vec k, \omega)$ in Eq.~\eqref{eq_AEtot}.
When it comes to the limit $\omega\rightarrow 0$, these expressions have to be regularized.
As this can be done in different ways, we first consider the problem in real space, where the vector potential can be defined in terms of the electric and magnetic fields by an initial-value problem: it satisfies for $t > t_0$ the equations
\begin{align}
 -\partial_t \vec A(\vec x, t) & = \vec E(\vec x, t) \,, \label{eq_diff1} \\[5pt]
 \nabla \times \vec A(\vec x, t) & = \vec B(\vec x, t) \,, \label{eq_diff2}
\end{align}
and is subject to the initial condition
\begin{equation}
 \vec A(\vec x, t_0) = \vec A_0(\vec x) \,.
\end{equation}
The latter cannot be chosen arbitrarily, but is constrained by
\begin{equation} \label{eq_diff3}
 \nabla \times \vec A_0(\vec x) = \vec B_0(\vec x) \equiv \vec B(\vec x, t_0) \,.
\end{equation}
We may choose $\vec A_0$ purely transverse, such that it is given explicitly in terms of $\vec B_0$ by
\begin{equation} \label{use_this}
 \vec A_0(\vec x) = \frac{1}{4\pi} \, \nabla \times \int \! \de^3 \vec x' \, \frac{\vec B_0(\vec x')}{|\vec x - \vec x'|} \,. \smallskip
\end{equation}
The unique solution to the initial value problem \eqref{eq_diff1}--\eqref{eq_diff3} is then given by
\begin{equation}
\vec A(\vec x,t)=-\int_{t_0}^t\de t'\,\vec E(\vec x, t')+\vec A_0(\vec x) \,. \label{eq_sol} \smallskip
\end{equation}
While Eq.~\eqref{eq_diff1} is obviously fulfilled,  Eq.~\eqref{eq_diff2} follows by Faraday's law,
\begin{align}
(\nabla\times\vec A)(\vec x, t)&=-\int^t_{t_0}\de t'\,(\nabla\times\vec E)(\vec x, t')+\nabla\times\vec A_0(\vec x) \\[1pt]
&=\int^t_{t_0}\de t'\,\partial_{t'}\vec B(\vec x, t')+\vec B(\vec x, t_0) \\[6pt]
&=\vec B(\vec x, t) \,.
\end{align}
This discussion shows that in general, the vector potential $\vec A$ cannot be expressed entirely in terms of the electric field $\vec E$. 
Instead, it is a functional of the electric field {\it and the initial magnetic field} $\vec B_0$. For a typical experimental setup where the external and induced quantities vanish before some initial switching-on, this does not pose any problem, because if we choose $t_0$ sufficiently early, the initial fields can be set to zero (see the discussion in Sec.~\ref{subsec_Maxwell_initial}). Unfortunately, however, this assumption precludes the possibility
of static fields which have the same value at all times. Strictly speaking, such static fields cannot be produced in the laboratory, but for many situations they are a suitable idealization. It is therefore desirable to allow more generally for external and induced fields which do not vanish before any initial time $t_0$. We are thus led to consider the more general case of fields with a nonvanishing {\itshape static contribution}
\begin{equation}
\vec B_0(\vec x) =\lim_{t_0 \rightarrow-\infty}\vec B(\vec x,t_0) \,,
\end{equation}
and analogously for the vector potential. The solution \eqref{eq_sol} can be written in the limit $t_0 \to -\infty$ as
\begin{equation} \label{eq_thissol}
 \vec A(\vec x, t) = -\int_{-\infty}^{\infty} \de t' \, \varTheta(t - t') \, \vec E(\vec x, t') + \vec A_0(\vec x) \,.
\end{equation}
This equation shows that for the inital value problem at hand, the negative Heaviside step function plays the r\^{o}le of the retarded Green function, while the propagator is given by the identity operator (cf.~Sec.~\ref{subsec_scalar}). Using that
\begin{equation}
\varTheta(t-t')=\j\int\frac{\de\omega}{2\pi} \, \frac{\e^{-\j\omega(t-t')}}{\omega+\j\eta} \,,
\end{equation}
we can write this solution in the momentum and frequency domain as
\begin{equation} \label{eq_solFourier}
 \vec A(\vec k, \omega) = \frac{1}{\j(\omega + \j \eta)} \h \vec E(\vec k, \omega) + \frac{\j \vec k \times \vec B_0(\vec k)}{|\vec k|^2} \, \delta(\omega) \,.
\end{equation}
In fact, this is a distributional solution to the equation
\begin{equation}
 \j \omega \h \vec A(\vec k, \omega) = \vec E(\vec k, \omega) \,.
\end{equation}
We conclude that in order to incorporate external magnetic fields with a static contribution, Eq.~\eqref{eq_AEtot} should be replaced by Eq.~\eqref{eq_solFourier}.
By contrast, if we consider fields produced in the laboratory after some initial switching-on, these fields vanish in the limit $t \to -\infty$, and hence we recover precisely our original formula \eqref{eq_AEtot} provided we regularize it by the prescription \mbox{$\omega\mapsto\omega+\j\eta$\hh.}

Next, let us derive also the functional $\vec A[\vec E_{\mathrm L},\vec B]$ allowing for a static contribution to the magnetic field. As indicated by the Dirac delta distribution in Eq.~\eqref{eq_solFourier}, for fields which do not vanish in the limit $t \rightarrow -\infty$, the Fourier transform becomes in general singular. Since formal manipulations of singular Fourier transforms may easily lead to nonsensical results, we have to proceed more carefully. We therefore avoid the frequency domain and instead start again from Eq.~\eqref{eq_sol} in the time domain (where we let $t_0 \to -\infty$). We decompose the electric field under the integral into its longitudinal
and transverse parts, and write the latter as
\begin{equation}
\vec E_{\mathrm T}(\vec x,t)= \frac{1}{4\pi} \, \nabla \times \int \! \de^3 \vec x' \, \frac{-\partial_t \vec B(\vec x', t)}{|\vec x - \vec x'|} \,, \smallskip
\end{equation}
which is equivalent to  Faraday's law. Plugging this into Eq.~\eqref{eq_sol} and using Eq.~\eqref{use_this}, we observe that the two contributions from $\vec B_0(\vec x)$ cancel, and we arrive at
\begin{equation}
 \vec A(\vec x, t) = -\int_{-\infty}^t \de t' \, \vec E_{\mathrm L}(\vec x, t') + \frac{1}{4\pi} \, \nabla \times \int \! \de^3 \vec x' \, \frac{\vec B(\vec x', t)}{|\vec x - \vec x'|} \,.
\end{equation}
In Fourier space, we can now rewrite this equation as
\begin{equation}
 \vec A(\vec k, \omega) = \frac{1}{\j(\omega + \j \eta)} \h \vec E_{\mathrm L}(\vec k, \omega) + \frac{\j\vec k \times \vec B(\vec k, \omega)}{|\vec k|^2} \,.
\end{equation}
This expression coincides precisely with our original formula \eqref{eq_AELB} under the regularization prescription $\omega \mapsto \omega + \j \eta$. Hence, in contrast to Eq.~\eqref{eq_AEtot}, this equation remains valid even in the presence of static magnetic fields.

In summary, we have introduced three ways to represent the vector potential in terms of the electric and magnetic fields:
\begin{align}
\vec A[\vec E,\vec B] & = \frac{\delta \vec A}{\delta \vec E} \, \vec E + \frac{\delta \vec A}{\delta \vec B} \, \vec B \,, \label{eq_exp_A_1} \\[7pt]
\vec A[\vec E,\vec B_0] & = \frac{\de \vec A}{\de \vec E} \, \vec E + \frac{\de\vec A}{\de\vec B} \, \vec B_0\,, \label{eq_exp_A_2} \\[7pt]
\vec A[\vec E_{\mathrm L},\vec B] & = \frac{\de \vec A}{\de \vec E} \, \vec E_{\mathrm L} + \frac{\de \vec A}{\de \vec B} \, \vec B \,, \label{eq_exp_A_3}
\end{align}

\vspace{3pt} \noindent
Here, we use the notation for $3\times 3$ matrices
\begin{equation}
 \left( \frac{\delta \vec A}{\delta \vec E} \right)_{\!\!ij} = \frac{\delta A_i}{\delta E_j} \,,  \smallskip
\end{equation}
and $\vec B_0 \equiv \vec B_0(\vec k,\omega) = \vec B_0(\vec k)\h \delta(\omega)$ denotes the static contribution to the magnetic field. The partial and the total functional derivatives appearing in the above expansions are given explicitly in the temporal gauge by Eqs.~\eqref{eq_partial_1}--\eqref{eq_partial_2} and by Eqs.~\eqref{eq_total_1}--\eqref{eq_total_2}, respectively. In the following section, we will use these expansions to derive electromagnetic material properties from the current response to an external vector potential. Moreover, we will show in Sec.~\ref{sec_comp} that the three different representations of the vector potential lead to three different but equivalent expansions of the induced fields in terms of the external electric and magnetic fields.

\section{Electrodynamics of materials} \label{sec_func}

\subsection{Fundamental response functions} \label{subsec_fund}

In a general experimental setup, the electromagnetic response of a material
probe is determined under an externally applied electromagnetic perturbation. For the whole system comprising the probe and the external perturbation, this means that all electromagnetic quantities (such as fields, charges and currents) are split into {\itshape internal} and {\itshape external} contributions. Thinking of the external perturbation as inducing a redistribution of charges and currents in the sample, a natural starting point for the theoretical description is the functional (see, e.g., \cite{Altland, Adler})
\begin{equation} \label{eq_func}
 j^\mu\itl = j^\mu\itl [ A^\nu\ext ] \,,
\end{equation}
where $j^\mu\itl$ is the four-current of the medium and $A^\nu\ext$ the four-potential of the perturbation. It is inherent in this definition that on the most fundamental level, the induced currents and charges are not functionals 
of the external electric and magnetic fields but of the electromagnetic potentials. 
The main arguments in favor of this viewpoint are: (i) In general, the electric and magnetic fields are not independent of each other, hence the functional $j_{\rm int}^\mu[\vec E\ext, \vec B\ext]$ is a priori not well-defined, (ii) the functional \eqref{eq_func} connects two Lorentz four-vectors and therefore leads to a manifestly Lorentz-covariant description of the electromagnetic response, and (iii) it is the external four-potential which couples to the microscopic degrees of freedom of the system, 
namely by the minimal coupling prescription $\partial^\mu \mapsto \partial^\mu + {\rm i} e A^\mu\ext/\hbar$ in the fundamental Hamiltonian of the material.

In general, the functional dependence \eqref{eq_func}
may be complicated, depend on past history (hysteresis), be nonlinear, etc.~\cite{Jackson}. Here we only assume that it is differentiable, hence up to first order the Taylor series expansion reads
\begin{equation}
 j^\mu\itl(x) = j^\mu_{\rm int,0}(x) + \int \! \mathrm d^4 x' \, \frac{\delta j^\mu\itl(x)}{\delta A^\nu\ext(x')} \left( A^\nu\ext(x') - A^\nu_{\rm ext,0}(x') \h \right).
\end{equation}
In this equation, the functional derivative is to be evaluated at the reference potential $A^\nu_{\rm ext, 0}$. The difference between internal currents in the presence and in the absence of the external perturbation is called the {\itshape induced four-current},
\begin{equation}
 j^\mu\ind = j^\mu\itl - j^\mu_{\rm int,0} \,.
\end{equation}
The above expansion is typically performed around $A^\nu_{\rm ext,0} \equiv 0$, but it is also possible to consider small perturbations around a finite reference potential. (In this way one can describe, for example, the current response to an applied voltage in the presence of a constant external magnetic field, i.e.~the Hall conductivity \cite{Fukuyama}.) From the functional point of view, linear response theory restricts attention to the first order terms, i.e.~to the calculation of the functional derivatives
\begin{equation} \label{eq_FRF}
 \chi\indices{^\mu_\nu}(x,x') = \frac{\delta j^\mu\ind(x)}{\delta A^\nu\ext(x')} \,, \smallskip
\end{equation}
which by the discussion in Sec.~\ref{subsec_Maxwell_initial} are causal (i.e.~retarded) response functions. We will refer to this $4 \! \times \! 4$ tensor as the {\itshape fundamental response tensor}, and to its components as the {\itshape fundamental response functions.} We do not presume that the system under consideration actually behaves linearly, but instead we will make generally valid statements about the first order derivatives of induced quantities with respect to external perturbations.

\begin{figure}[t]
\begin{center}
\includegraphics[width=0.65\textwidth]{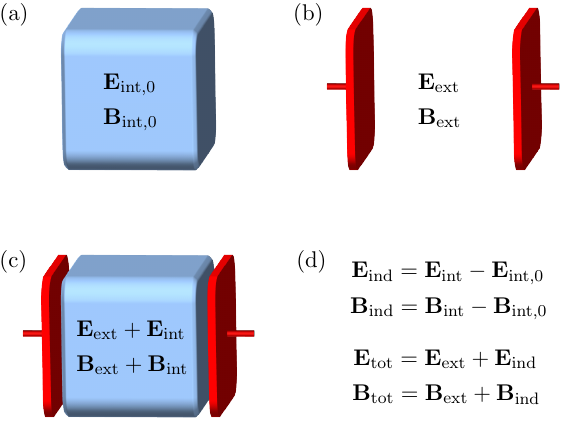}
\caption{Definition of external, internal, induced and total fields (cf.~\cite{Fliessbach, Smith}). (a) ($\vec E_{\rm int,0}$, $\vec B_{\rm int,0}$) are the fields produced by the internal sources, i.e., the charges and currents constituting the material sample (without the external perturbation). (b) The {\itshape external fields} are produced by the external sources in the absence of the medium. (c) In the presence of the external sources, the fields produced by the (redistributed) internal sources are a superposition of the external and the {\itshape internal fields}. (d) Subtracting from the latter the fields $(\vec E_{\rm int,0}, \vec B_{\rm int,0})$ yields the {\itshape induced fields}. Finally, the {\itshape total fields} are defined as the sum of the external and the induced fields.
\label{fig_fields}}
\end{center}
\end{figure}

The following constraints on $\chi\indices{^\mu_\nu}$ are required by the continuity equation and by the gauge invariance of the induced current, respectively \cite{Altland, Fradkin}:
\begin{align}
 \partial_\mu\chi\indices{^\mu_\nu}(x,x') & = 0 \,, \label{eq_gaugeinv_1} \\[5pt]
 \quad \partial'^\nu \chi\indices{^\mu_\nu}(x, x') & = 0 \,. \label{eq_gaugeinv_2}
\end{align}
Assuming homogeneity in time, these constraints are equivalent to
\begin{align}
 \frac{1}{c} \frac{\partial}{\partial t} \, \chi\indices{^0_\nu}(\vec x, \vec x'; t - t') + \sum_i \frac{\partial}{\partial x^i} \, \chi\indices{^i_\nu}(\vec x, \vec x'; t - t') & = 0 \,, \\[5pt]
 -\frac{1}{c} \frac{\partial}{\partial t'} \, \chi\indices{^\mu_0}(\vec x, \vec x'; t - t') + \sum_j \frac{\partial}{\partial x'^j} \, \chi\indices{^\mu_j}(\vec x, \vec x'; t - t') & = 0 \,.
\end{align}
Therefore, at most 9 of the 16 fundamental response functions are independent of each other (cf.~\cite{Giuliani, Cho08, Cho10}). 
Explicitly, we can express all components $\chi\indices{^\mu_\nu}$ in terms of the spatial components $\chi\indices{^i_j} \equiv \chi_{ij}$ as
\begin{align}
 \chi\indices{^0_j}(\vec x, \vec x'; \omega) & = \frac{c}{\j\omega} \, \sum_{i} \frac{\partial}{\partial x^i} \, \chi_{ij}(\vec x, \vec x'; \omega) \,, \label{eq_recon_1} \\[5pt]
 \chi\indices{^i_0}(\vec x, \vec x'; \omega) & = \frac{c}{\j\omega} \, \sum_{j} \frac{\partial}{\partial x'^j} \, \chi_{ij}(\vec x, \vec x'; \omega) \,, \label{eq_recon_2} \\[5pt]
 \chi\indices{^0_0}(\vec x, \vec x'; \omega) & = -\frac{c^2}{\omega^2} \, \sum_{i, \h j} \frac{\partial}{\partial x^i} \, \frac{\partial}{\partial x'^j} \, \chi_{ij}(\vec x, \vec x'; \omega) \,. \label{eq_recon_3}
\end{align}
In Fourier space, the fundamental response tensor can therefore be written compactly as
\begin{equation}\label{generalform}
\chi^\mu_{~\nu}(\vec k,\vec k',\omega)=
\left( \!
\begin{array}{rr} -\lar{\frac{c^2}{\omega^2}} \, \vec k^{\rm T} \, \tsr \chi \, \vec k' & \lar{\frac{c}{\omega}} \, \vec k^{\rm T} \, \tsr \chi \, \\[10pt] -\lar{{\frac{c}{\omega}}} \, \tsr \chi \, \vec k' & \, \tsr \chi \, 
\end{array} \right). \smallskip
\end{equation}
Apart from reducing the number of independent response functions, focusing on the fundamental response functions in the first place bears several advantages: They constitute a second-rank Lorentz tensor and thus yield a relativistic description of the electromagnetic response. This is especially relevant for the study of moving media \cite{Leonhardt, Mackay07, Mackay10} (see also~\cite{EDOhm} for a discussion of Ohm's law). The relativistic transformation law for the fundamental response functions reads explicitly
\begin{equation}
 \chi\indices{^\mu_\nu}(x,y) = \Lambda\indices{^\mu_{\mu'}} \, \Lambda\indices{_\nu^{\nu'}} \chi\indices{^{\mu'}_{\nu'}}(\Lambda^{-1} x, \Lambda^{-1} y) \,,
\end{equation}
with Lorentz transformations $\Lambda \in \mathrm O(3,1)$. 
Furthermore, the minimal coupling prescription implies that the fundamental response functions can be calculated directly in the Kubo formalism (see, e.g., \cite[Eqs.~(2.154)--(2.155)]{SchafWegener} and \cite{Altland, Kubo}).
However, the crucial question is whether one can derive closed expressions for the usual electromagnetic response functions (such as the conductivity, the dielectric tensor and the magnetic susceptibility) in terms of the fundamental response functions. This question will be addressed in Sections~\ref{subsec_emprop} and \ref{sec_univ}.

\subsection{Connection to Schwinger--Dyson and Hedin equations} \label{subsec_hedin}

As described in the previous subsection, the Functional Approach to electrodynamics of materials is based on the splitting of all electromagnetic quantities into external and internal (or induced) contributions. Each of these subsystems can be described equivalently by its charges and currents $j^\mu = (c\rho,\vec j)$, by its gauge potential $A^\mu = (\varphi/c, \vec A)$, or by its electric and magnetic fields $\{\vec E, \vec B\}$. The mutual functional dependencies of these different field quantities have been derived explicitly in the first part of this paper. In particular, the {\itshape induced electromagnetic fields} are uniquely determined as functionals of the induced charges and currents, which were defined in the previous subsection (see Sec.~\ref{subsec_Maxwell_initial}).
Schematically, the external, internal, induced and total fields are shown in Fig.~\ref{fig_fields}.

Now given the functional dependence of the induced four-current on the external four-potential, it follows that all other electromagnetic response functions (specifying, for example, the induced current in terms of the external electric field, or in terms of the external charges and currents) can be calculated. Moreover, Eq.~\eqref{eq_FRF} defines only one of a number of equivalent response tensors: instead of describing 
the response of the system in terms of the induced quantities, one may as well characterize it by the functional dependence of the total on the external, or even the induced on the total field quantities. Each of these possibilities gives rise to an equivalent set of response functions, all of which are mutually related by functional chain rules.
As an example, we consider the converse point of view to Eq.~\eqref{eq_FRF}, namely the dependence of the total four-potential on the external four-current. We define the {\itshape full tensorial Green function} as
\begin{equation} \label{eq_full_prop}
 D\indices{^\mu_\nu}(x,x') = \frac{\delta A^\mu\tot(x)}{\delta j^\nu\ext(x')} \,. \smallskip
\end{equation}
A straightforward application of the functional chain rule in the form
\begin{equation}
 \frac{\delta A\tot}{\delta j\ext} = \frac{\delta A\ext}{\delta j\ext}
 + \frac{\delta A\ind}{\delta j\ind} \frac{\delta j\ind}{\delta A\ext} \frac{\delta A\ext}{\delta j\ext}
\end{equation}
leads to the equation
\begin{equation}
 D = D_0 + D_0 \h \chi D_0 \,. \smallskip
\end{equation}
This relates $D$ to the fundamental response tensor $\chi$ via the free tensorial Green function $D_0$, which was defined in Sec.~\ref{sec_pre_propagators}. Here we have used that $A=D_0\h j$ \h implies in particular $D_0=\delta A/\delta j$ (cf.~\cite[p.~21]{Sibold}). Note that the formal products refer to the natural multiplication in the linear space of $4 \times 4$ tensorial integral kernels.

Furthermore, we introduce the {\itshape proper response functions} $\widetilde\chi$ which specify the response of the induced four-current to the total (instead of the external) four-potential,
\begin{equation}
 \widetilde\chi\indices{^\mu_\nu}(x, x') = \frac{\delta j^\mu\ind(x)}{\delta A^\nu\tot(x')} \,. \smallskip
\end{equation}
Again, the functional chain rule
\begin{equation}
 \frac{\delta j\ind}{\delta A\ext} = \frac{\delta j\ind}{\delta A\tot} \frac{\delta A\tot}{\delta A\ext}
 = \frac{\delta j\ind}{\delta A\tot} + \frac{\delta j\ind}{\delta A\tot} \frac{\delta A\ind}{\delta j\ind} \frac{\delta j\ind}{\delta A\ext}
\end{equation}
shows that these quantities are related to the fundamental response functions through
\begin{equation}\label{Dysonchi}
 \chi = \widetilde\chi + \widetilde\chi \h D_0 \h \chi \,. \smallskip
\end{equation}
In terms of these proper response functions, the full tensorial Green function $D$ satisfies
\begin{equation} \label{eq_Dyson}
 D = D_0 + D_0 \h \widetilde \chi \h D \,, \smallskip
\end{equation}
as follows again by a functional chain rule. This last equation has a Dyson-like structure and formally coincides with the Schwinger--Dyson equation for the full tensorial Green function in the sense of quantum electrodynamics \cite{Itzykson, Bjorken}, provided one identifies $\widetilde \chi$ with the irreducible photon self-energy. In fact, the photon self-energy is given by the current-current correlation function evaluated in the vacuum \cite{Roepstorff}, which corresponds to the Kubo formula for the fundamental response function $\chi$. This shows the literal analogy between solid state physics and quantum electrodynamics, where the vacuum plays the r\^{o}le of the polarizable material. Moreover, our analysis provides via Eq.~\eqref{eq_full_prop} a classical interpretation of the full tensorial Green function: just as $D_0$ yields the electromagnetic four-potential in terms of its own sources,
$D$ yields the total four-potential in terms of the external four-current, i.e.~to linear order,
\begin{equation}
 A^\mu\tot = D\indices{^\mu_\nu} \, j^\nu\ext \,. \smallskip
\end{equation}
We note, however, that in quantum electrodynamics one usually works with time-ordered instead of retarded response functions.

A non-relativistic version of the Schwinger--Dyson equations is known in electronic structure theory as {\itshape Hedin's equations} \cite{Hedin65, Hedin69}. These are among the most important first-principles techniques used nowadays for describing single-electron excitations in real materials (see \cite{Aryasetiawan09, Held}, or \cite{Starke} for a recent discussion in the context of Green function theory). They can be obtained from the relativistic Schwinger--Dyson equations by replacing the electronic Green function of the Dirac equation by the Green function of the Schr\"{o}dinger or the Pauli equation and approximating the full tensorial Green function in the Coulomb gauge by its $00$-component. Indeed, by keeping in Eq.~\eqref{eq_Dyson} only the Coulomb potential (cf.~Eq.~\eqref{eq_CoulGF})
\begin{equation}
 v = c^2 \h (D_0)\indices{^0_0}
\end{equation}
and the proper density response function
\begin{equation}
 \widetilde \upchi = \frac{1}{c^2} \h \widetilde \chi\indices{^0_0} \,,
\end{equation}
we obtain the approximate relation for $w = c^2 D\indices{^0_0}$\hh:
\begin{equation}
 w = v + v \, \widetilde \upchi \h w \,.
\end{equation}
This is precisely Hedin's equation for the {\itshape screened potential} $w$, where $\widetilde \upchi$ is known in electronic structure theory as the {\itshape (irreducible) polarizability} \cite{Botti, Bussi}.

\subsection{Material properties from fundamental response functions} \label{subsec_emprop} 

In electrodynamics of materials, the response of the polarization $\vec P$ and the magnetization~$\vec M$ to the external electric and magnetic fields $\vec E_{\rm ext}$ and $\vec B_{\rm ext}$ is especially relevant. Hence, it is desirable to derive these material properties directly from the fundamental response functions. Here, we are facing two basic questions:
(i) How is the response of $\vec P$ and $\vec M$ related to the induced four-current $j^\mu\ind$\h, and 
(ii) how do we express the response to $\vec E\ext$ and $\vec B\ext$ in terms of the response to the external four-potential $A^\nu\ext$\h. The first problem is solved under the identifications \eqref{eq_PE}--\eqref{eq_BB}: The Maxwell equations for the induced fields \eqref{eq_maxwell_induced_1}--\eqref{eq_maxwell_induced_4} imply the equations of motion \vspace{-3pt}
\begin{align}
 \Box\vec P(\vec x, t) & = \nabla \rho\ind(\vec x, t) + \frac{1}{c^2} \frac{\partial}{\partial t} \h \vec j\ind(\vec x, t) \,, \label{eq_eom_P} \\[5pt]
 \Box\vec M(\vec x, t) & = \nabla \times \vec j\ind(\vec x, t) \,.
\end{align}
By using the retarded Green function for the scalar wave equation, Eq.~\eqref{eq_ret_gf},
we thus obtain the induced fields in terms of the induced currents.
As for the second problem, it is natural to use the functional chain rules
\begin{align}
 \frac{\delta j^\mu\ind(x)}{\delta E^\ell\ext(x')} & = \int \! \mathrm d^4 y \, \frac{\delta j^\mu\ind(x)}{\delta A^\alpha\ext(y)} \, \frac{\delta A\ext^\alpha(y)}{\delta E\ext^\ell(x')} \,, \label{partOhm_1} \\[5pt]
 \frac{\delta j^\mu\ind(x)}{\delta B^\ell\ext(x')} & = \int \! \mathrm d^4 y \, \frac{\delta j^\mu\ind(x)}{\delta A^\alpha\ext(y)} \, \frac{\delta A\ext^\alpha(y)}{\delta B\ext^\ell(x')} \,, \label{partOhm_2}
\end{align}
and to evaluate these formulae using the functional defined in Sec.~\ref{subsec_Hamgauge},
 \begin{equation} \label{eq_Afunc}
 A^\mu[\vec E, \vec B] = \frac{1}{2\mu_0} \h \big( \partial_\lambda (D_0)\indices{^\mu_\nu} - \partial_\nu (D_0)\indices{^\mu_\lambda} \h \big) F^{\nu\lambda} + (\partial^\mu \hspace{-1pt} f)[\vec E, \vec B] \,,
\end{equation}
where $F^{\nu\lambda}$ is expressed explicitly in terms of $\vec E$ and $\vec B$ by Eq.~\eqref{eq_FEB}. However, as argued before, this functional is not uniquely determined for two reasons: (i) It contains an arbitrary pure gauge $\partial^\mu \hspace{-1pt} f$ (where the function $f$ may itself depend on $\vec E$ and $\vec B$), and (ii) there are infinitely many choices for the tensorial Green function $D_0$. We now have to show that despite \linebreak  this ambiguity of the functional $A^\mu[\vec E, \vec B]$, the electromagnetic response \linebreak

\pagebreak \noindent
functions \eqref{partOhm_1}--\eqref{partOhm_2} are well-defined. In order to do this, we first note that no matter how we choose $f[\vec E, \vec B]$ and $D_0$, the resulting four-potential
produces the same electromagnetic fields, namely for $A^\mu = (\varphi/c, \vec A)$:
\begin{align}
 (-\nabla \varphi - \partial_t \vec A)[\vec E, \vec B] & = \vec E \,, \\[3pt]
 (\nabla \times \vec A)[\vec E, \vec B] & = \vec B \,.
\end{align}
Since any two four-potentials $A^\mu$ and $A'^\mu$ which produce the same electromagnetic fields differ by a pure gauge,
\begin{equation}
 A^\mu[\vec E, \vec B] = A'^\mu[\vec E, \vec B] + ( \partial^\mu \hspace{-1pt} \tilde f \h )[\vec E, \vec B] \,,
\end{equation}
the arbitrariness in the choice of $D_0$ can be absorbed in the gauge freedom. Hence, it only remains to show 
that Eqs.~\eqref{partOhm_1}--\eqref{partOhm_2} yield gauge-independent results. For this purpose, it suffices to convince oneself that for an arbitrary function $\tilde f[\vec E, \vec B]$ the following equation holds:
\begin{equation}
 \int \! \mathrm d^4 y \,\h \chi\indices{^\mu_\alpha}(x, y) \, \frac{\delta (\partial^\alpha \hspace{-1pt} \tilde f \h )(y)}{\delta E\ext^\ell(x')} = 0
\end{equation}
(and the analogous equation for the magnetic field). Using that the functional
derivative commutes with the partial derivative and performing a partial integration, this condition follows directly from the constraint \eqref{eq_gaugeinv_2} on the fundamental response functions.

\subsection{Field strength response tensor} \label{subsec_fsrt}

We now derive an explicit expression for the partial functional derivatives of
the induced fields with respect to the external fields in terms
of the fundamental response tensor. These partial derivatives will be related to the constitutive tensor of bi-ansiotropic materials at the end of this subsection. We proceed in a manifestly relativistic framework:
In fact, the response functions relating induced to external fields
can be grouped into the {\it field strength response tensor}
\begin{equation}
\chi\indices{^{\mu\nu}_{\alpha\beta}}(x,x')
=\frac{\delta F^{\mu\nu}\ind(x)}{\delta F^{\alpha\beta}\ext(x')} \,.  \smallskip \label{eq_fsrt}
\end{equation}
By a functional chain rule, we can express this fourth rank tensor through the
fundamental response tensor as
\begin{equation} \label{eq_zw0}
\chi\indices{^{\mu\nu}_{\alpha\beta}}(x, x') =
\int \! \de^4 y \int \! \de^4 y' \
\frac{\delta F^{\mu\nu}\ind(x)}{\delta j^\lambda\ind(y)} \,
\chi\indices{^\lambda_\rho}(y, y') \,
\frac{\delta A^\rho\ext(y')}{\delta F^{\alpha\beta}\ext(x')} \,.
\end{equation}
We calculate the functional derivatives in the integrand as described in the previous subsection: From Eq.~\eqref{eq_FscDj},  we first obtain
\begin{equation} \label{eq_zw1}
\frac{\delta F^{\mu\nu}(x)}{\delta j^\lambda(y)} = (\partial^\mu \mathbbmsl D_0)(x-y) \, \eta\indices{^\nu_\lambda} -
(\partial^\nu \mathbbmsl D_0)(x-y) \, \eta\indices{^\mu_\lambda} \,.
\end{equation}
On the other hand, Eq.~\eqref{eq_FA} yields
\begin{equation}
 \frac{\delta A^\rho(y')}{\delta F^{\alpha\beta}(x')} = \frac{1}{2\mu_0} \h \big( \partial'_\beta (D_0)\indices{^\rho_\alpha}(y'-x') - \partial'_\alpha (D_0)\indices{^\rho_\beta}(y'-x') \h \big) \,,
\end{equation}
where the partial derivatives act on the $y'$ variables. As shown above, the result of Eq.~\eqref{eq_zw0} does not depend on the choice of the Green function $D_0$ in the functional $A^\rho[F^{\alpha\beta}]$, and by choosing the Feynman Green function \eqref{eq_Feynman} we obtain the simpler expression
\begin{align}
\frac{\delta A^\rho(y')}{\delta F^{\alpha\beta}(x')} & = \frac{1}{2\mu_0} \h \big( \partial'_\beta \mathbbmsl D_0(y'-x') \, \eta\indices{^\rho_\alpha} - \partial'_\alpha \mathbbmsl D_0(y'-x') \, \eta\indices{^\rho_\beta} \h \big) \,. \label{eq_zw2}
\end{align}
Inserting Eqs.~\eqref{eq_zw1} and \eqref{eq_zw2} into Eq.~\eqref{eq_zw0}  and performing partial integrations with respect to the $y$ and $y'$ variables, we arrive at the formula
\begin{equation}
\begin{aligned}
\chi\indices{^{\mu\nu}_{\alpha\beta}}(x, x') & =
\frac{1}{2\mu_0} \int \! \de^4 y \int \! \de^4 y' \, \mathbbmsl D_0(x-y) \\[5pt]
& \quad \times \Big(
\partial^\mu \h \partial'_\alpha \h \chi\indices{^\nu_\beta}(y,y')-
\partial^\nu \h \partial'_\alpha \h \chi\indices{^\mu_\beta}(y,y') \\[5pt]
& \quad \quad \quad - \, \partial^\mu \h \partial'_\beta \h \chi\indices{^\nu_\alpha}(y,y')+
\partial^\nu \h \partial'_\beta \h \chi\indices{^\mu_\alpha}(y,y')
\Big) \, \mathbbmsl D_0(y'-x') \,.
\end{aligned}
\end{equation}
Equivalently, we can write this in momentum space as
\begin{equation}
\begin{aligned}
\chi\indices{^{\mu\nu}_{\alpha\beta}}(k, k') & =
\frac{1}{2\mu_0} \, \mathbbmsl D_0(k) \, \Big(
k^\mu \h \chi\indices{^\nu_\beta}(k,k') \h k'_\alpha -
k^\nu \h \chi\indices{^\mu_\beta}(k,k') \h k'_\alpha \\[3pt]
& \quad - \, k^\mu \h \chi\indices{^\nu_\alpha}(k,k') \h k'_\beta +
k^\nu \h \chi\indices{^\mu_\alpha}(k,k') \h k'_\beta
\Big) \, \mathbbmsl D_0(k') \,. \label{eq_chichi}
\end{aligned}
\end{equation}
These general, manifestly Lorentz-covariant equations can already be found in Ref.~\cite[Eq.~(1.5.16)]{Melrose1Book}, where they have been discovered apparently for the first time. They allow for the calculation of all partial derivatives of the induced fields with respect to the external fields in terms of the fundamental response functions. Moreover, they fully take into account
the inhomogeneity, anisotropy and magnetoelectric coupling of the material as well as relativistic retardation effects, and they establish the connection to the experiment through the linear expansion
\begin{equation}\label{fundexp}
F^{\mu\nu}\ind(k) = \int \! \de^4 k' \, \chi\indices{^{\mu\nu}_{\alpha\beta}}(k, k') \, F^{\alpha\beta}\ext(k') \,.
\end{equation}
Spelling out the electric and magnetic field components of the (induced and external) field strength tensors, we obtain the equivalent expansions
\begin{align}
 \vec E\ind(k) & = \int \! \de^4 k' \, \tsr \chi{}\EE\p (k, k') \, \vec E\ext(k') + \int \! \de^4 k' \, \tsr \chi{}\EB\p (k, k') \, c \h \vec B\ext(k') \,, \label{eq_part_exp1} \\[5pt]
 c \h \vec B\ind(k) & = \int \! \de^4 k' \, \tsr \chi{}\BE\p (k, k') \, \vec E\ext(k') + \int \! \de^4 k' \, \tsr \chi{}\BB\p (k, k') \, c \h \vec B\ext(k') \,. \label{eq_part_exp2}
\end{align}
Here, the coefficient functions are given in terms of the field strength response tensor by
\begin{align}
 \frac{\delta E_{\rm ind}^i(k)}{\delta E_{\rm ext}^j(k')} & \, \equiv \, \big(\chi{}\EE\p\big){}\indices{^i_j}(k, k') \, = \, 2 \h \chi\indices{^{0i}_{0j}}(k, k') \,, \label{eq_par_1} \\[5pt]
 \frac{1}{c} \h \frac{\delta E_{\rm ind}^i(k)}{\delta B_{\rm ext}^j(k')} & \, \equiv \, \big(\chi{}\EB\p\big){}\indices{^i_j}(k, k') \, = \, \epsilon_{j\ell n} \, \chi\indices{^{0i}_{\ell n}}(k, k') \,, \\[5pt]
 c \, \frac{\delta B_{\rm ind}^i(k)}{\delta E_{\rm ext}^j(k')} & \, \equiv \, \big(\chi{}\BE\p\big){}\indices{^i_j}(k, k') \, = \, \epsilon_{ikm} \, \chi\indices{^{km}_{0j}}(k, k') \,, \\[5pt]
 \frac{\delta B_{\rm ind}^i(k)}{\delta B_{\rm ext}^j(k')} & \, \equiv \, \big(\chi{}\BB\p\big){}\indices{^i_j}(k, k') \, = \, \frac 1 2 \, \epsilon_{ikm} \, \epsilon_{j \ell n} \, \chi\indices{^{km}_{\ell n}}(k,k') \,. \label{eq_par_4}
\end{align}
The superscript $\mathrm p$ indicates {\itshape partial} functional derivatives, meaning that the external electric and magnetic fields are varied independently of each other. By contrast, any physical response function should be identified with a {\itshape total} functional derivative, as we will argue in Sec.~\ref{subsec_physical}.

By our formula \eqref{eq_chichi} we have derived a closed expression for the field strength response tensor in terms of the fundamental response tensor. As shown in Sec.~\ref{subsec_fund}, the fundamental response tensor has only 9 independent components. Therefore, it is also possible to express all components of the field strength response tensor in terms of only 9 independent current response functions. We will now derive explicit expressions for the 36 partial derivatives \eqref{eq_par_1}--\eqref{eq_par_4} in terms of the 9 spatial components $\chi_{mn}$ of the fundamental response tensor. Consider, for example, the electric field response
\begin{equation}
\begin{aligned}
 \big(\chi{}\EE\p\big){}\indices{^i_j}(k, k') & = \frac{1}{\mu_0} \, \mathbbmsl D_0(k) \, \Big(
k^0 \h \chi\indices{^i_j}(k,k') \h k'_0 -
k^i \h \chi\indices{^0_j}(k,k') \h k'_0 \\[3pt]
& \quad - \, k^0 \h \chi\indices{^i_0}(k,k') \h k'_j +
k^i \h \chi\indices{^0_0}(k,k') \h k'_j
\Big) \, \mathbbmsl D_0(k') \,.
\end{aligned}
\end{equation}
Using the constraints \eqref{eq_recon_1}--\eqref{eq_recon_3} on the fundamental response functions to express $\chi\indices{^0_j}$, $\chi\indices{^i_0}$ and $\chi\indices{^0_0}$ in terms of the spatial components $\chi_{mn}$, we obtain
\begin{equation}
\begin{aligned}
 \big(\chi{}\EE\p\big){}\indices{^i_j}(\vec k, \vec k'; \omega) & = \frac{1}{\mu_0} \, \mathbbmsl D_0(\vec k, \omega) \left( -\frac{\omega}{c} \, \delta_{im} + \frac{c \h k_i k_m}{\omega} \right) \chi_{mn}(\vec k, \vec k'; \omega) \\[3pt]
 & \quad \times \left( \frac{\omega}{c} \, \delta_{nj} - \frac{c \h k'_n k'_j}{\omega} \right) \mathbbmsl D_0(\vec k', \omega) \,.
\end{aligned}
\end{equation}
In the same way, we can express also the other partial derivatives in terms of $\chi_{mn}$. 
We summarize all results by the following set of equations:
 \begin{align}
\frac{\delta E^i_{\rm ind}(\vec k, \omega)}{\delta E^j_{\rm ext}(\vec k', \omega)} & =
 -\frac{1}{\varepsilon_0 \h \omega^2} \,
 \frac{\omega^2 \delta_{im} - c^2 k_i k_m}{\omega^2 - c^2 |\vec k|^2} \ 
 \chi_{mn}(\vec k, \vec k'; \omega) \
 \frac{\omega^2 \delta_{nj} - c^2 k'_n k'_j}{\omega^2 - c^2 |\vec k'|^2} \,, \label{eq_part_der1} \\[-5pt]
\frac{1}{c} \, \frac{\delta E^i_{\rm ind}(\vec k, \omega)}{\delta B^j_{\rm ext}(\vec k', \omega)} & =
 -\frac{1}{\varepsilon_0 \h \omega^2} \,
 \frac{\omega^2 \delta_{im} - c^2 k_i k_m}{\omega^2 - c^2 |\vec k|^2} \
 \chi_{mn}(\vec k, \vec k'; \omega) \
 \frac{\epsilon_{n\ell j} \, \omega \, c \h k'_\ell}{\omega^2 - c^2 |\vec k'|^2} \,, \label{eq_part_der2} \\[10pt]
c \, \frac{\delta B^i_{\rm ind}(\vec k, \omega)}{\delta E^j_{\rm ext}(\vec k', \omega)} & =
 -\frac{1}{\varepsilon_0 \h \omega^2} \,
 \frac{\epsilon_{ikm} \, \omega \, c \h k_k}{\omega^2 - c^2 |\vec k|^2} \
 \chi_{mn}(\vec k, \vec k'; \omega) \
 \frac{\omega^2 \delta_{nj} - c^2 k'_n k'_j}{\omega^2 - c^2 |\vec k'|^2} \,, \label{eq_part_der3} \\[10pt]
\frac{\delta B^i_{\rm ind}(\vec k, \omega)}{\delta B^j_{\rm ext}(\vec k', \omega)} & =
 -\frac{1}{\varepsilon_0 \h \omega^2} \,
 \frac{\epsilon_{ikm} \, \omega \, c \h k_k}{\omega^2 - c^2 |\vec k|^2} \
 \chi_{mn}(\vec k, \vec k'; \omega) \
 \frac{\epsilon_{n\ell j} \, \omega \, c \h k'_\ell}{\omega^2 - c^2 |\vec k'|^2} \,. \label{eq_part_der4}
\end{align}

\vspace{2pt} \noindent
With the help of the electric and magnetic solution generators defined in Sec.~\ref{subsec_Hamgauge}, these relations can be written compactly as
\begin{align}
 \tsr \chi{}\EE\p(\vec k, \vec k'; \omega) & = -\frac{1}{\varepsilon_0 \h \omega^2} \, \tsr{\mathbbmsl E}(\vec k, \omega) \, \tsr \chi(\vec k, \vec k';\omega) \, \tsr{\mathbbmsl E}(\vec k', \omega) \,, \label{eq_lor_1} \\[8pt]
 \tsr \chi{}\EB\p(\vec k, \vec k'; \omega) & = -\frac{1}{\varepsilon_0 \h \omega^2} \, \tsr{\mathbbmsl E}(\vec k, \omega) \, \tsr \chi(\vec k, \vec k';\omega) \, \tsr{\mathbbmsl B}(\vec k', \omega) \,, \label{eq_lor_2} \\[8pt]
 \tsr \chi{}\BE\p(\vec k, \vec k'; \omega) & = -\frac{1}{\varepsilon_0 \h \omega^2} \, \tsr{\mathbbmsl B}(\vec k, \omega) \, \tsr \chi(\vec k, \vec k';\omega) \, \tsr{\mathbbmsl E}(\vec k', \omega) \,, \label{eq_lor_3} \\[8pt]
 \tsr \chi{}\BB\p(\vec k, \vec k'; \omega) & = -\frac{1}{\varepsilon_0 \h \omega^2} \, \tsr{\mathbbmsl B}(\vec k, \omega) \, \tsr \chi(\vec k, \vec k';\omega) \, \tsr{\mathbbmsl B}(\vec k', \omega) \,. \label{eq_lor_4}
\end{align}
Thus, we have expressed all components of the field strength response tensor, i.e., the 36 partial derivatives of the induced electric/magnetic fields with respect to the external electric/magnetic fields, in terms of only 9 independent current response functions.

Finally, we remark that a covariant approach to the electromagnetic response functions
has already been suggested by Hehl et al.~\cite{Fechner, Hehl1, Hehltensor}. The linear constitutive relations used in this approach, see \cite[Eq.~(15)]{Hehltensor}, can be interpreted as
\begin{equation} \label{eq_Hehltens}
 F\ext^{\mu\nu} = \widetilde \chi\indices{^{\mu\nu}_{\alpha\beta}} \h F\tot^{\alpha\beta} \,, \medskip
\end{equation}
with a fourth-rank {\itshape constitutive tensor} $\widetilde \chi$. These constitutive relations generalize the equation $\vec D = \varepsilon_0 \h \varepsilon_{\mathrm r} \h \vec E$ to the case of linear bi-anisotropic materials \cite{Mackay10}. Within linear response theory, the 36 components of the constitutive tensor are to be identified with the functional derivatives
\begin{equation}
\widetilde\chi\indices{^{\mu\nu}_{\alpha\beta}}(x,x')
= \frac{\delta F^{\mu\nu}\ext(x)}{\delta F^{\alpha\beta}\tot(x')} \,.
\end{equation}
Writing the total fields as a sum of the external and the induced fields, we see that
\begin{equation}
 \frac{\delta F^{\mu\nu}\tot(x)}{\delta F^{\alpha\beta}\ext(x')} = \frac 1 2 \h \big( \delta\indices{^\mu_\alpha} \delta\indices{^\nu_\beta} - \delta\indices{^\mu_\beta} \delta\indices{^\nu_\alpha} \big) \h \delta^4(x-x') + \frac{\delta F^{\mu\nu}\ind(x)}{\delta F^{\alpha\beta}\ext(x')} \,.
\end{equation}
Consequently, the following formal relation holds between the constitutive tensor and the field strength response tensor:
\begin{equation}
(\widetilde\chi^{-1})\indices{^{\mu\nu}_{\alpha\beta}}(x, x') = \frac{1}{2} \h \big( \delta\indices{^\mu_\alpha} \delta\indices{^\nu_\beta} - \delta\indices{^\mu_\beta} \delta\indices{^\nu_\alpha} \big) \h \delta^4(x-x') + \chi\indices{^{\mu\nu}_{\alpha\beta}}(x, x') \,, \label{eq_connection}
\end{equation}
where $\delta\indices{^\mu_\nu}\equiv\eta\indices{^\mu_\nu}$.
This equation combined with Eq. \eqref{eq_chichi} allows to express all 36 components of the constitutive tensor analytically in terms of only 9 independent components of the fundamental response tensor.

\section{Universal response relations} \label{sec_univ}

\subsection{Physical response functions} \label{subsec_physical}

Within the linear regime, the electromagnetic properties of any material are typically characterized by the electric susceptibility tensor $\chi_{\rm e}$,
which gives the polarization in terms of the total electric field, 
\begin{equation}
\vec P = \varepsilon_0 \h \tsr\chi_{\rm e} \h \vec E \,, 
\end{equation}
by the magnetic susceptibility $\chi_{\rm m}$ specifying the magnetization in terms of the external magnetic field, 
\begin{equation}
\vec M = \tsr\chi_{\rm m} \h \vec H \,, 
\end{equation}
and by the conductivity $\sigma$ giving the response of the induced current to the external electric field,\footnote{The conductivity $\sigma$ must not be confused with the {\itshape proper conductivity} $\widetilde\sigma$, which is defined by
\begin{equation}
 \vec j\ind = \tsr{\widetilde\sigma} \vec E\tot \,. \smallskip
\end{equation}
These two notions of conductivity are related through
\begin{equation}
 \tsr{\widetilde\sigma} = \tsr \sigma \h \tsr{\varepsilon_{\mathrm r}} \,,
\end{equation}
with the dielectric tensor $\varepsilon_{\mathrm r}$ given by Eq.~\eqref{eq_epsilon} (cf.~\cite{Smith}).} 
\begin{equation}
\vec j\ind = \tsr\sigma \vec E\ext \,. \smallskip
\end{equation}
Other common response functions are the relative permittivity or dielectric tensor $\varepsilon_{\mathrm r}$ \h defined by 
\begin{equation} \label{eq_epsilon}
\vec D = \varepsilon_0 \h \tsr{\varepsilon_{\mathrm r}} \h \vec E \,, \smallskip
\end{equation}
and the relative magnetic permeability $\mu_{\mathrm r}$ defined by 
\begin{equation}
\vec B = \mu_0 \h \tsr{\mu_{\mathrm r}} \h \vec H. 
\end{equation}
They are related to the corresponding susceptibilities by 
\begin{align}
\tsr{\varepsilon_{\mathrm r}} &= 1 + \tsr\chi_{\rm e} \,, \label{eq_eps} \\[3pt]
\tsr\mu_{\mathrm r} &= 1 + \tsr\chi_{\rm m} \,. \label{eq_mu}
\end{align}
In principle, all these response functions represent $3 \times 3$ tensorial integral kernels with respect to the space and time variables. 

Naively, one could expect that the components of the physical response tensors $\chi_{\rm e}$ and $\chi_{\rm m}$
correspond to functional derivatives which can be read out directly from the field strength response tensor of Sec.~\ref{subsec_fsrt}. It turns out, however, that this is not the case: although the components of the induced (and external) field strength tensors $F^{\mu\nu}\ind$ (and $F^{\mu\nu}\ext$) correspond to the induced (and external) electric and magnetic fields, which are observable, the components of the field strength {\itshape response} tensor $\chi\indices{^{\mu\nu}_{\alpha\beta}}$ do not correspond 
to physical response functions (cf.~\cite{Lange}). In fact, the components of the field strength response tensor are not directly observable, and it is only the entire expansion \eqref{fundexp} of the induced fields in terms of the external
fields which has a physical meaning. The reason for this is that the field strength response tensor contains
{\it partial} functional derivatives of the induced fields with respect to the external fields. 
Mathematically, such partial derivatives are obtained by varying the external electric
and magnetic field independently of each other. In reality, however, the external electromagnetic fields have to fulfill the Maxwell equations and hence cannot be varied independently: partial derivatives
do not correspond to physical perturbations.

Consider, for example, a typical physical response relation like Ohm's law,
\begin{equation} \label{eq_Ohm}
 \vec j\ind(\vec k, \omega) = \int \! \de^3 \vec k' \, \tsr \sigma(\vec k, \vec k'; \omega) \h \vec E\ext(\vec k', \omega) \,, 
\end{equation}
where $\vec E\ext$ is the external electric field acting on the medium. In general, a frequency-dependent
electric field is also accompanied by a magnetic field $\vec B\ext$, and the actual current $\vec j\ind$ in the medium is induced under the combined action of 
external electric and magnetic fields. Mathematically, this means that the measured conductivity has to be identified with the {\it total} functional derivative with respect to the external electric field:
\begin{align}
\tsr\sigma(\vec k, \vec k'; \omega) & = \frac{\de\vec j\ind(\vec k, \omega)}{\de\vec E\ext(\vec k', \omega)} \label{eq_totcond} \\[3pt]
 & \equiv \frac{\delta\vec j\ind(\vec k, \omega)}{\delta\vec E\ext(\vec k', \omega)}
+\frac{\delta\vec j\ind(\vec k, \omega)}{\delta\vec B\ext(\vec k', \omega)}\frac{\delta\vec B\ext(\vec k', \omega)}{\delta\vec E\ext(\vec k',\omega)} \,.
\end{align}

\vspace{2pt} \noindent
By contrast, determining only the first term on the right hand side of this equation would require to introduce a hypothetical ``partial'' current which is induced exclusively by the external electric field. We conclude that only the total functional derivative is a directly observable response function. Moreover, we will see in the next subsection that the total functional derivative is directly related to the fundamental response tensor, which can be computed in the Kubo formalism. Mutatis mutandis, the same argument applies to any frequency-dependent {\itshape physical} (i.e.~directly observable) response function, which should therefore be identified with a total functional derivative. In the following subsections, we will derive concrete expressions for the physical response functions and their general interrelations.

\subsection{Optical conductivity} \label{subsec_cond}

We first express the microscopic (frequency and wave-vector dependent) conductivity tensor in terms of the fundamental response tensor. 
In accordance with the above considerations, we define the microscopic conductivity tensor in real space as
\begin{equation}
 \sigma_{k\ell}(\vec x, \vec x'; t - t') \stackrel{\rm def}{=} \frac{\mathrm d j\ind^k(\vec x, t)}{\mathrm d E\ext^\ell(\vec x', t')} \,.
\end{equation}
By a functional chain rule, this is equivalent to
\begin{equation}
 \sigma_{k\ell}(\vec x, \vec x'; t - t') = \int \! \mathrm d^3 \vec y \int \! c\, \mathrm d s \ \frac{\delta j\ind^k(\vec x, t)}{\delta A\ext^\mu(\vec y, s)} \,
 \frac{\mathrm d A\ext^\mu(\vec y, s)}{\mathrm d E\ext^\ell(\vec x', t')} \,.
\end{equation}
We note that the total derivative has no bearing on the first term, because the induced current is per definitionem a functional of the external four-potential, while only the latter is regarded as a functional 
of the electromagnetic fields $\{\vec E\ext,\vec B\ext\}$. The functional derivative of the four-potential is calculated
in the temporal gauge $\varphi\ext=0$. Using the result \eqref{eq_put_3} from Sec.~\ref{sec_pre_total},
we obtain
\begin{equation}
 \sigma_{k\ell}(\vec x, \vec x'; t - t') = - \int_{-\infty}^{\infty} \mathrm d s \, \chi_{k\ell}(\vec x, \vec x'; t - s) \, \varTheta(s - t') \,.
\end{equation}
By Fourier transforming with respect to the time variables, this is equivalent to
\begin{equation} \label{eq_chi_sigma}
 \sigma_{k\ell}(\vec x, \vec x'; \omega) = \frac{1}{\j\omega} \, \chi_{k\ell}(\vec x, \vec x'; \omega) \,. \smallskip \vspace{1pt}
\end{equation}
In matrix notation, we can write this identity as
\begin{equation}\label{currrespcond}
 \tsr\sigma = \frac{1}{\j\omega} \, \tsr\chi \,,
\end{equation}
where
\begin{equation}\label{currresp}
\tsr\chi=\frac{\delta\vec j\ind}{\delta\vec A\ext} \smallskip 
\end{equation}
is the spatial part of the fundamental response tensor (see Eq.~\eqref{generalform}). 
In fact, Eq.~\eqref{currrespcond} is a standard relation (see, e.g., \cite[Sec.~3.4.4]{Giuliani} or \cite[Eq.~(4.13)]{Nam}), which in conjunction with the Kubo formula for the fundamental response tensor
forms the basis for microscopic calculations of the optical conductivity \cite{Basov, Perlov, Lee}. We stress, however, that in this standard relation it is imperative
to interpret the conductivity as the total derivative of the induced current with respect to the external
electric field. A similar calculation of the \textit{partial conductivity}
\begin{equation}
 (\sigma\p)_{k\ell}(\vec k, \vec k'; \omega) \stackrel{\rm def}{=} \frac{\delta j^k_{\rm ind}(\vec k, \omega)}{\delta E_{\rm ext}^\ell(\vec k', \omega)}
\end{equation}
yields instead the relation
\begin{equation}
(\sigma\p)_{k\ell}(\vec k, \vec k'; \omega) = \frac{1}{\j\omega}\, \chi_{km}(\vec k, \vec k'; \omega) \, \frac{ \omega^2 \delta_{m\ell} - c^2 k'_m k'_\ell}{\omega^2 - c^2 |\vec k'|^2} \,. \vspace{2pt}
 \end{equation}
Consequently, the partial and the total conductivity are related by \vspace{1pt}
\begin{equation} \label{eq_partial_electric}
 \tsr \sigma{}\p(\vec k, \vec k'; \omega) = \tsr \sigma(\vec k, \vec k'; \omega) \, \tsr{\mathbbmsl E}(\vec k', \omega)  \,,
\end{equation}
where $\mathbbmsl E$ is the electric solution generator defined in Sec.~\ref{subsec_Hamgauge}. Finally, we note that by the constraints \eqref{eq_recon_1}--\eqref{eq_recon_3} on the fundamental response functions, we can express also $\chi\indices{^k_0}$, $\chi\indices{^0_\ell}$ and $\chi\indices{^0_0}$ in terms of $\chi\indices{^k_\ell}$ and consequently in terms of $\sigma_{k\ell}$. Since $\chi\indices{^\mu_\nu}$ can be reconstructed entirely from $\sigma_{k\ell}$, it follows that the microscopic conductivity tensor contains the complete information about the linear electromagnetic response (cf.~\cite{Melrose}).

\subsection{Dielectric tensor}

We now derive the most general relation between the dielectric tensor (defined by Eq.~\eqref{eq_epsilon}) and the microscopic conductivity tensor. The former determines the (magneto)optical properties of materials \cite{Zvezdin} and contains the entire information about elementary excitations and collective modes in solids \cite{Hwang}. With $\vec E \equiv \vec E\tot = \vec E\ext + \vec E\ind$ and $\vec D \equiv \varepsilon_0 \vec E\ext$, we have
\begin{align}
 \big[(\tsr{\varepsilon_{\mathrm r}})^{-1}\big]_{ij}(x, x') & = \frac{\mathrm d E\tot^i(x)}{\mathrm d E\ext^j(x')} \\[5pt]
 & = \delta_{ij} \h \delta^4(x - x') + \big(\chi\EE\big)_{ij}(x, x') \,, \label{eq_hlpr_1}
\end{align}
where the second term determines the response of the induced to the external electric field. This is given by the fundamental relation (with $\vec P \equiv -\varepsilon_0 \vec E\ind$)
\begin{align}
 \big(\chi\EE\big)_{ij}(x, x') & = -\frac{1}{\varepsilon_0} \frac{\mathrm d P^i(x)}{\mathrm d E\ext^j(x')} \\[5pt]
 & = -\frac{1}{\varepsilon_0} \int \! \mathrm d^4 y \int \! \mathrm d^4 y' \, \frac{\delta P^i(x)}{\delta j^\mu\ind(y)} \, \chi\indices{^\mu_\nu}(y,y') \, \frac{\mathrm d A\ext^\nu(y')}{\mathrm d E\ext^j(x')} \,. \label{eq_hlpr_2}
\end{align}
The first term in the integrand is evaluated as described in Sec.~\ref{subsec_emprop} by solving the equation of motion \eqref{eq_eom_P} for the induced field in terms of the induced sources: With the retarded Green function $\mathbbmsl D_0$ for the scalar wave equation, we can express this as
\begin{equation}
\begin{aligned}
 P^i(\vec x, t) & = \int \! \mathrm d^3 \vec y \int \! c\,\mathrm d s \, \frac{1}{\mu_0} \h \mathbbmsl D_0(\vec x - \vec y, t - s) \\[5pt]
 & \quad \times \left( \frac{\partial}{\partial y^i} \h \rho\ind(\vec y, s) + \frac{1}{c^2} \h \frac{\partial}{\partial s} \h j^i\ind(\vec y, s) \right).
\end{aligned}
\end{equation}
Hence after partial integration, we obtain the functional derivatives
\begin{align}
 \frac 1 c \h \frac{\delta P^i(\vec x, t)}{\delta \rho\ind(\vec y, s)} & = - \frac{1}{c \h \mu_0} \h \frac{\partial}{\partial y^i} \h \mathbbmsl D_0(\vec x - \vec y, t - s) \,, \label{eq_put_1} \\[5pt]
 \frac{\delta P^i(\vec x, t)}{\delta j^m\ind(\vec y, s)} & = -\delta_{im} \h \frac{1}{c^2 \mu_0} \h \frac{\partial}{\partial s} \h \mathbbmsl D_0(\vec x - \vec y, t - s) \,. \label{eq_put_2}
\end{align}
For the last term in the integrand of Eq.~\eqref{eq_hlpr_2} we use again Eq.~\eqref{eq_put_3}.
By putting these expressions into Eq.~\eqref{eq_hlpr_2}, performing partial integrations and reexpressing the fundamental response tensor in terms of the conductivity tensor, we arrive at the relation
\begin{equation}
\begin{aligned}
 & \big[(\tsr{\varepsilon_{\mathrm r}})^{-1}\big]_{ij}(\vec x, \vec x'; t - t') = 
 \frac{1}{c} \, \delta_{ij} \h \delta(t - t') \h \delta^3(\vec x - \vec x') \\[5pt]
 & \quad - c \int \! \mathrm d^3 \vec y \int \! \mathrm d s \int \! \mathrm d s' \, \mathbbmsl D_0(\vec x - \vec y; t - s) \\[3pt]
 & \quad \times \left( \delta_{im} \h \frac{\partial^2}{\partial s^2} - c^2 \frac{\partial}{\partial y^i} \frac{\partial}{\partial y^m} \right) \sigma_{mj}(\vec y, \vec x'; s - s') \, \varTheta(s' - t') \,. \label{eq_univ_eps_pos}
\end{aligned}
\end{equation}
This is the most general relation between the dielectric tensor and the microscopic conductivity tensor (assuming only homogeneity in time). 
By a Fourier transformation with respect to the space and time variables, it is equivalent to
\begin{equation}
\begin{aligned}
 \big[(\tsr{\varepsilon_{\mathrm r}})^{-1}\big]_{ij}(\vec k, \vec k'; \omega) & = \delta_{ij} \, \delta^3(\vec k-\vec k') \\[3pt]
 & \quad - \frac{\j}{\omega} \, \mathbbmsl D_0(\vec k, \omega) \, (c^2 k_i k_m - \omega^2 \delta_{im} ) \, \sigma_{mj}(\vec k, \vec k'; \omega) \,. \smallskip
\end{aligned}
\end{equation}
In matrix notation, we can write this compactly as
\begin{equation}
\begin{aligned}
 (\tsr{\varepsilon_{\mathrm r}})^{-1}(\vec k, \vec k'; \omega) & = \tsr 1 \, \delta^3(\vec k-\vec k') \\[3pt]
 & \quad -\frac{\j}{\omega} \, \mathbbmsl D_0(\vec k, \omega) \, \left( c^2 |\vec k|^2 \tsr P_{\mathrm L} (\vec k) - \omega^2 \tsr 1 \, \right) \tsr \sigma(\vec k, \vec k'; \omega) \,. \label{eq_univ_eps}
\end{aligned}
\end{equation}
In Sec.~\ref{subsec_nonrel} we will show that this general formula reduces to a well-known identity in the special case of homogeneous, isotropic materials and in a non-relativistic approximation.

\subsection{Magnetic susceptibility}

Next we consider the magnetic susceptibility, which contains the description of magnetic orderings and excitation spectra 
and is directly related to the neutron scattering cross section \cite{Lovesey}.\footnote{If $\chi\indices{^\mu_\nu}$ is associated with the response of a microscopic {\itshape charge} current to the electromagnetic potential, the corresponding response function $\chi_{\rm m}$ refers only to the {\itshape orbital} contribution to the magnetic susceptibility \cite{Thonhauser}. Spin contributions can be included in our picture by a divergence free contribution to the current of the Pauli equation, as it is motivated by the non-relativistic limit of the Dirac equation (cf. \cite[chap. XX, $\S$29]{Messiah}).}
In analogy to the electric case, we identify the physical response function with the total functional derivative
\begin{align}
 (\chi_{\rm m})_{ij}(x,x') & = \mu_0 \h \frac{\mathrm d M^i(x)}{\mathrm d B\ext^j(x')} \\[5pt]
 & = \mu_0 \int \! \mathrm d^4 y \int \! \mathrm d^4 y' \, \frac{\delta M^i(x)}{\delta j^\mu\ind(y)} \, \chi\indices{^\mu_\nu}(y, y') \, \frac{\mathrm d A\ext^\nu(y')}{\mathrm d B\ext^j(x')} \,. \label{eq_fund}
\end{align}
Here again, the total derivative displays its effect only in the functional derivative of the four-potential.
Similarly as in the electric case, we obtain for the first term in the integrand
\begin{equation} \label{eq_first}
 \frac{\delta M^i(\vec x, t)}{\delta j^m\ind(\vec y, s)} = -\frac{1}{\mu_0} \, \epsilon_{ikm} \, \frac{\partial}{\partial y^k} \h \mathbbmsl D_0(\vec x - \vec y, t - s) \,.
\end{equation}
For the last term in the integrand of Eq.~\eqref{eq_fund} we use the result \eqref{eq_third} from Sec.~\ref{sec_pre_total}. By putting these results into Eq.~\eqref{eq_fund} and performing partial integrations with respect to the $y$ and $y'$ variables, we arrive at the relation
\begin{equation}
\begin{aligned}
 & (\chi_{\rm m})_{ij}(\vec x, \vec x'; t - t') = c \int \! \mathrm d^3 \vec y \int \! \mathrm d s \int \! \mathrm d^3 \vec y' \int \! \mathrm d s' \, \mathbbmsl D_0(\vec x - \vec y; t - s) \\[3pt]
 & \qquad \times \left( \epsilon_{i k m} \h \epsilon_{j \ell n} \h \frac{\partial}{\partial y^k} \frac{\partial}{\partial y'^\ell} \h \chi_{mn}(\vec y, \vec y'; s - s') \right)
 \frac{1}{4\pi} \frac{\delta(s' - t')}{|\vec y' - \vec x'|} \,.
\end{aligned}
\end{equation}
By a Fourier transformation, this is equivalent to
\begin{equation}
 (\chi_{\rm m})_{ij}(\vec k, \vec k'; \omega) = \mathbbmsl D_0(\vec k, \omega) \left( \epsilon_{ikm} \h \epsilon_{j \ell n} \h k_k \h k'_\ell \h \chi_{mn}(\vec k, \vec k'; \omega) \right) \frac{1}{|\vec k'|^2} \,. \label{eq_univ_chi}
\end{equation}
We will show in Sec.~\ref{subsec_nonrel} that this general expression reduces to a well-known formula in the special case of homogeneous materials and in a non-relativistic approximation.

\subsection{Magnetoelectric coupling and synopsis} \label{subsec_cross}

For general materials, there are besides electric and magnetic susceptibilities also cross-coupling (magnetoelectric) 
coefficients, which specify the response of an induced magnetic field to an external electric field, and vice versa \cite{Mackay10}.
These can be calculated in close analogy to the examples above, and we summarize all results 
by the following set of equations:
\begin{align}
 \frac{\mathrm d E^i_{\rm ind}(\vec k, \omega)}{\mathrm d E^j_{\rm ext}(\vec k', \omega)} & = -\frac{1}{\varepsilon_0 \h \omega^2} \, \frac{\omega^2 \delta_{im} - c^2 k_i k_m}{\omega^2 - c^2 |\vec k|^2} \ \j\omega \h \sigma_{mj}(\vec k, \vec k'; \omega) \,, \label{eq_tot_der_1} \\[10pt]
 \frac{1}{c} \, \frac{\mathrm d E^i_{\rm ind}(\vec k, \omega)}{\mathrm d B^j_{\rm ext}(\vec k', \omega)} & = -\frac{1}{\varepsilon_0 \h \omega^2} \, \frac{\omega^2 \delta_{im} - c^2 k_i k_m}{\omega^2 - c^2 |\vec k|^2} \ \j \omega  \h \sigma_{mn}(\vec k, \vec k'; \omega) \ \frac{\epsilon_{n\ell j} \, \omega \, c \h k'_\ell}{- c^2 |\vec k'|^2} \,, \label{eq_tot_der_2} \\[10pt]
 c \, \frac{\mathrm d B^i_{\rm ind}(\vec k, \omega)}{\mathrm d E^j_{\rm ext}(\vec k', \omega)} & =-\frac{1}{\varepsilon_0 \h \omega^2} \, \frac{\epsilon_{ikm} \, \omega \, c \h k_k}{\omega^2 - c^2 |\vec k|^2} \ \j \omega \h \sigma_{mj}(\vec k, \vec k'; \omega) \,, \label{eq_tot_der_3} \\[10pt]
 \frac{\mathrm d B^i_{\rm ind}(\vec k, \omega)}{\mathrm d B^j_{\rm ext}(\vec k', \omega)} & = -\frac{1}{\varepsilon_0 \h \omega^2} \, \frac{\epsilon_{ikm} \, \omega \, c \h k_k}{\omega^2 - c^2 |\vec k|^2} \ \j \omega \h \sigma_{mn}(\vec k, \vec k'; \omega) \ \frac{\epsilon_{n\ell j} \, \omega \, c \h k'_\ell}{-c^2 |\vec k'|^2} \,. \label{eq_tot_der_4}
\end{align}
Using the electric and magnetic solution generators from Sec.~\ref{subsec_Hamgauge}, we can write these relations in matrix form as
\begin{align}
 \tsr \chi\EE(\vec k, \vec k'; \omega) & = -\frac{1}{\varepsilon_0 \h \omega^2} \, \tsr{\mathbbmsl E}(\vec k, \omega) \, \h \j\omega \h \tsr \sigma(\vec k, \vec k'; \omega) \,, \label{eq_univ_1} \\[8pt]
 \tsr \chi\EB(\vec k, \vec k'; \omega) & = \tsr \chi\EE(\vec k, \vec k'; \omega) \left( -\frac{\omega}{c|\vec k'|} \, \tsr R_{\mathrm T}(\vec k') \right), \label{eq_univ_2} \\[8pt]
 \tsr \chi\BE(\vec k, \vec k'; \omega) & = -\frac{1}{\varepsilon_0 \h \omega^2} \, \tsr{\mathbbmsl B}(\vec k, \omega) \, \h \j\omega \h \tsr \sigma(\vec k, \vec k'; \omega) \,, \label{eq_univ_3} \\[8pt]
 \tsr \chi\BB(\vec k, \vec k'; \omega) & = \tsr \chi\BE(\vec k, \vec k'; \omega) \left( -\frac{\omega}{c|\vec k'|} \, \tsr R_{\mathrm T}(\vec k') \right). \label{eq_univ_4} 
\end{align}
To connect these physical response functions with those defined in Sec.~\ref{subsec_physical}, we note that the magnetic susceptibility coincides with \eqref{eq_univ_4}, i.e.,
\begin{equation}
 \tsr\chi_{\rm m} = \tsr\chi\BB \,, \label{eq_magnsusc}
\end{equation}
whereas the electric susceptibility is given in terms of \eqref{eq_univ_1} by
\begin{equation}
\tsr\chi_{\rm e} = -\tsr\chi\EE \h ( \h 1 + \tsr\chi\EE \h )^{-1} \,.
\end{equation}

\pagebreak \noindent
For the permittivity and permeability tensors we further have
\begin{align}
 (\tsr {\varepsilon_{\mathrm r}})^{-1} & = \tsr 1 + \tsr \chi\EE \,, \label{eq_permittivity} \\[3pt]
 \tsr{\mu_{\mathrm r}} & = \tsr 1 + \tsr \chi\BB \,. \label{eq_permeability}
\end{align}
The above relations between electromagnetic response functions are {\itshape universal}, i.e., valid on the microscopic and on the macroscopic scale and
in any material, because of our model-independent definitions \eqref{eq_PE} and \eqref{eq_MB} of the electric and magnetic polarizations.\footnote{In applying the universal relations on a macroscopic scale, one must keep in mind the following limitation: Each macroscopic material property is usually dominated by a 
certain class of degrees of freedom. For example, the conductivity may be associated with the conduction electrons' orbital motion, whereas the magnetic susceptibility is often dominated by localized spin degrees of freedom. Thus, by describing each macroscopic response function only in terms of its most relevant degrees of freedom, the universal relations get lost. By contrast, the microscopic current response in principle refers to all degrees of freedom contributing to the electromagnetic four-current.} In the homogeneous limit, the universal relations hold in any inertial frame, as they are derived from the Lorentz tensor of fundamental response functions. Given a particular microscopic model of the medium, Eqs.~\eqref{eq_univ_1}--\eqref{eq_univ_4} can be evaluated in the Kubo formalism, where the conductivity tensor is expressed by a current-current correlation function \cite{Altland, Kubo}.

In concluding this subsection, we note that it is not yet obvious how the induced fields can be expressed in terms of the external perturbation by means of the physical response functions. The naive expansion, which would be analogous to the expansion \eqref{eq_part_exp1}--\eqref{eq_part_exp2} in terms of the partial derivatives,
is in general not correct, i.e.,
\begin{align}
 \vec E\ind \not = \tsr \chi\EE \, \vec E\ext + \tsr \chi\EB \, c \h \vec B\ext \,, \label{eq_naive_1} \\[3pt]
 c \h \vec B\ind \not = \tsr \chi\BE \, \vec E\ext + \tsr \chi\BB \, c \h \vec B\ext \,. \label{eq_naive_2} 
\end{align}
This is because the physical response functions correspond to total functional derivatives, hence part of the magnetic response is already contained in the electric response and vice versa. To solve this problem, we now take recourse to the canonical functional of Sec.~\ref{sec_CanFun}.

\subsection{Field expansions} \label{sec_comp}

In Sec.~\ref{sec_CanFun} we have introduced three different representations of the vector potential in the temporal gauge in terms of the electric and magnetic fields: (i) in terms of $\vec E$ and $\vec B$ as given by the canonical functional, (ii) only in terms of $\vec E$ (and possibly a static magnetic field $\vec B_0$), and (iii) in terms of $\vec E_{\mathrm L}$ and $\vec B$ (cf.~Eqs.~\eqref{eq_exp_A_1}, \eqref{eq_exp_A_2} and \eqref{eq_exp_A_3}).
As we are now going to show, these representations translate directly into three different electromagnetic field expansions.
Consider, for example, the induced electric field, which can be expanded to first order in the external perturbation as
\begin{align}
 \vec E_{\rm ind}(\vec k, \omega) & = \frac{1}{\varepsilon_0} \, \frac{1}{\j\omega} \, \tsr{\mathbbmsl E}(\vec k, \omega) \, \vec j_{\rm ind}(\vec k, \omega) \\[5pt]
 & = \frac{1}{\varepsilon_0} \, \frac{1}{\j\omega} \, \tsr{\mathbbmsl E}(\vec k, \omega) \int \! \de^3 \vec k' \, \tsr{\chi}(\vec k, \vec k'; \omega) \h \vec A_{\rm ext}(\vec k', \omega) \,. \label{eq_zw3}
\end{align}
Here $\vec A_{\rm ext}$ is the external vector potential in the temporal gauge, and we have used Eq.~\eqref{eq_tsrE}.
Representing the external vector potential in terms of the electric and magnetic fields by the canonical functional \eqref{eq_can_tsr}, we obtain
\begin{align}
 \vec E_{\rm ind}(\vec k, \omega) & = -\frac{1}{\varepsilon_0 \h \omega^2} \, \tsr{\mathbbmsl E}(\vec k, \omega) \int \! \de^3 \vec k' \, \tsr{\chi}(\vec k, \vec k'; \omega)\nonumber \\[5pt]
 & \quad \, \times \Big\{ \tsr{\mathbbmsl E}(\vec k', \omega') \h \vec E_{\rm ext}(\vec k', \omega) + \tsr{\mathbbmsl B}(\vec k', \omega') \h c \h \vec B_{\rm ext}(\vec k', \omega) \Big\} \\[10pt]
 & = \int \! \de^3 \vec k' \ \tsr \chi{}\EE\p(\vec k, \vec k'; \omega) \, \vec E\ext(\vec k', \omega) \nonumber \\[5pt]
 & \quad \, + \int \! \de^3 \vec k' \ \tsr \chi{}\EB\p(\vec k, \vec k'; \omega) \, c \h \vec B\ext(\vec k', \omega) \,,
\end{align}
where the partial functional derivatives are given by the expressions \eqref{eq_lor_1}--\eqref{eq_lor_2}.
On the other hand, using Eq.~\eqref{eq_AEtot} to express $\vec A\ext$ solely in terms
of the electric field leads to
\begin{align}
 \vec E_{\rm ind}(\vec k, \omega) & = -\frac{1}{\varepsilon_0 \h \omega^2} \, \tsr{\mathbbmsl E}(\vec k, \omega) \int \! \de^3 \vec k' \, \tsr{\chi}(\vec k, \vec k'; \omega) \h \vec E_{\rm ext}(\vec k', \omega) \\[8pt]
 & = \int \! \de^3 \vec k' \, \tsr\chi\EE(\vec k, \vec k'; \omega) \h \vec E\ext(\vec k', \omega) \,,
\end{align}
with the total functional derivative given by the universal relation \eqref{eq_univ_1}. In the presence of 
a static magnetic field, the situation is slightly more complicated in that we earn the additional contribution
\begin{equation}
\int \! \de^3\vec k'\,\tsr\chi_{EB}(\vec k,\vec k';\omega=0) \, c \h \vec B_{\rm ext, 0}(\vec k') \h \delta(\omega) \,,
\end{equation}
with $\vec B_{\rm ext, 0}(\vec x) = \lim_{\,t \to -\infty} \vec B\ext(\vec x, t)$ (see Sec.~\ref{sec_zero_freq}).
Finally, expressing $\vec A\ext$ in Eq.~\eqref{eq_zw3} in terms of the longitudinal part of the electric field and the magnetic field as in Eq.~\eqref{eq_ABtot} yields
\begin{align}
 \vec E_{\rm ind}(\vec k, \omega) & = -\frac{1}{\varepsilon_0 \h \omega^2} \, \tsr{\mathbbmsl E}(\vec k, \omega) \int \! \de^3 \vec k' \, \tsr{\chi}(\vec k, \vec k'; \omega) \nonumber \\[5pt]
 & \quad \, \times \left\{ (\vec E_{\rm ext})_{\mathrm L} (\vec k', \omega) + \left( -\frac{\omega}{c|\vec k'|} \right) \tsr R_{\mathrm T}(\vec k') \, c \h \vec B\ext(\vec k', \omega) \right\} \\[8pt]
 & = \int \! \de^3 \vec k' \ \tsr \chi{}\EE(\vec k, \vec k'; \omega) \, (\vec E\ext)_{\rm L}(\vec k', \omega) \nonumber \\[5pt]
 & \quad \, + \int \! \de^3 \vec k' \ \tsr \chi{}\EB(\vec k, \vec k'; \omega) \, c \h \vec B\ext(\vec k', \omega) \,,
\end{align}
where the cross-coupling tensor in the second line is given by the universal relation \eqref{eq_univ_2}.

We conclude that as a matter of principle, there are three different but equivalent field expansions. First, in terms of the partial functional derivatives: \vspace{-5pt}
\begin{align}
\vec E_{\rm ind} & = \tsr \chi{}\EE\p \, \vec E\ext + \tsr \chi{}\EB\p \, c \h \vec B\ext \,, \label{eq_exp_1} \\[5pt]
c \h \vec B\ind & = \tsr\chi{}\BE\p \, \vec E\ext + \tsr \chi{}\BB\p \, c \h \vec B\ext \,. \label{eq_exp_1b}
\end{align}

\vspace{2pt} \noindent
This expansion is componentwise equivalent to the expansion \eqref{fundexp} of the induced field strength tensor in terms of the external field strength tensor.
Second, in terms of the external electric field, the static contribution to the magnetic field and the physical response functions:
\begin{align}
\vec E_{\rm ind} & = \tsr \chi\EE \, \vec E_{\rm ext} + \tsr\chi\EB \, c \h \vec B_{\rm ext,0}\,, \label{eq_exp_2} \\[5pt]
c \h \vec B\ind &= \tsr \chi\BE \, \vec E\ext +\tsr\chi\BB \, c \h \vec B_{\rm ext,0}\,. \label{eq_exp_2b}
\end{align}

\vspace{2pt} \noindent
Third, we have a mixed expansion in terms of the longitudinal electric field, the magnetic field and the physical response functions:
\begin{align}
\vec E_{\rm ind} & = \tsr \chi\EE \h (\vec E_{\rm ext})_{\rm L} + \tsr \chi\EB \, c \h \vec B\ext \,, \label{eq_exp_3}\\[5pt]
c \h \vec B\ind &= \tsr \chi\BE \h (\vec E\ext)_{\rm L} + \tsr\chi\BB \, c \h \vec B\ext \,. \label{eq_exp_3b}
\end{align}
The total and the partial functional derivatives are related through functional chain rules, e.g.,
\begin{align}
 \tsr \chi\EE & = \frac{\de \vec E\ind}{\de \vec E\ext} = \frac{\delta \vec E\ind}{\delta \vec E\ext} + \frac{\delta \vec E\ind}{\delta \vec B\ext} \frac{\delta \vec B\ext}{\delta \vec E\ext} \\[8pt]
 & = \tsr\chi{}\EE\p + \tsr\chi{}\EB\p \left( \frac{c|\vec k'|}{\omega} \, \tsr R_{\mathrm T}(\vec k') \right), \label{eq_rel1}
\end{align}
and similarly,
\begin{equation}
 \tsr \chi\EB = \tsr\chi{}\EE\p \left( -\frac{\omega}{c|\vec k'|} \, \tsr R_{\mathrm T}(\vec k') \right) + \tsr\chi{}\EB\p \,. \medskip \label{eq_rel2}
\end{equation}
Here we have used Eqs.~\eqref{eq_BE_comp} and \eqref{eq_EB_comp} respectively. We remark that for {\itshape longitudinal} electric perturbations, the second term in Eq.~\eqref{eq_rel1} vanishes and therefore,
\begin{equation}
 \vec E\ind = \tsr\chi\EE (\vec E\ext)_{\mathrm L} = \tsr\chi{}\EE\p (\vec E\ext)_{\mathrm L} \,. \medskip
\end{equation}
Hence in this case the expansions in terms of the total and the partial functional derivatives coincide.

\subsection{Magnetic conductivity}

To complete our discussion of the physical response functions, we now turn our attention to the induced current $\vec j\ind$, which can be expanded in an analogous way in terms of the external electric and magnetic fields. We define the {\itshape magnetic conductivity} as
\begin{equation}
\tsr\kappa=\frac{1}{c} \h \frac{\de\vec j\ind}{\de\vec B\ext} \,.
\end{equation}
This is related to the fundamental response functions by
\begin{align}
 \tsr\kappa(\vec k, \vec k'; \omega) & = \frac 1 c \h \frac{\delta\vec j\ind(\vec k, \omega)}{\delta\vec A\ext(\vec k', \omega)} \, \frac{\de\vec A\ext(\vec k', \omega)}{\de\vec B\ext(\vec k', \omega)} \\[5pt]
 & = \tsr\chi(\vec k, \vec k'; \omega) \, \frac{1}{\j\omega} \left(-\frac{\omega}{c|\vec k'|} \, \tsr R_{\mathrm T}(\vec k') \right).
\end{align}
Consequently, the following universal relation holds:
\begin{equation} \label{eq_kappa}
 \tsr \kappa(\vec k, \vec k'; \omega) = \tsr\sigma(\vec k, \vec k'; \omega) \left(-\frac{\omega}{c|\vec k'|} \, \tsr R_{\mathrm T}(\vec k') \right).
\end{equation}
The corresponding {\itshape partial} magnetic conductivity is defined by
\begin{equation}
\tsr\kappa{}\p=\frac{1}{c} \h \frac{\delta\vec j\ind}{\delta\vec B\ext} \,.
\end{equation}
The partial electric and magnetic conductivities can be expressed in terms of the electric and magnetic solution generators from Sec.~\ref{subsec_Hamgauge} as
\begin{align}
 \tsr \sigma{}\p(\vec k, \vec k';\omega) = \frac{1}{\j\omega} \, \tsr \chi(\vec k, \vec k';\omega) \, \tsr{\mathbbmsl E}(\vec k', \omega) \,, \\[5pt]
 \tsr \kappa{}\p(\vec k, \vec k';\omega) = \frac{1}{\j\omega} \, \tsr \chi(\vec k, \vec k';\omega) \, \tsr{\mathbbmsl B}(\vec k', \omega) \,.
\end{align}
Analogous to the case of the induced electromagnetic fields, the following three different but equivalent expansions of the induced current hold to first order in the external perturbation:
\begin{align}
\vec j\ind & = \tsr \sigma{}\p \, \vec E\ext + \tsr\kappa{}\p \, c \h \vec B\ext \\[5pt]
& = \tsr\sigma \, \vec E\ext + \tsr\kappa \, c \h \vec B_{\rm ext,0} \label{eq_preferred} \\[5pt]
& = \tsr\sigma \, (\vec E\ext)_{\mathrm L} + \tsr\kappa \, c \h \vec B\ext \,.
\end{align}
In this case, however, the second expansion \eqref{eq_preferred} is practically preferred: It induces a 
decomposition of the induced current into a dynamical part responding to the external electric field and a temporally constant current $\vec j_{\rm ind, 0}$ induced by a static magnetic field,
\begin{equation}
 \vec j_{\rm ind, 0} = \tsr \kappa \, c \h \vec B_{\rm ext,0} \,.
\end{equation}
It is natural to split off this contribution from the dynamical current by making
the transition to the corresponding static magnetization $\vec M_0$, which is defined~by
\begin{align}
 \nabla \times \vec M_0 & = \vec j_{\rm ind, 0} \,, \\[5pt]
 \nabla \cdot \vec M_0 & = 0 \,.
\end{align}
Microscopically, such a transition can be motivated by the fact that often the magnetization is dominated by the spin
degrees of freedom, while the dynamical current is dominated by the conduction electron's orbital motion. The response of $\vec M_0$ to the static external magnetic field is now given by the instantaneous limit of the magnetic susceptibility,
\begin{equation}
 \vec M_0 = \frac{\tsr \chi_{\rm m}(\omega = 0)}{\mu_0} \, \vec B_{\rm ext, 0} \,.
\end{equation}
Here, $\chi_{\rm m}$ is related to the magnetic conductivity by
\begin{align}
 \tsr\chi_{\rm m}(\vec k, \vec k'; \omega) & = \frac{1}{\varepsilon_0} \, \frac{1}{\j\omega} \, \tsr{\mathbbmsl B}(\vec k, \omega) \, \tsr \kappa(\vec k, \vec k'; \omega) \\[5pt]
 & = \j \h \mathbbmsl D_0(\vec k, \omega) \, c|\vec k| \, \tsr R_{\mathrm T}(\vec k) \, \tsr \kappa(\vec k, \vec k'; \omega) \,,
\end{align}
which follows by comparing the universal relations for $\chi_{\rm m}$ and $\kappa$, Eq.~\eqref{eq_magnsusc} together with \eqref{eq_univ_3}--\eqref{eq_univ_4} and Eq.~\eqref{eq_kappa}, respectively. The 
remaining dynamical current then fulfills exactly
\begin{equation}
 \vec j\ind - \vec j_{\rm ind, 0} = \tsr \sigma \h \vec E\ext \,. 
\end{equation}
This shows that to first order in the external perturbation, Ohm's law can be upheld in the presence of magnetic fields. In fact, Ohm's law is even relativistically covariant (see \cite{EDOhm} for a discussion).

\section{Empirical limiting cases}\label{sec_emp}

In their most general form, the universal response relations \eqref{eq_univ_1}--\eqref{eq_univ_4} 
are not always particularly useful. In practice, one should take them to several empirically motivated limits. Foremost among these are:

\smallskip
\begin{enumerate}
 \setlength{\itemsep}{1em}
 \item[(i)] The {\itshape homogeneous limit}, where response functions are 
 proportional to $\delta^3(\vec k - \vec k')$ and hence essentially depend only on one momentum,
 \item[(ii)] the {\itshape isotropic limit}, where spatial response tensors are of the form \vspace{-3pt}
 \begin{equation}
\tsr\chi(\vec k, \omega) = \chi_{\mathrm L} (\vec k, \omega) \tsr P_{\mathrm L} (\vec k)
 + \chi_{\mathrm T}(\vec k, \omega) \tsr P_{\mathrm T}(\vec k) \label{eq_iso}
\end{equation}
with longitudinal and transverse response functions $\chi_{\mathrm L}$ and $\chi_{\mathrm T}$,
 \item[(iii)] the {\itshape ultra-relativistic limit}, where $\omega = c|\vec k|$,
 \item[(iv)] the {\itshape non-relativistic limit}, where $\omega\ll c|\vec k|$,
 \item[(v)] the {\itshape instantaneous limit}, where $\omega = 0$.
\end{enumerate}

\pagebreak \noindent
In this section, we will derive concrete expressions for both the partial functional derivatives of Sec.~\ref{subsec_fsrt} and the physical response functions of Sec.~\ref{sec_univ} in these limiting cases.  

\subsection{Homogeneous and isotropic limit}

We equate $\vec k = \vec k'$ in the expressions~\eqref{eq_lor_1}--\eqref{eq_lor_4} for the partial functional derivatives and assume an isotropic current response as described by Eq.~\eqref{eq_iso}. By using the algebra of the projection and  rotation operators shown in Table \ref{table}, the components of the field strength response tensor then simplify as
\begin{align}
\tsr \chi{}\EE\p(\vec k, \omega) & = -\frac{1}{\varepsilon_0 \h \omega^2} \, \chi_{\mathrm L}(\vec k, \omega) \, \tsr P_{\rm L}(\vec k) - \tsr \chi{}\BB\p(\vec k, \omega) \,, \label{eq_simpl_fsrt_1} \\[7pt]
\tsr \chi{}\EB\p(\vec k, \omega) & = \tsr \chi{}\BE\p(\vec k, \omega) \,, \\[7pt]
\tsr \chi{}\BE\p(\vec k, \omega) & = -\varepsilon_0 \h \omega^2 \h \mathbbmsl D_0(\vec k, \omega) \, \frac{c|\vec k|}{\omega} \, \chi_{\rm T}(\vec k, \omega) \, \mathbbmsl D_0(\vec k, \omega) \, \tsr R_{\mathrm T}(\vec k) \,, \\[5pt]
\tsr \chi{}\BB\p(\vec k, \omega) & = \varepsilon_0 \h \omega^2 \h \mathbbmsl D_0(\vec k,\omega) \, \frac{c^2 |\vec k|^2}{\omega^2} \, \chi_{\rm T}(\vec k, \omega) \, \mathbbmsl D_0(\vec k, \omega) \, \tsr P_{\rm T}(\vec k) \,. \label{eq_simpl_fsrt_4}
\end{align}
Similarly, the universal response relations \eqref{eq_univ_1}--\eqref{eq_univ_4} reduce to
\begin{align}
 \tsr \chi\EE(\vec k, \omega) & = -\frac{1}{\varepsilon_0 \h \omega^2} \, \chi_{\mathrm L}(\vec k, \omega) \, \tsr P_{\mathrm L}(\vec k) + \tsr \chi\BB(\vec k, \omega) \,, \label{eq_simpl_1} \\[5pt]
 \tsr \chi\EB(\vec k, \omega) & = - \mathbbmsl D_0(\vec k, \omega) \, \frac{\omega}{c|\vec k|} \, \chi_{\mathrm T}(\vec k, \omega) \, \tsr R_{\mathrm T}(\vec k) \,, \label{eq_simpl_2} \\[3pt]
 \tsr \chi\BE(\vec k, \omega) & = \mathbbmsl D_0(\vec k, \omega) \, \frac{c|\vec k|}{\omega} \, \chi_{\mathrm T}(\vec k, \omega) \, \tsr R_{\mathrm T}(\vec k) \,, \label{eq_simpl_3} \\[7pt]
 \tsr \chi\BB(\vec k, \omega) & = \mathbbmsl D_0(\vec k, \omega) \, \chi_{\mathrm T}(\vec k, \omega) \, \tsr P_{\mathrm T}(\vec k) \label{eq_simpl_4} \,.
\end{align}

\vspace{5pt} \noindent
For later purposes, we further note that in the homogeneous and isotropic 
limit, the {\itshape density response function}
\begin{equation}
 \upchi = \frac{\delta\rho\ind}{\delta\varphi\ext} = \frac{1}{c^2} \, \chi^0_{~0}
\end{equation}

\pagebreak \noindent
is related to the longitudinal current response function $\chi_{\mathrm L}$ by
\begin{equation} \label{eq_chiEEchi}
 \upchi(\vec k,\omega) = -\frac{k_i \h k_j}{\omega^2} \, \chi_{ij}(\vec k, \omega) = -\frac{|\vec k|^2}{\omega^2} \, \chi_{\mathrm L} (\vec k,\omega) \,,
\end{equation}
as follows from the representation \eqref{generalform} of the fundamental response tensor.
In the following subsections, we will discuss the ultra-relativistic limit $\omega = c |\vec k|$, the non-relativistic limit $\omega \ll c|\vec k|$ and the instantaneous limit $\omega = 0$ presuming homogeneity and isotropy.

\subsection{Ultra-relativistic limit} \label{subsec_ultra}

In the ultra-relativistic limit where $\omega = c|\vec k|$, the universal response relations read as follows:
\begin{align}
\tsr\chi\EE(\vec k,\omega) & = \tsr\chi\BB(\vec k,\omega) \,, \label{eq_ultra_1} \\[7pt]
\tsr\chi\EB(\vec k,\omega) & = - \tsr\chi\BE(\vec k,\omega) \,, \label{eq_ultra_2} \\[5pt]
\tsr\chi\BE(\vec k,\omega) & = \mathbbmsl D_0(\vec k,\omega) \, \chi_{\mathrm T}(\vec k,\omega) \, \tsr R_{\mathrm T}(\vec k) \,, \label{eq_ultra_3} \\[5pt]
\tsr\chi\BB(\vec k,\omega) & = \mathbbmsl D_0(\vec k,\omega) \, \chi_{\mathrm T}(\vec k,\omega) \, \tsr P_{\mathrm T}(\vec k) \,. \label{eq_ultra_4} 
\end{align}

\vspace{2pt} \noindent
These equations can formally be obtained by the replacement $\omega\mapsto c|\vec k|$ in Eqs.~\eqref{eq_simpl_1}--\eqref{eq_simpl_4} and by omitting the first term on the right hand side of Eq.~\eqref{eq_simpl_1}. However, as the response functions are singular distributions and hence only their action on (external) fields is well-defined, we must discuss these formulae more thoroughly: The dispersion relation $\omega=c|\vec k|$ means that the external fields are vacuum solutions of the Maxwell equations, which are purely transverse. In particular, this implies that $\rho\ext = \varepsilon_0 \nabla \cdot \vec E\ext = 0$. By isotropy, the induced fields are also transverse, and consequently the longitudinal electric field response vanishes in the ulta-relativistic limit.
The induced fields satisfy the equations of motion
\begin{align}
\Box\vec E\ind & = - \mu_0 \, \partial_t \h \vec j\ind \,, \label{eq_zw_1} \\[5pt]
\Box\vec B\ind & = \mu_0 \nabla \times \vec j\ind \,, \label{eq_zw_2}
\end{align}
where we have used \h$\rho\ind = 0$ \h by the transversality of the induced electric field. \linebreak

\pagebreak \noindent
With $\vec E\ext = -\partial_t \vec A\ext$\h, the vector potential $\vec A\ext$ is also transverse. Hence the induced current is given to linear order by
\begin{equation}
 \vec j\ind = \chi_{\mathrm T} \vec A\ext \,,
\end{equation}
with the transverse response function $\chi_{\mathrm T}$. Putting this result into Eqs.~\eqref{eq_zw_1}--\eqref{eq_zw_2}, performing partial integrations and reexpressing $\vec A\ext$ in terms of the electric and magnetic fields yields
\begin{align}
 \Box \vec E\ind & = \mu_0 \h \chi_{\mathrm T} \h \vec E\ext \,, \\[5pt]
 \Box \vec B\ind & = \mu_0 \h \chi_{\mathrm T} \h \vec B\ext \,.
\end{align}
Using the retarded Green function $\mathbbmsl D_0$ of the scalar wave equation, these equations can be inverted as\begin{align}
\vec E\ind & = \mathbbmsl D_0 \, \chi_{\mathrm T} \h \vec E\ext \,, \label{eE} \\[5pt]
\vec B\ind & = \mathbbmsl D_0 \, \chi_{\mathrm T} \h \vec B\ext \,. \label{bB}
\end{align}
Thus, we have proven Eqs.~\eqref{eq_ultra_1} and \eqref{eq_ultra_4}. Furthermore, for $\omega = c |\vec k|$ the external electric and magnetic fields are related by Faraday's law as
\begin{align}
 c \h \vec B = \frac{c\vec k \times \vec E}{\omega} = \frac{\vec k \times \vec E}{|\vec k|} = \tsr R_{\mathrm T}(\vec k) \vec E \,,
\end{align}
or equivalently,
\begin{equation}
 \vec E = -\tsr R_{\mathrm T}(\vec k) \h c \h \vec B \,. \smallskip
\end{equation}
Therefore, we obtain also the relations \vspace{-2pt}
\begin{align}
 \vec E\ind & = -\mathbbmsl D_0 \, \chi_{\mathrm T}\h \tsr R_{\mathrm T}(\vec k) \h c \h \vec B\ext \,, \\[5pt]
 c \h \vec B\ind & = \mathbbmsl D_0 \, \chi_{\mathrm T}\h \tsr R_{\mathrm T}(\vec k) \h \vec E\ext \,,
\end{align}
which yield Eqs.~\eqref{eq_ultra_2}--\eqref{eq_ultra_3} for the cross-coupling coefficients. We note that in the ultra-relativistic limit, the redundancy of the field expansion in 
terms of the physical response functions becomes particularly evident. In \linebreak

\pagebreak \noindent
fact, we have
\begin{align}
\vec E\ind & = \tsr \chi\EE \h \vec E\ext = \tsr \chi\EB \h c \h \vec B\ext \,, \label{eq_nonnaive_1} \\[5pt]
c \h \vec B\ind & = \tsr \chi\BE \h \vec E\ext = \tsr \chi\BB \h c \h \vec B\ext \,, \label{eq_nonnaive_2}
\end{align}
instead of the naive expansion \eqref{eq_naive_1}--\eqref{eq_naive_2} in terms of $\vec E\ext$ {\itshape and} $\vec B\ext$\h.

We remark that the partial functional derivatives become ill-defined in the ultra-relativistic limit. The reason for this is that the partial derivatives rely on the canonical functional \eqref{eq_can_comp}, which had been constructed for retarded fields generated by sources and therefore does not apply to vacuum fields. Correspondingly, Eq.~\eqref{eq_can_comp} is formally divergent at $\omega = c|\vec k|$. By contrast, as argued in Sec.~\ref{sec_pre_total}, the functionals  \eqref{eq_AE} and \eqref{eq_AELB}, on which the physical response functions are based, are valid in the vacuum as well and can be evaluated at $\omega = c|\vec k|$.

\subsection{Non-relativistic limit} \label{subsec_nonrel}

The non-relativistic limit where $\omega \ll c|\vec k|$ is treated by approximating the scalar Green function $\mathbbmsl D_0(\vec k, \omega)$ by its zero-frequency limit,
\begin{equation} \label{eq_nrl}
\mathbbmsl D_0(\vec k,\omega=0) = \frac{\mu_0}{|\vec k|^2} \equiv \frac{v(\vec k)}{c^2} \,,
\end{equation}
while keeping the frequency dependence of the fundamental response functions. Here, $v(\vec k)$ denotes the Coulomb potential given by
\begin{equation} \label{eq_coul}
v(\vec k) \equiv v(\vec k,\omega)=\frac{1}{\varepsilon_0|\vec k|^2} \,,
\end{equation}
or in real space by
\begin{equation}
v(\vec x-\vec x', t- t')=\frac{1}{4\pi\varepsilon_0} \, \frac{\delta(c \h t-c \h t')}{|\vec x-\vec x'|} \,.
\end{equation}
Hence, in this limit relativistic retardation effects are neglected.
The non-relativistic limit is especially interesting in the case of the dielectric tensor. In a homogeneous medium and for $\omega \ll c|\vec k|$, Eq.~(\ref{eq_univ_eps}) simplifies as
\begin{equation}
 (\tsr{\varepsilon_{\mathrm r}})^{-1}(\vec k, \omega) = \tsr 1 - \frac{\j}{\omega\varepsilon_0} \tsr{P_{\mathrm L} }(\vec k) \h \tsr\sigma(\vec k, \omega) \,.
\end{equation}
Specializing to an isotropic medium, where 
\begin{equation}
\tsr\sigma(\vec k, \omega) = \sigma_{\mathrm L} (\vec k, \omega) \tsr{P_{\mathrm L} }(\vec k) + \sigma_{\mathrm T} (\vec k, \omega) \tsr{P_{\mathrm T} }(\vec k)
\end{equation}
with longitudinal and transverse conductivities $\sigma_{\mathrm L}$ and $\sigma_{\mathrm T} $, we obtain the simpler relation
\begin{equation}
 (\tsr{\varepsilon_{\mathrm r}})^{-1}(\vec k, \omega) = \tsr 1 - \frac{\j}{\omega \varepsilon_0} \, \sigma_{\mathrm L} (\vec k, \omega) \tsr{P_{\mathrm L} }(\vec k) \,.
\end{equation}
In particular, the {\itshape longitudinal dielectric constant} $\varepsilon_{\mathrm r, \mathrm L}$ is related to the longitudinal conductivity by
\begin{equation} \label{eq_long}
 \varepsilon_{\mathrm r, \mathrm L} ^{-1}(\vec k, \omega) = 1 - \frac{\j}{\omega \varepsilon_0} \, \sigma_{\mathrm L} (\vec k, \omega) \,.
\end{equation}
This is indeed a well-known identity \cite[Eq.~(6.51)]{Bruus}. 
We have shown that it follows from the general relation \eqref{eq_univ_eps} in a non-relativistic 
approximation and by restricting to homogeneous, isotropic materials.

Similarly, for the magnetic susceptibility $\chi_{\rm m}$ 
we obtain from Eq.~\eqref{eq_univ_chi} in the homogeneous and non-relativistic limit
\begin{equation}
 (\chi_{\rm m})_{ij}(\vec k, \omega) = \mu_0 \, \frac{1}{|\vec k|^4} \, \epsilon_{ikm} \h \epsilon_{j \ell n} \h k_k \h k_\ell \h \chi_{mn}(\vec k, \omega) \,.
\end{equation}
This formula agrees for $\omega = 0$ with \cite[Eq.~(AI.6)]{Takimoto}. In isotropic media, it further simplifies as
\begin{equation}
 (\chi_{\rm m})_{ij}(\vec k, \omega) = \mu_0 \, \frac{\chi_{\mathrm T} (\vec k, \omega)}{|\vec k|^2} \, \bigg( \delta_{ij} - \frac{k_i k_j}{|\vec k|^2} \bigg) \,. \medskip
\end{equation}
Since $\chi_{\rm m}$ acts on magnetic fields, which are purely transverse anyway, 
the transverse projection operator in brackets can be omitted and we arrive at
\begin{equation}\label{eq_chim_lim}
 \chi_{\rm m}(\vec k, \omega) = \mu_0 \, \frac{\chi_{\mathrm T} (\vec k, \omega)}{|\vec k|^2} \,.
\end{equation}
The two equations \eqref{eq_long} and \eqref{eq_chim_lim} can be written in an analogous way if we express the longitudinal conductivity $\sigma_{\mathrm L}$ through the longitudinal current response function $\chi_{\mathrm L}$. We then get the relations for the longitudinal permittivity and the permeability
\begin{align}
 \varepsilon_{\mathrm r, \mathrm L}^{-1}(\vec k, \omega) & = 1 - \frac{1}{\varepsilon_0 \h \omega^2} \, \chi_{\mathrm L}(\vec k, \omega) \,, \\[5pt]
 \mu_{\mathrm r}(\vec k, \omega) & = 1 + \frac{\mu_0}{|\vec k|^2} \, \chi_{\mathrm T}(\vec k, \omega) \,.
\end{align}
Further relating $\chi_{\rm L}$ to the density response function $\upchi$ via Eq.~\eqref{eq_chiEEchi} and using Eq.~\eqref{eq_coul}, we arrive at the standard formulae~\cite{Bruus, Giuliani}
\begin{align}
 \varepsilon_{\mathrm r, \mathrm L}^{-1}(\vec k, \omega) & = 1 + v(\vec k) \h \upchi(\vec k, \omega) \,, \label{eq_erste} \\[5pt]
 \mu_{\mathrm r}(\vec k, \omega) & = 1 + \frac{1}{c^2} \, v(\vec k) \h \chi_{\rm T}(\vec k, \omega) \label{eq_zweite} \,.
\end{align}
Finally, we comment on the cross-coupling coefficients in the non-relativistic limit. 
One sees immediately from Eq.~\eqref{eq_simpl_2} that $\chi_{EB}$ becomes small in this limit due to the factor $\omega / c |\vec k|$. 
By contrast, $\chi_{BE}$ becomes large because of the inverse factor in Eq.~\eqref{eq_simpl_3}. However, this does not have any physical consequences, because in the expansion
\begin{equation}
c \h \vec B\ind=\tsr\chi_{BE} \h (\vec E\ext)_{\mathrm L} + \tsr\chi_{BB} \h c \h \vec B\ext \,,
\end{equation}
this response function acts on the longitudinal electric field. As $\chi_{BE}$ contains the transverse rotation operator, this yields zero.

\subsection{Instantaneous limit} \label{subsec_inst}

As mentioned in Sec.~\ref{sec_pre_notations}, the limit $\omega \to 0$ corresponds to static (induced and external) field quantities and to instantaneous response functions. Consider again Eqs.~\eqref{eq_simpl_1}--\eqref{eq_simpl_4} for the physical response functions in the homogeneous and isotropic limit. By Faraday's law, static electric fields are 
purely longitudinal. Since $\chi\BE$ contains the transverse rotation operator, this response function effectively vanishes. For $\chi\EB$ the limit $\omega \rightarrow 0$ can be performed trivially giving also $\chi\EB=0$.
The instantaneous limit of the longitudinal dielectric constant $\varepsilon_{\mathrm r, \mathrm L}$ and the magnetic susceptibility $\chi_{\rm m}$ can be inferred from Eqs.~\eqref{eq_erste}--\eqref{eq_zweite} in the previous subsection, 
where it remains to set $\omega=0$. In particular, we recover the well-known formula for \linebreak

\pagebreak \noindent
the (orbital) magnetic susceptibility \cite[Eq.~(3.183)]{Giuliani},
\begin{equation}
\chi_{\rm m} = \mu_0 \lim_{|\vec k| \rightarrow 0} \frac{\chi_{\mathrm T} (\vec k, \omega = 0)}{|\vec k|^2} \,,
\end{equation}
where the limit $|\vec k| \rightarrow 0$ corresponds to integrating out the spatial dependence. 
We conclude that in the homogeneous, isotropic and instantaneous limit the universal response relations simplify as
\begin{align}
\tsr \chi\EE(\vec k, 0) & = v(\vec k) \, \upchi(\vec k, 0) \h \tsr P_{\mathrm L}(\vec k) \,, \label{eq_static_1} \\[6pt]
\tsr \chi\EB(\vec k, 0) & = 0 \,, \\[6pt]
\tsr \chi\BE(\vec k, 0) & = 0 \,, \\[1pt]
\tsr \chi\BB(\vec k, 0) & = \frac{1}{c^2} \, v(\vec k) \h \chi_{\mathrm T}(\vec k, 0) \h \tsr P_{\mathrm T}(\vec k) \,. \label{eq_static_4}
\end{align}
Since static electric fields are always longitudinal and magnetic fields are always transverse, we may omit the projection operators in Eqs.~\eqref{eq_static_1} and \eqref{eq_static_4} and write shorthand
\begin{align}
 \chi\EE(\vec k) & = v(\vec k) \, \upchi(\vec k) \,, \\[5pt]
 \chi\BB(\vec k) & = \frac{1}{c^2} \, v(\vec k) \h \chi_{\mathrm T}(\vec k) \,.
\end{align}
Thus, in the instantaneous limit the medium is effectively described
by only two independent scalar response functions. These relate the induced electric or magnetic field to the external electric or magnetic field respectively, while the cross couplings vanish. The reason for this is indeed simple: if the response func\-{}tion is isotropic, a static (i.e.~longitudinal) electric field can only induce a longitudinal electric field but not a transverse magnetic field, and the converse holds true for a magnetic (i.e.~transverse) perturbation.
Finally, we note that the same expressions \eqref{eq_static_1}--\eqref{eq_static_4} can be derived even more directly for the partial response functions starting from Eqs.~\eqref{eq_simpl_fsrt_1}--\eqref{eq_simpl_fsrt_4},
which means that in the instantaneous limit the total and the partial derivatives coincide. This is again consistent with the fact that static electric fields are purely longitudinal and hence decouple completely from the transverse magnetic fields.

\section{Conclusion}

The Functional Approach developed in this paper disentangles electrodynamics of media from macroscopic electrodynamics: while the latter corresponds to spatial and/or temporal averaging of the fields (which can be performed with or without media), electrodynamics of media is set up by regarding induced quantities as functionals of external perturbations (on the microscopic or macroscopic scale). In terms of the first order functional derivatives, our approach gives a model- and material-independent, 
relativistic account of electromagnetic response functions that fully incorporates the effects of retardation,
inhomogeneity, anisotropy and magnetoelectric coupling. Our core findings are:

\medskip \noindent
\begin{enumerate}
 \setlength{\itemsep}{1em}
 \item A generalized expression for the Green function of the electromagnetic four-potential, Eq.~\eqref{eq_gen_GF_dimless}.
 \item The connection between classical electrodynamics of media and the Schwinger--Dyson or Hedin equations (Sec.~\ref{subsec_hedin}).
 \item The explicit formulae \eqref{eq_lor_1}--\eqref{eq_lor_4}, by which one can calculate the 36 components of the field strength response tensor in terms of only 9 independent components of the fundamental response tensor. 
 \item The identification of physical response functions with {\itshape total} functional derivatives (Sections~\ref{sec_pre_total} and \ref{subsec_physical}).
 \item The universal relations between the physical response functions, cf. Eqs.~\eqref{eq_univ_1}--\eqref{eq_univ_4} together with \eqref{eq_magnsusc}--\eqref{eq_permeability}. 
 \item The existence of different but equivalent field expansions in terms of partial or total functional derivatives (Sec.~\ref{sec_comp}).
 \item The rederivation of well-known response relations as limiting cases of the universal response relations (Sec.~\ref{sec_emp}).
\end{enumerate}

\medskip \noindent
The Functional Approach to electrodynamics of materials is exclusively based on the Maxwell equations and the interpretation of induced fields as functionals of the external perturbations.
From this, all our conclusions and formulae follow analytically. As the Functional Approach is free of any assumption about the medium, it is particularly useful for studying conceptual issues, such as general conditions for magnetoelectric coupling. On the other hand, the Functional Approach is
computationally based on the explicit expression of all linear electromagnetic material properties in terms of the fundamental response tensor. In microscopic calculations, this fundamental response tensor is directly
accessible from the four-point Green function through the Kubo formalism~\cite{Bruus, Giuliani}. Since typical first-principles methods such as the Hartree-Fock
or the $GW$ approximation can be grouped into a hierarchy of approximations for this four-point Green function \cite{Starke}, 
the Functional Approach is ideally suited for the ab initio calculation of electromagnetic material properties.

\bigskip

\section*{Acknowledgements}
This work was supported by the Austrian Science Fund (FWF) within the SFB ViCoM, Grant No.~F41, and by the DFG Reseach Unit FOR 723. We are grateful to Manfred Salmhofer for valuable comments on an earlier draft of this article. We further thank Alex Gottlieb, Styles Sass and Heinrich-Gregor Zirnstein for helpful discussions and feedback on the manuscript. R.\,S.~thanks the Institute for Theoretical Physics at the TU Bergakademie Freiberg for its hospitality.

%% The Appendices part is started with the command \appendix;
%% appendix sections are then done as normal sections
%% \appendix

%% \section{}
%% \label{}

%% References
%%
%% Following citation commands can be used in the body text:
%% Usage of \cite is as follows:
%% \cite{key} = = >> [#]
%% \cite[chap. 2]{key} = = >> [#, chap. 2]
%% \citet{key} = = >> Author [#]

%% References with bibTeX database:

\bibliographystyle{model1-num-names}
\bibliography{/net/home/lxtsfs1/tpc/schober/Ronald/masterbib}

%% Authors are advised to submit their bibtex database files. They are
%% requested to list a bibtex style file in the manuscript if they do
%% not want to use model1-num-names.bst.

%% References without bibTeX database:

% \begin{thebibliography}{00}

%% \bibitem must have the following form:
%% \bibitem{key}...
%%

% \bibitem{}

% \end{thebibliography}

\end{document}